\documentclass[twocolumn,prd,aps,nofootinbib,showpacs,superscriptaddress]{revtex4}
\usepackage{multirow}
\usepackage{graphicx}
\usepackage[dvips]{color}
\usepackage{amssymb,amsmath}

\begin{document}

\title{Lattice QCD study of the $s$-wave $\pi\pi $ scattering lengths
in the $I=0$ and $2$ channels}

\author{Ziwen Fu }
\email{fuziwen@scu.edu.cn}
\affiliation{
Key Laboratory for Radiation Physics and Technology of  Education Ministry;
Institute of Nuclear Science and Technology, College of Physical Science and Technology,
Sichuan University, Chengdu 610064, People's Republic of China
}
\date{\today}

\begin{abstract}
The $s$-wave pion-pion ($\pi\pi$) scattering lengths
are computed below the inelastic threshold by the L\"uscher technique
with pion masses ranging from $240$~MeV to $463$~MeV.
In the Asqtad-improved staggered fermion formulation,
we calculate the $\pi\pi$ four-point functions
for the $I=0$ and $2$ channels with ``moving'' wall sources
without gauge fixing, and analyze them
at the next-to-leading order
in the continuum three-flavor chiral perturbation theory.
At the physical pion mass, we secure the $s$-wave $\pi\pi$
scattering lengths as
$m_\pi a_{\pi\pi}^{I=0} =  0.214(4)(7)$ and
$m_\pi a_{\pi\pi}^{I=2} = -0.04430(25)(40)$
for the $I=0$ and $2$ channels, respectively,
where the first uncertainties are statistical
and second ones are our estimates of several systematic effects.
Our lattice results for the $s$-wave $\pi\pi$ scattering lengths
are in well accordance with available experimental reports
and theoretical forecasts at low momentum.
A basic ingredient in our study for the $I=0$ case is
properly incorporating disconnected diagram.
These lattice computations are carried out with the MILC $2+1$ flavor
gauge configurations at two lattice spacings $a \approx 0.15$ and $0.12$~fm.
\end{abstract}

\pacs{12.38.Gc, 11.15.Ha}

\maketitle

\section{Introduction}
The research on the $\pi\pi$ scattering is a basic and classical subject
in the field of strong hadronic interactions.
Its handleability and simplicity essentially stem from
the pseudo Nambu-Goldstone boson nature of pion,
a natural aftermath of the spontaneous chiral symmetry breaking
in quantum chromodynamics (QCD),
which imposes rigid  constraints on the $\pi\pi$  low-energy interactions.
Moreover, the $s$-wave $\pi\pi$ scattering lengths vanish 
in the chiral limit when the momentum of the pions approaches zero.
Since these quantities stand for a sensitive probe of
the chiral symmetry breaking generated by the quark masses,
the lattice QCD study, an objective of this paper,
is a non-perturbative method in an effort to comprehend
the low-energy nature of QCD.

With small pion masses and low-momenta,
the $s$-wave $\pi\pi$ scattering lengths
can be solely predicted at leading order (LO)
in chiral perturbation theory ($\chi$PT)~\cite{Weinberg:1966kf}.
The next-to-leading order (NLO) and next-to-next-to-leading order (NNLO)
corrections in the chiral expansion~\cite{Bijnens:1997vq,
Colangelo:2000jc,Colangelo:2001df} lead to perturbative
deviations from the LO  and
involve both computable nonanalytic contributions and
analytic terms with some unknown low-energy constants (LEC's),
which can be obtained from lattice simulations or experimental measurements.

A combination of some experimental and theoretical inputs from
CGL~\cite{Colangelo:2000jc,Colangelo:2001df}, along with the Roy-equation~\cite{Roy:1971tc,Ananthanarayan:2000ht},
produced a precise result of the $s$-wave $\pi\pi$ scattering lengths.
Zhou et al. studied the pole structure
of the low-energy $\pi\pi$ scattering amplitudes
using a proper chiral unitarization method
in addition to the crossing symmetry and low-energy phase shift data,
and estimated the $s$-wave $\pi\pi$ scattering lengths~\cite{Zhou:2004ms}.
K. Sasaki and N. Ishizuka found that the scattering phase
can be obtained
directly from the $\pi\pi$ wave function~\cite{Sasaki:2008sv}.
Guo et al. provided a reliable and solid estimation
of all part of the ${\cal{O}}(p^6)$ calculation~\cite{Guo:2009hi},
and some resonance contributions were added in
to the former phenomenological calculations~\cite{Colangelo:2000jc,Colangelo:2001df},
and obtained the slight differences with respect to
previous results in Refs.~\cite{Colangelo:2000jc,Colangelo:2001df}.
Using the NLO $SU(2)$ unitary chiral perturbation theory to
examine the $\pi\pi$ scattering, Albaladejo and Oller
obtained a good reproduction of the $s$-wave $\pi\pi$ scattering
lengths~\cite{Albaladejo:2012te}.

In conjunction with the strict $\chi$PT constraints in the analysis,
the considerably improved accuracy for the $s$-wave $\pi\pi$ scattering lengths
has been obtained from the experimental measurement of the semileptonic $K_{e4}$
decay by E865~\cite{Pislak:2003sv}.
With the independent experimental uncertainties and different theoretical
inputs~\cite{Colangelo:2001df,Colangelo:2000jc,GarciaMartin:2011cn},
the NA48/2 high-precision analyses of the $K_{e4}$ and
$K_{3\pi}$ decays~\cite{Batley:2007zz,Batley:2010zza,Bizzeti:2011zza,Wanke:2011zz}
gave rise to the complementary information on the $s$-wave $\pi\pi$ scattering lengths~\cite{Bizzeti:2011zza}.
All of these theoretical (or phenomenological) predictions and
experimental determinations are consistent with each other.

Lattice studies on the $\pi\pi$ scattering have been conducted
in quenched QCD by various groups~\cite{Sharpe:1992pp,Kuramashi:1993ka,
Fukugita:1994ve,Gupta:1993rn,Li:2007ey,Aoki:2002in,Du:2004ib}.
The full lattice study of the $s$-wave $I=2$ $\pi\pi$ scattering length
was first carried out by CP-PACS~\cite{Yamazaki:2004qb}.
Fully-dynamical computation of the $I=2$ $\pi\pi$ scattering
was explored by NPLQCD with the domain-wall valence quarks on
a fourth-rooted staggered sea quarks~\cite{Beane:2005rj,Beane:2007xs}.
Using the $N_f=2$ maximally twisted mass fermion ensembles,
Xu et al. employed the lightest pion mass at that time,
conducted an explicit check for lattice artifacts~\cite{Feng:2009ij}.
With an anisotropic $N_f = 2 +1$ clover fermion discretization,
the $I=2$ $\pi\pi$ scattering phase shift is calculated
by NPLQCD to determine all the threshold parameters~\cite{Beane:2011sc}.
Moreover, efforts were made to first secure the $d$-wave
$I = 2$ $\pi\pi$  phase shift in some nice works by HSC~\cite{Dudek:2010ew,Dudek:2012gj}.
Using overlap fermion formulation, Yagi et al.
examined the consistency of the lattice data with the NNLO $\chi$PT prediction
after correcting finite volume effects~\cite{Yagi:2011jn}.

Nevertheless, only a couple of lattice studies in the $I=0$ channel
are reported so far, whose computations are hindered by the so-called
``disconnected diagram''.
Using the quenched approximation, Kuramashi et~al.
carried out the pioneering work for isospin-$0$,
however, the vacuum diagram was disregarded
assuming that vacuum amplitude remains small
for large $t$~\cite{Kuramashi:1993ka}.
Additionally, for the rectangular and vacuum diagrams,
quark loops are required to make the scattering amplitudes unitary,
otherwise, the basic part of the physics is lost due to quenched approximation~\cite{Sharpe:1992pp}.
Liu conducted the first full QCD calculation for the $I=0$  channel
including the vacuum graph, however the error of
the extracted scattering length is remarkably large
due to the usage of big pion masses (small one is $430$~MeV)~\cite{Liu:2009uw}.
With the presence of the vacuum diagram,
we have attempted to crudely calculate the $\pi\pi$ scattering for isospin-$0$,
and made a first lattice calculation for ${l_{\pi\pi}^{I=0}}(\mu)$,
which is a LEC appearing in the $\chi$PT expression
of the $\pi\pi$ scattering length for isospin-$0$~\cite{Fu:2011bz}.
Nonetheless, we used the partially quenched QCD to save computational cost,
and worked with large quark masses~\cite{Fu:2011bz}.
Moreover, the statistical errors are underestimated
since we only considered the primary one~\cite{Fu:2011bz}.
Furthermore, we neglected the obvious oscillating term
due to the staggered scheme.
We understood that the statistical errors for the ratio of vacuum amplitude
grow  as $\displaystyle e^{2m_\pi t}$~\cite{Lepage:1989hd}.
Consequently, using the small quark mass is very important for the $I=0$ channel.
As presented later, our lattice results will indeed quantitatively
confirm this argument, and we acquire the good signals of
vacuum diagram for the lattice ensembles
with small pion masses.

To overcome the Maiani-Testa theorem~\cite{Maiani:1990ca},
people usually calculate the energy levels of two-(many-)particle system
enclosed in a torus, and its scattering amplitudes can be recovered~\cite{Luscher:1986pf,Luscher:1990ux,Luscher:1990ck,Beane:2003da,
Rummukainen:1995vs,Kim:2005gf,Christ:2005gi,Feng:2004ua,Lang:2011mn,
Fu:2011xz,Leskovec:2012gb,Doring:2012eu,Briceno:2012rv,Guo:2012hv}.
In this work, L\"uscher's technique~\cite{Luscher:1986pf,Luscher:1990ux,Luscher:1990ck}
is employed to extract the scattering phase shift
with the lattice-calculated energy eigenstates.

We here use the MILC gauge configurations~\cite{Aubin:2004wf,Bernard:2001av}
with the $2+1$ flavors of the
Asqtad-improved staggered dynamical quarks~\cite{stag_fermion}
to compute the $s$-wave $\pi\pi$ scattering lengths.
The technique of the ``moving'' wall source without gauge fixing~\cite{Fu:2011wc}
first introduced in Refs.~\cite{Kuramashi:1993ka,Fukugita:1994ve}
are exploited to calculate all the four diagrams classified
in Refs.~\cite{Sharpe:1992pp,Kuramashi:1993ka,Fukugita:1994ve},
and special effort is payed to the disconnect diagram.
Our lightest pion mass is about $240$~MeV, which is
lighter than those of the former lattice studies on the $\pi\pi$ scattering
and enables us to further explore the chiral limit.
Consequently, the signals of vacuum diagram are remarkably improved.
Moreover, due to the nature of staggered fermion,
our computations are automatically precise to ${\cal{O}}(a^2)$~\cite{Sharpe:1992pp}.
Additionally, we used the continuum three-flavor $\chi$PT at NLO
to extrapolate our lattice-measured $\pi\pi$ scattering lengths to the physical point.
As presented later, we find
\begin{equation}
m_\pi a_{\pi\pi}^{I=2} = -0.04430(25)(40); \quad
{l_{\pi\pi}^{I=2}}     =  3.27(.77)(1.12) ,
\nonumber
\end{equation}
where $a_{\pi\pi}^{I=2}$ denote the $s$-wave $\pi\pi$ scattering lengths
in the $I=2$  channel and
${l_{\pi\pi}^{I=2}}(\mu)$ is a LEC evaluated
at the physical pion decay constant.
These results are in well agreement with the experimental measurements and
theoretical (or phenomenological) determinations as well as
previous lattice calculations.
Most of all, we obtain
\begin{equation}
m_\pi a_{\pi\pi}^{I=0} = 0.214(4)(7);  \quad
{l_{\pi\pi}^{I=0}}     = 43.2(3.5)(5.6),
\nonumber
\end{equation}
which are in fair accordance with the experimental reports
and theoretical (or phenomenological) predictions,
and significantly improve our former study in Ref.~\cite{Fu:2011bz}.

The paper is organized as follows.
In Sec.~\ref{sec:Methods} we will review
the basic formalism for the calculation of the $s$-wave $\pi\pi$ scattering~\cite{Luscher:1990ux,Luscher:1990ck}.
The simulation parameters and our concrete lattice calculations
are shown in Sec.~\ref{sec:latticeCal}.
We will give the results of the lattice simulation data
in Sec.~\ref{sec:Results}, fitting analyses in Sec.~\ref{sec:fittingResults},
and chiral extrapolation along with the comparisons of different results in Sec.~\ref{sec:chiralExtrapolation}.
Finally, a summary of our conclusions and outlooks
are arrived at in Sec.~\ref{sec:conclude}.
The compact continuum three-flavor $\chi$PT forms at NNLO
for the $\pi\pi$ scattering lengths
are courteously dedicated in Appendix~\ref{app:ChPT NNLO}.

\section{Method}
\label{sec:Methods}
On the basis of the original  derivations and notations
in Refs.~\cite{Sharpe:1992pp,Kuramashi:1993ka,Fukugita:1994ve},
we reviewed the indispensable formulas for the lattice QCD evaluation
of the $s$-wave $\pi\pi$ scattering lengths in a torus.
The formulas and the notations adopted here are actually the same as
those in Refs.~\cite{Fu:2011bz,Fu:2012gf}.
But, to make this paper self-supporting,
all the essential parts will be reiterated subsequently.

Let us review scattering of two Nambu-Goldstone pions 
in the Asqtad-improved staggered fermion formalism.
For the $s$-wave $\pi\pi$ scattering, only the isospin $I=0$ and
$2$ channels are permitted owing to Bose symmetry.
We build these $\pi\pi$ isospin eigenchannels
using the following interpolating operators~\cite{Kuramashi:1993ka,Fukugita:1994ve}
\begin{eqnarray}
\label{EQ:op_pipi}
 {\cal O}_{\pi\pi}^{I=0} (t) &=&  \frac{1}{\sqrt{3}}
     \Bigl\{  \pi^{-}(t) \pi^{+}(t+1) + \pi^{+}(t) \pi^{-}(t+1) - \cr
              &&\pi^{0}(t) \pi^{0}(t+1) \Bigl\} , \cr
 {\cal O}_{\pi\pi}^{I=2} (t) &=& \pi^{+}(t) \pi^{+}(t+1) ,
\end{eqnarray}
with the interpolating pion operators denoted by
\begin{eqnarray}
{\pi^+}(t) &=& -\sum_{\bf{x}} \bar{d}({\bf{x}}, t)\gamma_5 u({\bf{x}},t) , \cr
{\pi^-}(t) &=& \sum_{\bf{x}} \bar{u}({\bf{x}}, t)\gamma_5 d({\bf{x}},t), \cr
{\pi^0}(t) &=&
\frac{1}{\sqrt{2}}\sum_{\bf{x}} [
\bar{u}({\bf{x}},t)\gamma_5 u({\bf{x}},t) -
\bar{d}({\bf{x}},t)\gamma_5 d({\bf{x}},t) ] , \nonumber
\end{eqnarray}
then we express the $\pi\pi$ four-point function in the zero momentum
state as
\begin{eqnarray}
C_{\pi\pi}(t_1,t_2,t_3,t_4) \hspace{-0.2cm}&=&\hspace{-0.2cm}
\sum_{{\bf{x}}_1}\sum_{{\bf{x}}_2}\sum_{{\bf{x}}_3}\sum_{{\bf{x}}_4}
\langle
{\cal O}_{\pi}({\bf{x}}_4, t_4)
{\cal O}_{\pi}({\bf{x}}_3, t_3) \cr
\hspace{-0.2cm}&&\hspace{-0.2cm}\times \,
{\cal O}_{\pi}^{\dag}({\bf{x}}_2, t_2)
{\cal O}_{\pi}^{\dag}({\bf{x}}_1, t_1) \rangle, \nonumber
\end{eqnarray}
where, to prevent the intricate color Fierz rearrangement of the
quark lines~\cite{Kuramashi:1993ka,Fukugita:1994ve},~\footnote{
Fierz contributions force us to overcome the obstacle due to the
staggered-flavor symmetry breaking~\cite{Sharpe:1992pp}.
The same problem is also met for the $\pi K$ scattering,
which is addressed by Lang et al. in Ref.~\cite{Lang:2012sv}.
In principle, they can be disentangled by the way
discussed in Ref.~\cite{Sharpe:1992pp},
but strenuous  in practice.
Fortunately, it can be trivially handled by the way introduced
in Refs.~\cite{Kuramashi:1993ka,Fukugita:1994ve}, i.e.,
$\pi$ meson operators are separated by a unit time slice.
}
we familiarly select $t_1 =0$, $t_2=1$, $t_3=t$,
and $t_4 = t+1$~\cite{Kuramashi:1993ka,Fukugita:1994ve}.

\begin{figure}[h!]
\includegraphics[width=8.0cm,clip]{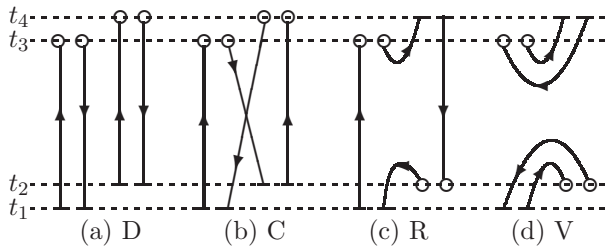}
\vspace{-0.20cm}
\caption{ \label{fig:diagram}
Quark-line diagrams contributing to $\pi\pi$ four-point functions.
Short bars indicate the wall sources.
The wall sinks for local pion operators are represented by open circles.
}
\end{figure}

In the isospin limit, only four diagrams contribute to
$\pi\pi$ scattering amplitudes,
in Fig.~\ref{fig:diagram}, we show these quark-line diagrams,
which are identified as direct ($D$), crossed ($C$), rectangular ($R$),
and vacuum ($V$) diagrams, respectively~\cite{Kuramashi:1993ka,Fukugita:1994ve}.
The reliable evaluation of the rectangular diagram is
challenging and the rigorous computation
of the vacuum diagram is pretty hard~\cite{Kuramashi:1993ka,Fukugita:1994ve}.

In our former works~\cite{Fu:2011bz,Fu:2012gf},
we calculated these four diagrams via
evaluating $T$ quark propagators~\cite{Kuramashi:1993ka,Fukugita:1994ve}
$$
\sum_{x} D_{n,x}G_t(x) = \sum_{\mathbf{x}}
\delta_{n,({\mathbf{x}},t)}, \quad 0 \leq t \leq T-1.
$$
The combination of $G_t(n)$ which we apply for the $\pi\pi$ four-point
functions is schematically illustrated in Fig.~\ref{fig:diagram},
and these diagrams can be described
by means of $G$,
\begin{eqnarray}
\label{eq:dcr}
C^D(t_1,t_2,t_3,t_4) &=&
\sum_{ {\bf{x}}_3}\sum_{ {\bf{x}}_4 } \langle \mbox{Tr}
[G_{t_1}^{\dag}({\bf{x}}_3, t_3) G_{t_1}({\bf{x}}_3, t_3)] \cr
&&\times \,
\mbox{Tr} [G_{t_2}^{\dag}({\bf{x}}_4, t_4) G_{t_2}({\bf{x}}_4, t_4) ] \rangle,\cr
C^C(t_1,t_2,t_3,t_4) &=&
\sum_{ {\bf{x}}_3}\sum_{{\bf{x}}_4 } \langle \mbox{Tr}
[G_{t_1}^{\dag}({\bf{x}}_3, t_3) G_{t_2}({\bf{x}}_3, t_3) \cr
&& \times \, G_{t_2}^{\dag}({\bf{x}}_4, t_4) G_{t_1}({\bf{x}}_4, t_4) ] \rangle,\cr
C^{R}(t_1,t_2,t_3,t_4) &=&  \sum_{ {\bf{x}}_2,{\bf{x}}_3 }
\langle \mbox{Tr}
[G_{t_1}^{\dag}({\bf{x}}_2, t_2) G_{t_4}({\bf{x}}_2, t_2) \cr
&& \times \, G_{t_4}^{\dag}({\bf{x}}_3, t_3) G_{t_1}({\bf{x}}_3, t_3) ]
\rangle, \cr
C^V(t_1,t_2,t_3,t_4) &=&
\sum_{ {\bf{x}}_2}\sum_{ {\bf{x}}_3}  \bigg\{ \langle \mbox{Tr}
[G_{t_1}^{\dag}({\bf{x}}_2, t_2) G_{t_1}({\bf{x}}_2, t_2)] \cr
&&\times \,
\mbox{Tr} [G_{t_4}^{\dag}({\bf{x}}_3, t_3) G_{t_4}({\bf{x}}_3, t_3) ] \rangle \cr
&& - \, \langle \mbox{Tr}
[G_{t_1}^{\dag}({\bf{x}}_2, t_2) G_{t_1}({\bf{x}}_2, t_2) \rangle \cr
&& \times \, \langle \mbox{Tr}
[G_{t_4}^{\dag}({\bf{x}}_3, t_3) G_{t_4}({\bf{x}}_3, t_3) ]
\rangle\bigg\} ,
\end{eqnarray}
where the indicated traces are conducted over color
and the $\gamma^5$ factors are neatly removed
using the Hermiticity attributes of the propagator $G$,
and a vacuum deduction is a natural companion to
the vacuum diagram~\cite{Blum:2011pu}.

The rectangular and vacuum diagrams inevitably
create the gauge-variant noise~\cite{Kuramashi:1993ka,Fukugita:1994ve},
which are neatly diminished by executing the gauge field average
without gauge fixing as we practiced
in Refs.~\cite{Fu:2011xw,Fu:2012gf,Fu:2011wc,Fu:2011bz,Fu:2012tj,Fu:2012ng,Fu:2012jb}.
In the isospin limit, the $\pi\pi$ four-point functions
for the $I=0$ and $2$ channels can be expressed on the strength of four diagrams~\cite{Sharpe:1992pp,Kuramashi:1993ka,Fukugita:1994ve},
\begin{eqnarray}
\label{EQ:phy_I0_2}
C_{\pi\pi}^{I=0}(t) \hspace{-0.15cm}&\equiv&\hspace{-0.15cm}
\left\langle {\cal O}_{\pi\pi}^{I=0} (t) | {\cal O}_{\pi\pi}^{I=0} (0)
\right\rangle  \hspace{-0.1cm}=\hspace{-0.1cm}
D + \frac{N_f}{2} C - 3N_f R + \frac{3}{2}V , \cr
C_{\pi\pi}^{I=2}(t) \hspace{-0.15cm}&\equiv&\hspace{-0.15cm}
\left\langle {\cal O}_{\pi\pi}^{I=2} (t) | {\cal O}_{\pi\pi}^{I=2} (0) \right\rangle
\hspace{-0.1cm}=\hspace{-0.1cm}
D - N_f C ,
\end{eqnarray}
where the staggered-flavor factor $N_f$ is inserted to rectify for
the extra factor $N_f$ in the valence fermion loops~\cite{Sharpe:1992pp}.
The four-fold degeneracy of the staggered sea quarks
is removed by conducting the quadruple root of the fermion determinant~\cite{Sharpe:1992pp,DeGrand:2006zz}.
The fourth-root recipeis assumed to be able to
restore the right continuum limit of QCD~\cite{DeGrand:2006zz},
and our results rest on this hypothesis.
See Ref.~\cite{Durr:2004as}
for more discussions about the fourth-root trick.

It is customary to make use of the effective range expansion
for parameterizing the low-momentum
behavior of the $s$-wave $\pi\pi$ scattering phase $\delta_0$,
\begin{equation}
\label{eq:exact}
k  \cot \delta_0(k) = \frac{1}{a} + \frac{1}{2} r k^2 + {\cal O}(k^4) ,
\end{equation}
where $a$ is the $s$-wave $\pi\pi$  scattering length,
$r$ is the effective range,
and $k$ is the magnitude of the center-of-mass scattering momentum
related to the energy $E_{\pi\pi}^I$ of the $\pi\pi$ system
with total isospin $I$ in a torus of size $L$  by
\begin{equation}
\label{eq:MF_k_e}
k^2 = \frac{1}{4} \left( E_{\pi\pi}^I \right)^2 - m_\pi^2,
\quad k = \frac{2\pi}{L} q ,
\end{equation}
here the dimensionless momentum $q \in \mathbb{R}$.
The $s$-wave $\pi\pi$ scattering length
in the continuum limit is denoted by
$$
a_0 = \lim_{k\to 0} \frac{\tan\delta_0(k)}{k} ,
$$
which is purely elastic below inelastic thresholds.~\footnote{
We are only interested in the elastic region:
$2m_\pi < E_{\pi\pi}^I < 4m_\pi$,
where there is no $4\pi$ channel,
and not up to the opening of the
$K\overline{K}$ channel at around $1$~GeV yet~\cite{Albaladejo:2008qa},
where the $K\overline{K}$ channel
contributes remarkably  to the isoscalar $\pi\pi$ interactions~\cite{Hanhart:2012wi}.
}
We should keep in mind that the truncation of the effective range $r$
in Eq.~(\ref{eq:exact}) is considered as an important source
of systematic error, which appears as ${\cal O}(1/L^6)$.
The $\delta_0(k)$ can be computed by the
L\"uscher formula~\cite{Luscher:1990ux,Luscher:1990ck},
\begin{equation}
k \cot\delta_0(k) =
\frac{2\pi}{L} \pi^{-3/2} {\mathcal Z}_{00}(1, q^2),
\label{eq:luscher}
\end{equation}
where  the dimensionless momentum $q=kL/(2\pi)$
and the zeta function $\mathcal{Z}_{00}(1;q^2)$ is
formally expressed by
\begin{equation}
\label{eq:Z00d}
\mathcal{Z}_{00}(1;q^2) = \frac{1}{\sqrt{4\pi}}
\sum_{{\mathbf n}\in\mathbb{Z}^3} \frac{1}{n^2-q^2} .
\end{equation}
We generally compute the zeta function
by the method discussed in Ref.~\cite{Yamazaki:2004qb}.
Recently, an equivalent formula is established~\cite{Doring:2011vk}.
It allows us to avoid the subthreshold
singularities inherent to L\"uscher technique~\cite{Doring:2011vk}.

The energy $E_{\pi\pi}^I$ can be secured from the $\pi\pi$ four-point function
which manifests as~\cite{Barkai:1985gy}
\begin{eqnarray}
\label{eq:E_pionpion}
\hspace{-0.7cm} C_{\pi\pi}^I(t)  &=&
Z_{\pi\pi}\cosh\left[E_{\pi\pi}^I\left(t - \tfrac{1}{2}T\right)\right] \cr
&& +
(-1)^t Z_{\pi\pi}^{\prime}\cosh \left[E_{\pi\pi}^{I \prime} \left(t-\tfrac{1}{2}T\right)\right] + \cdots.
\end{eqnarray}
for a large $t$ to reduce the excited states,
the terms alternating in sign is a representative feature
of a staggered scheme~\cite{Barkai:1985gy},
and the ellipsis indicates the contributions from the excited states
which are suppressed exponentially.
In practice, the pollution
due to the ``wraparound'' effects~\cite{Gupta:1993rn,Umeda:2007hy,Feng:2009ij}
should be taken into account.

It should be worth to stress that,
even if we project onto the Goldstone pions at source and sink timeslices,
pions with all $16$ staggered-flavors still emerge at the
intermediate times~\cite{Sharpe:1992pp}.
However, in large $t$,
the contributions of non-Goldstone pions in the intermediate states
is exponentially reduced due to their heavier masses
in contrast with those of the Goldstone
pions~\cite{Sharpe:1992pp,Kuramashi:1993ka,Fukugita:1994ve}.

In practice, for the sake of a more intuitive presentation of our results,
we compute the ratios~\footnote{
In principle, when $t \ll T/2$, even if placing
the periodic boundary condition in the temporal direction,
the energy shift of the $\pi\pi$ system can be still roughly
evaluated from these ratios.
In this work we do not use these ratios to
quantitatively calculate any physical quantities,
nonetheless, these ratios will indeed help us comprehend qualitatively
or intuitively some physical quantities.
}
\begin{equation}
\label{EQ:ratio}
R^X(t) = \frac{ C_{\pi\pi}^X(0,1,t,t+1) }
{ C_\pi (0,t) C_\pi(1,t+1) },
\quad  X = D, C, R,  {\rm and} \, V,
\end{equation}
where $C_\pi (0,t)$ and $C_\pi (1,t+1)$ are pion correlators with zero momentum.
With the consideration of Eq.~(\ref{EQ:phy_I0_2}),
we can depict the $\pi\pi$ scattering  amplitudes which project out
the $I=0$ and $2$ isospin eigenstates as
\begin{eqnarray}
\label{EQ:proj_I0I2}
R_{I=0}(t) &=&
R^D(t) + \frac{1}{2} N_f R^C(t) - 3N_f R^R(t) + \frac{3}{2}R^V(t), \cr
R_{I=2}(t) &=&
R^D(t) - N_f R^C(t) .
\end{eqnarray}

\begin{table*}[htb]
\caption{\label{tab:MILC_configs}
The parameters of MILC gauge configurations used in the present work.
The lattice dimensions are expressed in lattice units in the second block
with spatial ($L$) and temporal ($T$) size.
The gauge coupling  $\beta = 10/g^2$ is shown in Column three.
The fourth and fifth blocks give the bare masses of the light and
strange quark masses in terms of $am_l$ and $am_s$, respectively.
Tadpole-improvement factor $u_0$ is listed in Column six.
The ratio $r_1/a$ is provided in Column seven
(see Ref.~\cite{Bernard:2000gd} for the MILC definition of $r_1$).
The inverse lattice spacing $a^{-1}$ is recapitulated in Column eight
(for the ($0.00484, 0.0484$) ensemble, we obtain the value of $r_1/a$
from Ref.~\cite{Bazavov:2009bb}, then calculate $a^{-1}$).
In the last column the number of gauge configurations  is given.
}
\begin{ruledtabular}
\begin{tabular}{llllllllll}
Ensemble &$L \times T$ &$\beta$ & $a m_l$ &  $a m_s$
& $u_0$ &$r_1/a$ & $a^{-1}{\rm GeV}$ & $N_{cf}$  \\
\hline
\multicolumn {9}{c}{$a \approx 0.12$ fm}        \\
2464f21b676m005m050    & $24^3\times64$ & $6.76$ &$0.005$  &$0.050$ &
$0.8678$& $2.647(3)$ & $1.679^{+43}_{-16}$ & 156 \\
2064f21b676m007m050    & $20^3\times64$ & $6.76$ &$0.007$  &$0.050$ &
$0.8678$& $2.635(3)$ & $1.672^{+43}_{-16}$ & 200 \\
2064f21b676m010m050    & $20^3\times64$  &$6.76$ &$0.010$  &$0.050$ &
$0.8677$& $2.619(3)$ & $1.663^{+43}_{-16}$ & 200 \\
\hline
\multicolumn {9}{c}{$a \approx 0.15$ fm}        \\
2048f21b6566m00484m0484& $20^3\times48$  &$6.566$&$0.00484$&$0.0484$&
$0.8602$& $2.162(5)$ & $1.373^{+34}_{-14}$ & 560 \\
1648f21b6572m0097m0484 & $16^3\times48$  &$6.572$&$0.0097$ &$0.0484$&
$0.8604$& $2.140(4)$ & $1.358^{+35}_{-13}$ & 250 \\
1648f21b6586m0194m0484 & $16^3\times48$  &$6.586$&$0.0194$ &$0.0484$&
$0.8609$& $2.129(3)$ & $1.352^{+35}_{-13}$ & 200 \\
\end{tabular}
\end{ruledtabular}
\end{table*}

In this work, we will employ two approaches to
calculate the pion mass $m_\pi$.
The first method is to use both the point-source and point-sink operator.
Nevertheless, the point operator has big overlap with excited states~\cite{Aubin:2004fs},
and in practice it is customary to use the wall-source operator
which efficiently reduces these overlaps, along with a point-sink~\cite{Aubin:2004fs}.
In addition we need both propagators to calculate pion
decay constant~\cite{Bernard:2001av,Aubin:2004wf}.

\section{Lattice calculation}
\label{sec:latticeCal}
We used the MILC gauge configurations~\cite{Aubin:2004wf,Bernard:2001av}
with the $2+1$ flavors of asqtad-improved
staggered sea quarks~\cite{stag_fermion} and a Symanzik-improved gluon action~\cite{Alford:1995hw}.
See detailed simulation parameters in Ref.~\cite{Bazavov:2009bb}.
It is worth mentioning that the MILC gauge configurations
are generated using the staggered formulation of
lattice fermions~\cite{Kaplan:1992bt}
with the fourth-root of fermion determinant
which are hypercubic-smeared (HYP-smeared)~\cite{Hasenfratz:2001hp}.
As shown in Ref.~\cite{Renner:2004ck},
the chiral symmetry are significantly enhanced
via the HYP-smeared gauge link.

The lattice simulation parameters of the MILC gauge configurations
used here are epitomized in Table~\ref{tab:MILC_configs}.
The simulated bare masses of light and strange sea quarks are
denoted by $am_l$ and $am_s$, respectively.
The masses of the $u$ and $d$ quarks are degenerate,
which are small enough, such that the physical up- and down-quark masses
can be attained by the chiral extrapolation.
The lattice spacing $a$ for first three lattice ensembles is about $0.12$~fm,
and that of last three lattice ensembles is around $0.15$~fm.
By MILC convention, the lattice ensembles are
referred to as ``coarse'' for $a\approx0.12$~fm,
and ``medium-coarse'' for $a\approx0.15$~fm.
For easy notation, it is convenient to use $(am_l, am_s)$ to
mark lattice ensembles,
e.g., ``the (0.01, 0.05) ensemble''.
The tadpole factors $u_0$~\cite{Lepage:1992xa} are utilized to enhance
the gauge configuration action~\cite{Bernard:2001av,Aubin:2004wf}.

To compute the $\pi\pi$ four-point functions $C_{\pi\pi}(t)$,
the standard conjugate gradient technique~\footnote{
The conjugate gradient residual used in this work
is $1.0\times 10^{-5}$, which is smaller than that used in generating
gauge configurations~\cite{Bernard:2001av,Aubin:2004wf}.
Moreover, to avoid the potential roundoff errors as much as possible,
all the numerical calculations are calculated in double precision.
} is used to acquire
the required matrix element of inverse fermion matrix.
The periodic boundary condition is applied to the three spatial directions
and temporal direction.
We compute $C_{\pi\pi}(t)$ on all the possible time slices,
and collect them at the end of the measurement,
namely,
$$
 C_{\pi\pi}(t) = \frac{1}{T}\sum_{t_s=0}^{T-1} \langle
\left(\pi\pi\right)(t+t_s)\left(\pi\pi\right)^\dag(t_s) \rangle .
$$
After averaging the correlators over all the $T$ possible values of
common time shift $t_s$,
as illustrated later, we found that the statistics are
indeed significantly improved.

For each time slice, three fermion matrix inversions are
needed corresponding to the  $3$ color choices
for the pion source, and each inversion takes about
$1000$ iterations (about $2000$
for the ($0.00484, 0.0484$) and ($0.005, 0.05$) ensembles)
during the conjugate gradient calculation.
Thus, totally we carry out $3 T$ inversions
on a single gauge configuration.
As shown later, this rather big number of the inversions
offers the substantial statistics needed to
get the $\pi\pi$ scattering amplitudes with high accuracy.

In practice, we calculate the pion correlators,
\begin{eqnarray}
C_\pi^{\rm PP}(t) &=& \frac{1}{T}\sum_{t_s=0}^{T-1}
\langle 0|\pi^\dag (t+t_s) \pi(t_s) |0\rangle, \cr
C_\pi^{\rm WP}(t) &=& \frac{1}{T}\sum_{t_s=0}^{T-1}
\langle 0|\pi^\dag (t+t_s) W_\pi(t_s) |0\rangle,
\end{eqnarray}
where $\pi$ is the pion point-source operator
and $W_\pi$ is the pion wall-source
operator~\cite{Bernard:2001av,Aubin:2004wf}.
To simplify the notation in this section,
the summation over the lattice space point in sink is not written out.
In this work, we will adopt the shorthand notation:
``PP'' for the point-source point-sink propagators, and
``WP'' for the wall-source point-sink propagators~\cite{Aubin:2004wf}.
We should stress that the summations are
also taken over all the time slice for the pion propagators,
and we found that the statistics are
indeed significantly improved.
This is very important to obtain pion mass with high accuracy.

Overlooking the excited state contributions, the pion mass $m_\pi$
can be secured at large $t$ with a single exponential
fit ansatz~\cite{DeGrand:2006zz,Bernard:2007ps,Bazavov:2009bb}
\begin{eqnarray}
\label{eq:pi_fit_PP}
C_\pi^{\rm PP}(t) &=& A_\pi^{\rm PP} \left[e^{-m_\pi t}+e^{-m_\pi(T-t)}\right], \\
\label{eq:pi_fit_WP}
C_\pi^{\rm WP}(t) &=& A_\pi^{\rm WP} \left[e^{-m_\pi t}+e^{-m_\pi(T-t)}\right],
\end{eqnarray}
where $T$ is the temporal extent of the lattice,
$A_\pi^{\rm PP}$ and $A_\pi^{\rm WP}$ are  overlapping amplitudes.
We will use these values to estimate the wrap-around contributions~\cite{Gupta:1993rn,Umeda:2007hy,Feng:2009ij} and
calculate the pion decay constant~\cite{Aubin:2004fs} as well.

We should remark at this point that, in the calculation of
the $\pi\pi$ four-point functions for the $I=0$ channel,
we try our best effort to compute the vacuum diagram,
since the other three diagrams can be relatively easily calculated.
We found that the vacuum diagram plays a critical role in this
correlator.~\footnote{
In our previous work~\cite{Fu:2011zzh}, we presented the detailed procedure
to calculate the disconnected diagram for the $f_0(600)$ meson.
It helps us a lot to implement
the evaluation of the vacuum diagram here, especially
for how to conduct a vacuum subtraction.
}

\section{Lattice simulation results}
\label{sec:Results}

\subsection{Pion mass and pion decay constant}

\begin{figure}[h]
\includegraphics[width=8cm,clip]{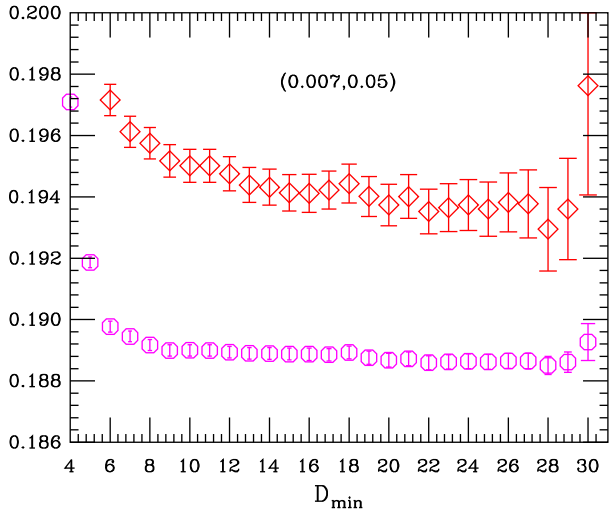}
\caption{(color online).
Pion masses (magenta octagons) and WP amplitudes $A_\pi^{\rm WP}$
(red diamonds) as a function of the minimum time
distance in the fit for the $(0.007,0.05)$ ensemble.
The amplitudes have been divided by $2060$.
\label{fig:pion_amp}
}
\end{figure}

In practice, the $\pi$ propagators were fit by
varying minimum fitting distances $\rm D_{min}$,
and with maximum distance $\rm D_{max}$ either at $T/2$
or where the fractional statistical errors surpassed about $20\%$
for two sequential time
slices~\cite{Bernard:2001av,Aubin:2004wf,DeGrand:2006zz}.~\footnote{
Since the lattice data points at the largest distances contain relatively
little information, the exact selection of large distance cutoff $\rm D_{max}$
is not very critical~\cite{Bernard:2001av,Aubin:2004wf,DeGrand:2006zz}.
}
In this work, pion masses were secured
from the ``effective mass'' plots
for each of the MILC lattice ensembles,
and they were strenuously opted by looking for a combination of a ``plateau''
in the mass as a function of  $\rm D_{min}$, good fit
quality (i.e., $\chi^2/{\rm dof} \leq 1$),
and $\rm D_{min}$ large enough to reduce the excited states~\cite{Beane:2005rj,Beane:2007xs,Feng:2009ij}.
The WP propagators were fit to Eq.~(\ref{eq:pi_fit_WP})
using a minimum time distance of $14a$ for the ``medium-coarse'' lattices
and $20a$ for the ``coarse'' lattices,
and the full covariance matrix is used to compute statistical errors.
At these large distances, the pollution from excited
states is at most comparable to the statistical errors~\cite{Aubin:2004fs}.
For example, Fig.~\ref{fig:pion_amp} exhibits the results
for pion masses and amplitudes as a function of $\rm D_{min}$ for
the $(0.007,0.05)$ ensemble.
Since the major objective of this work is to present the work for isospin-$0$,
as explained later, for the $\pi\pi$ scattering in the $I=0$ channel
at this work, the systematic error of the energy of the $\pi\pi$ system
is pretty large,
we can temporarily neglect the systematic effect for pion mass
due to excited states.
All of these fitted values of pion masses are listed in Table~\ref{tab:m_pi_pi}.

\begin{table*}[t]	
\caption{\label{tab:m_pi_pi}
Summaries of the pion masses and pion decay constants.
The third block shows the pion masses in lattice units and the fifth block
give the overlapping amplitude using WP propagators.
The product of $m_\pi L$ is presented in Column four.
The pion decay constants in lattice units are provided in sixth block,
where the star on the superscript indicate the MILC's determination.
Column two shows the values of pion mass in MeV,
where the errors are estimated from both the error on lattice spacing $a$
and statistical errors in Column three.
The seven block shows the dimensionless ratio $m_\pi/f_{\pi}$,
where the errors are estimated from the $am_\pi$ and $af_\pi$.
The last block shows the pion masses which are  measured
by the point-wall point-sink propagators, and only used as a consistency check.
}
\begin{ruledtabular}
\begin{tabular}{llllllll}
$\rm Ensemble$ & $m_\pi({\rm MeV})$  & $a m_\pi^{\rm WP}$  & $m_\pi^{\rm WP} L$ &
$A_\pi^{\rm WP}$ & $a f_{\pi}$       & $m_\pi/f_{\pi}$ & $a m_\pi^{\rm PP}$ \\
\hline
\hspace{-0.08cm}($0.00484, 0.0484$)  & $240(4)$  & $0.17503(09)$ & $3.5006(18)$ &$1000.18\pm1.580$  & $0.11767(45)$  & $1.4874(57)$       & $0.17504(09)$ \\
\hspace{-0.08cm}($0.005,   0.05$)    & $268(5)$  & $0.15970(15)$ & $3.8345(48)$
& $770.534\pm2.577$ & $0.09054(33)^\ast$ & $1.7639(66)$   & $0.15992(16)$ \\
\hspace{-0.08cm}($0.007,   0.05$)    & $315(6)$  & $0.18868(22)$ & $3.7736(44)$
& $399.094\pm1.393$ & $0.09364(20)^\ast$ & $2.0149(49)$   & $0.18871(24)$ \\
\hspace{-0.08cm}($0.0097,  0.0484$)  & $334(6)$  & $0.24566(18)$ & $3.9306(29)$
& $395.107\pm1.151$  & $0.12136(29)$ & $2.0242(51)$       & $0.24587(21)$  \\
\hspace{-0.08cm}($0.01,    0.05$)    & $373(7)$  & $0.22455(27)$ & $4.4910(54)$
& $365.595\pm2.039$ & $0.09805(14)^\ast$ & $2.2902(42)$  & $0.22447(17)$ \\
\hspace{-0.08cm}($0.0194,  0.0484$)  & $463(8)$  & $0.34279(19)$ & $5.4846(30)$
& $315.695\pm0.865$ & $0.13055(48)$ & $2.6258(98)$       & $0.34279(23)$
\end{tabular}
\end{ruledtabular}
\end{table*}

In our previous work~\cite{Fu:2011bz},
we used the method described in Ref.~\cite{Aubin:2004fs}
to extract the pion decay constant
for the ($0.0097, 0.0484$) ensemble~\cite{Fu:2011bz}.
In the light of same procedures,
we calculated the pion decay constants for other ``medium coarse'' ensembles.
All of these fitted values of pion decay constants
are listed in Table~\ref{tab:m_pi_pi}.

As a consistency check, the PP correlators
were reliably measured in this work.
Use these correlators, we can secure pion masses
via Eq.~(\ref{eq:pi_fit_PP}) which are listed in last block in Table~\ref{tab:m_pi_pi},
and these pion masses are found to be consistent with
its counterparts extracted with WP propagators,
which are summarized in Table~\ref{tab:m_pi_pi}.

Our fitted values of pion masses and pion decay constants
listed in Table~\ref{tab:m_pi_pi},
are in rather good agreement with the same quantities
which are computed on the same lattice ensembles
by MILC collaboration in Refs.~\cite{Bernard:2007ps,Bazavov:2009bb}.
For the ``coarse''  ensembles,
the MILC's determinations on
pion decay constants are directly quoted~\cite{Bernard:2007ps,Bazavov:2009bb},
which are also summarized in Table~\ref{tab:m_pi_pi}.

\subsection{Diagrams $D, C, R$, and $V$}
\begin{figure*}[p]
\includegraphics[width=8cm,clip]{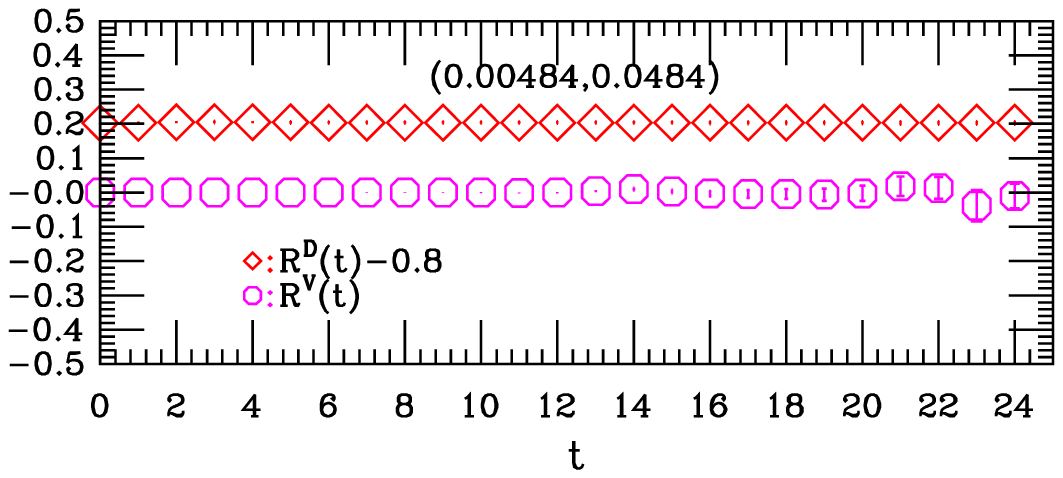} \hspace{0.5cm}
\includegraphics[width=8cm,clip]{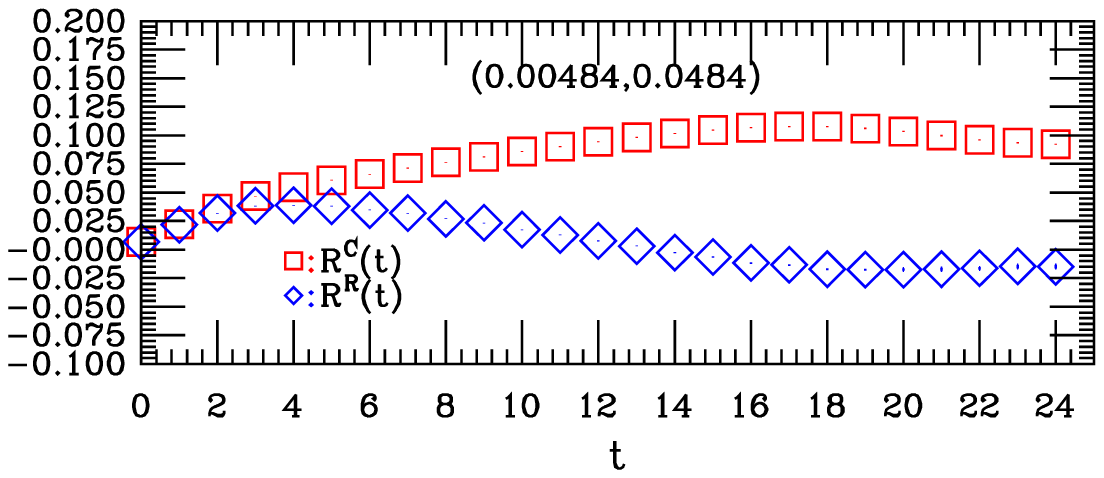}
\includegraphics[width=8cm,clip]{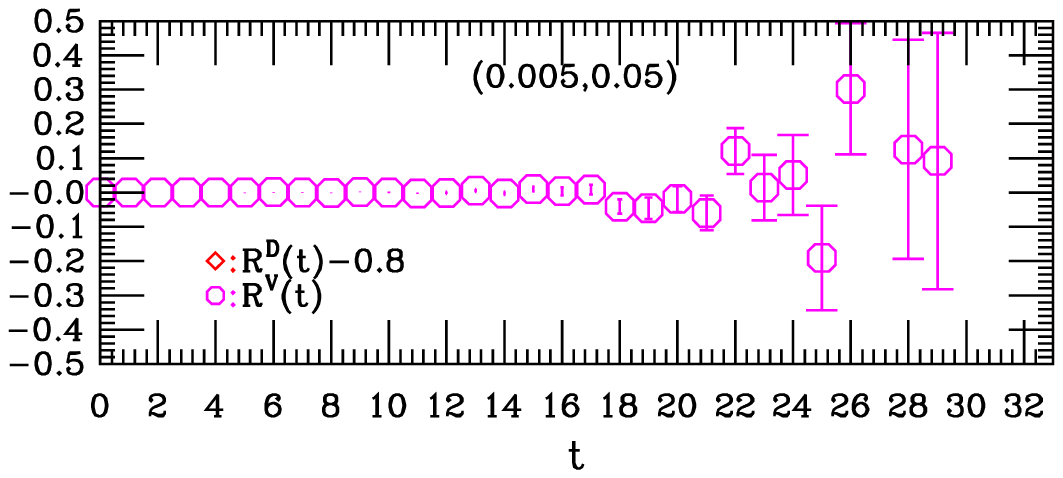}   \hspace{0.5cm}
\includegraphics[width=8cm,clip]{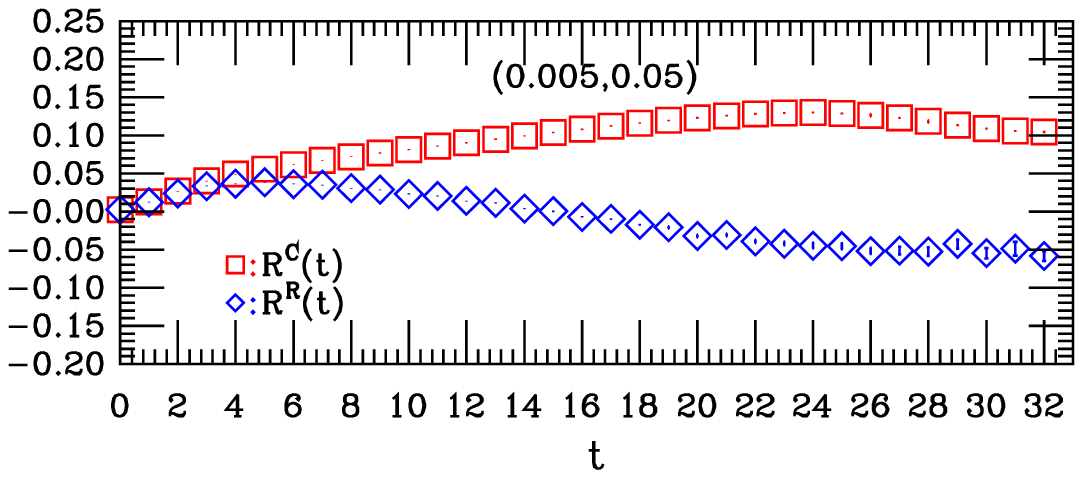}
\includegraphics[width=8cm,clip]{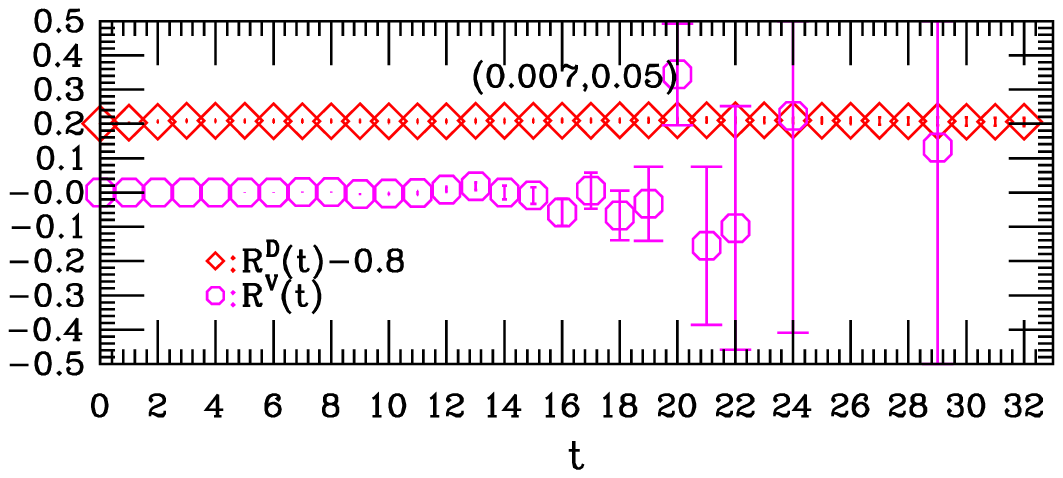}   \hspace{0.5cm}
\includegraphics[width=8cm,clip]{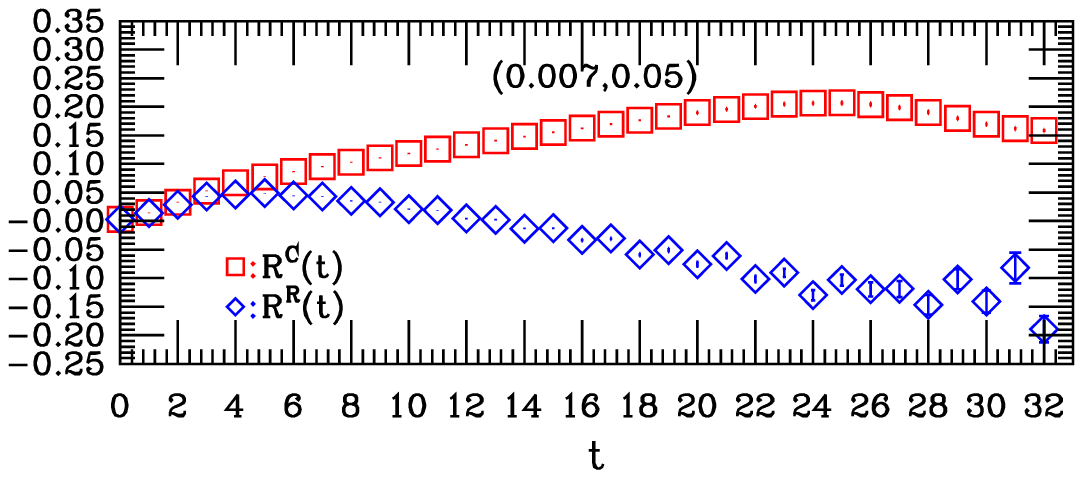}
\includegraphics[width=8cm,clip]{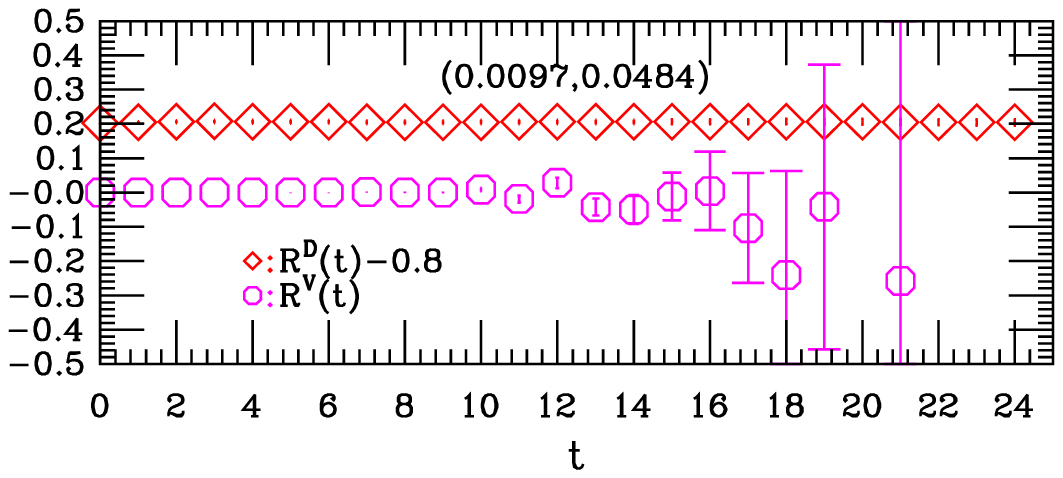}  \hspace{0.5cm}
\includegraphics[width=8cm,clip]{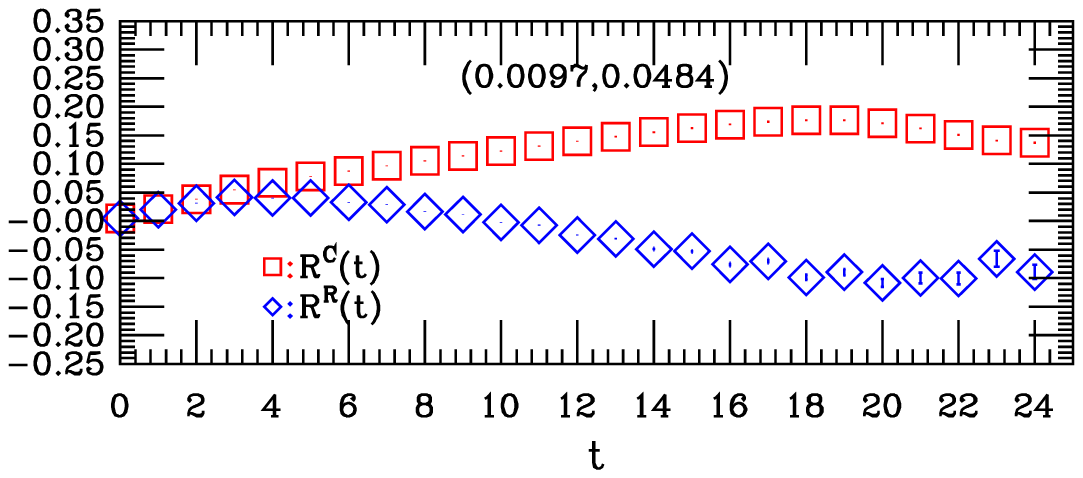}
\includegraphics[width=8cm,clip]{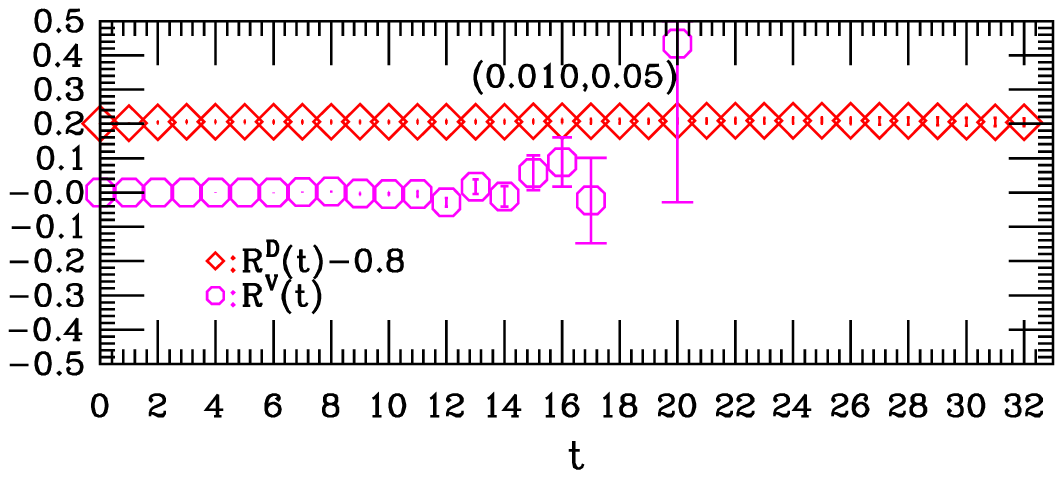}   \hspace{0.5cm}
\includegraphics[width=8cm,clip]{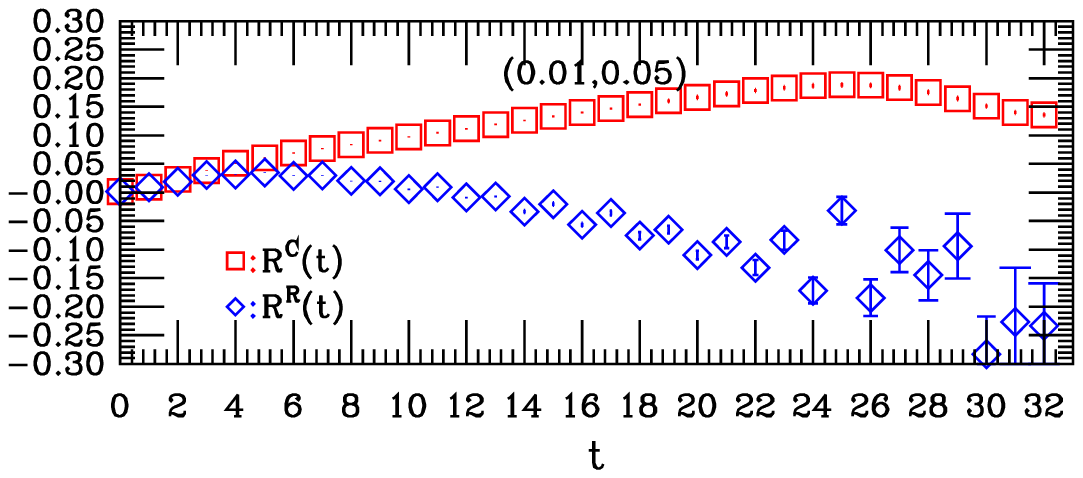}
\includegraphics[width=8cm,clip]{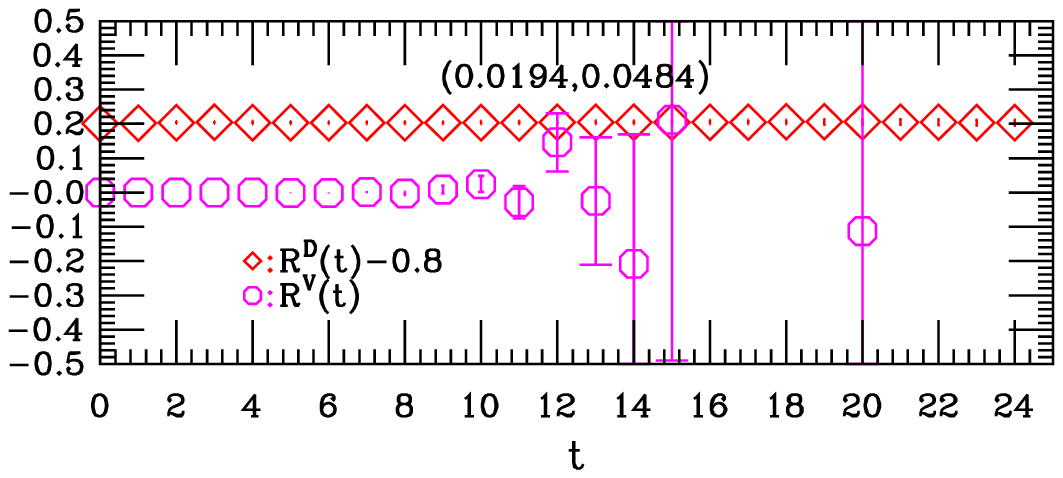}  \hspace{0.5cm}
\includegraphics[width=8cm,clip]{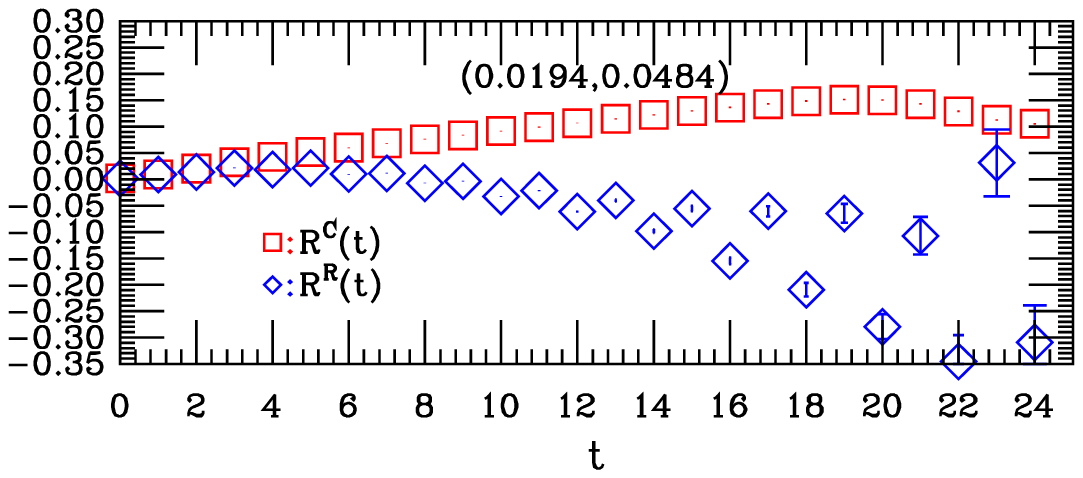}
\caption{(color online).
Individual amplitude ratios $R^X(t)$ of the
$\pi\pi$ four-point functions computed via the moving wall source
without gauge fixing as the functions of $t$ for six MILC lattice ensembles.
Direct diagram (red diamonds) displaced by $0.8$,
vacuum diagram (magenta octagons),
crossed diagram (red squares) and rectangular diagram (blue diamonds).
\label{fig:ratio}
}
\end{figure*}

As practiced in our former work~\cite{Fu:2011bz},
the $\pi\pi$ four-point functions are robustly calculated
on six MILC lattice ensembles listed in Table~\ref{tab:MILC_configs}
using the technique of the moving wall source
without gauge fixing~\cite{Kuramashi:1993ka,Fukugita:1994ve,Fu:2011wc}.
In Fig.~\ref{fig:ratio}, the individual ratios,
which are denoted in Eq.~(\ref{EQ:ratio}), $R^X$ ($X=D, C, R$ and $V$)
are illustrated as the functions of $t$
for each lattice ensemble.

The ratio values of the direct amplitude $R^D$ are quite close to oneness,
indicating a pretty weak interaction in this channel.
The crossed amplitude, in  another aspect, increases linearly up to
$t \sim 18$  for the ``medium coarse'' ensemble and
$t \sim 24$  for the ``coarse'' ensemble,
implying a repulsive interaction between two pions in the $I=2$ channel.
In contrast, after a beginning increase up until $t \sim 4$, the rectangular amplitude
demonstrates a roughly linear decrease up until
$t \sim 18$  for the ``medium coarse'' ensemble and
$t \sim 24$  for the ``coarse'' ensemble,
suggesting an attractive force among two pions for the $I=0$ channel.
Additionally, the magnitude of the slope is similar to
that of the crossed amplitude but with opposite sign.
Furthermore, we observe that the crossed and rectangular amplitudes have
the same values at $t=0$ and close ones for small $t$.
These characteristics are in good keeping with
the theoretical predictions~\cite{Sharpe:1992pp}.

It is extremely noisy in the amplitude of the vacuum diagram ($V$)
for the $(0.0194,0.0484)$ ensemble,
while we can see a good signal up to $t = 10$,
and loss of signals after that.
These characteristics are in well accordance
with the Okubo-Zweig-Iizuka (OZI) rule and $\chi$PT in leading order,
which expect the disappearing of the vacuum
amplitude~\cite{Sharpe:1992pp,Kuramashi:1993ka,Fukugita:1994ve}.
For the lattice ensembles with the pion mass changing smaller,
the signals of the vacuum diagram are becoming more and more tolerating.
For the $(0.00484,0.0484)$ ensemble,
the signals of the vacuum diagram is already tolerating.
This qualitatively confirmed the analytical arguments in Ref.~\cite{Lepage:1989hd}
which indicates that the error of the vacuum amplitude grows
exponentially as $\displaystyle e^{2m_\pi t}$.
The numerical calculation of the amplitude for the vacuum diagram
stands as our principal and distinctive accomplishment of this paper.

The systematically oscillating behavior of the rectangular amplitude
in large $t$ is evidently observed,
which is a typical feature of
the staggered formulation of lattice fermions and
corresponds to the contributions from the intermediate states
with opposite parity~\cite{Barkai:1985gy},
and for the lattice ensemble with large pion mass, this oscillating feature
become more obvious.
In contrast, for that with small pion mass, this feature is not appreciable,
and even not perceptible for the MILC $(0.00484,0.0484)$ ensemble.
The physical meaning of this fascinatingly oscillating behavior is
easily understood~\cite{Barkai:1985gy}.
Nevertheless, its quantitative mass dependence is not clear to us,
which are highly needed to be further investigated.

\subsection{The errors of $R^V(t)$ and $R^R(t)$}
According to the analytical arguments in Ref.~\cite{Lepage:1989hd},
the error of the ratio for the vacuum amplitude
increases exponentially as $\displaystyle e^{2m_\pi t}$.
Therefore, it is pretty difficult to secure the correct signal
for large $t$~\cite{Lepage:1989hd}.
Likewise, the ratio for the rectangular diagram has errors,
which grow exponentially as $\displaystyle e^{m_\pi t}$ for large $t$~\cite{Lepage:1989hd}.
Our lattice data indeed demonstrates such dependence
with the expected slopes.

\begin{figure}[h!]
\includegraphics[width=8cm,clip]{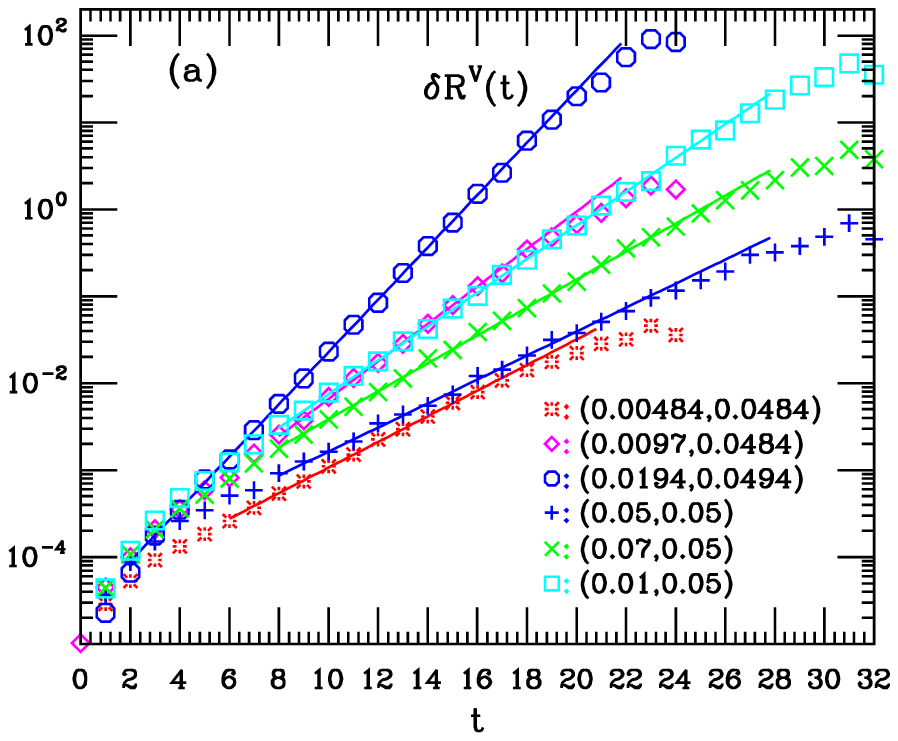}
\includegraphics[width=8cm,clip]{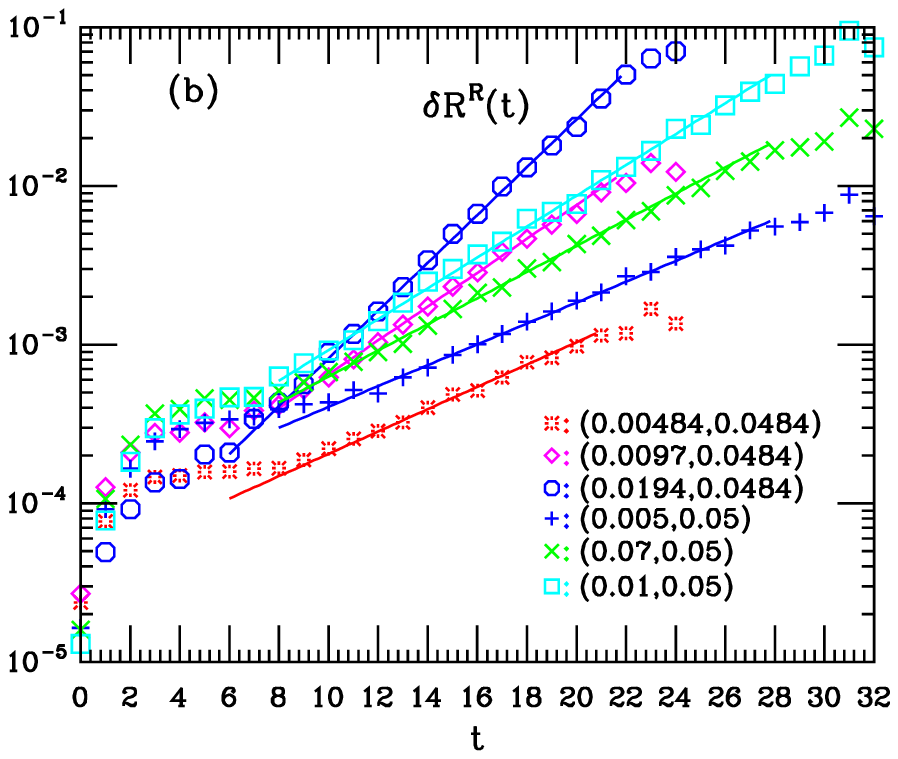}
\caption{(color online).
The errors of the amplitude ratios $R^X(t)(X=V,R)$ as the functions of $t$
for each of the MILC lattice ensembles.
Solid lines are single exponential fits, and the fitting ranges
are listed in Table~\ref{tab:error_fitting}.
(a) The errors of the amplitude ratios $R^V(t)$;
(b) Those of  $R^R(t)$
\label{fig:error_RX}
}
\end{figure}

The magnitudes of these errors are quantitatively in line
with these analytical predictions as demonstrated
in Fig.~\ref{fig:error_RX}.
Fitting the errors $\delta  R^V(t)$ and $\delta  R^R(t)$
by a single exponential fit ansatz
$\displaystyle \delta R^V(t) \sim  e^{\mu_V t}$
and $\displaystyle \delta R^R(t) \sim  e^{\mu_R t}$, respectively,
for six lattice ensembles,
we extract the fitted values
of $\mu_V$ and $\mu_R$, which are summarized in Table~\ref{tab:error_fitting},
together with their fitting ranges.
\begin{table}[h!]	
\caption{\label{tab:error_fitting}
Summaries of the fitted values for $\mu_V$ and $\mu_R$ in lattice units.
The second and third blocks show the fitted values of $\mu_V$,
and $\mu_R$, respectively,
and Column four gives the time range for the chosen fit.
}
\begin{ruledtabular}
\begin{tabular}{lccl}
${\rm Ensemble} $   & $a \mu_V$    & $a\mu_R$ & Range  \\
\hline
$(0.00484,0.0484)$  & $0.3392$   & $0.1621$  & $10-18$  \\
$(0.0097,0.0484) $  & $0.4956$   & $0.2457$  & $8-16$  \\
$(0.0194,0.0484) $  & $0.6927$   & $0.3487$  & $8-16$  \\
$(0.005,0.05)$      & $0.3178$   & $0.1513$  & $10-20$  \\
$(0.007,0.05)$      & $0.3701$   & $0.1895$  & $10-20$  \\
$(0.01,0.05) $      & $0.4463$   & $0.2237$  & $10-20$  \\
\end{tabular}
\end{ruledtabular}
\end{table}

From Table~\ref{tab:error_fitting},
we note that the fitted values of $\mu_R$  can be compared
with the pion masses $m_\pi$ listed in Table~\ref{tab:m_pi_pi},
and half of the fitted values of $\mu_V$  can also be reasonably
compared with these pion masses.
We here have numerically confirmed the Lepage's analytical
arguments~\cite{Lepage:1989hd} about the $\pi\pi$ scattering.
This testifies the practical applicability of
the moving wall source without gauge fixing
from another point of view.
Thus, we can reasonably assume that
the vacuum amplitude  remains small for large $t$.
In principle, we can overlook the vacuum amplitude in the rest of the analysis.
However, we will explicitly include it
for the sake of completeness of the lattice QCD calculation.

\subsection{$R_I$ projected onto the $I =0$ and $2$ channels}
\begin{figure}[ht]
\includegraphics[width=8.0cm,clip]{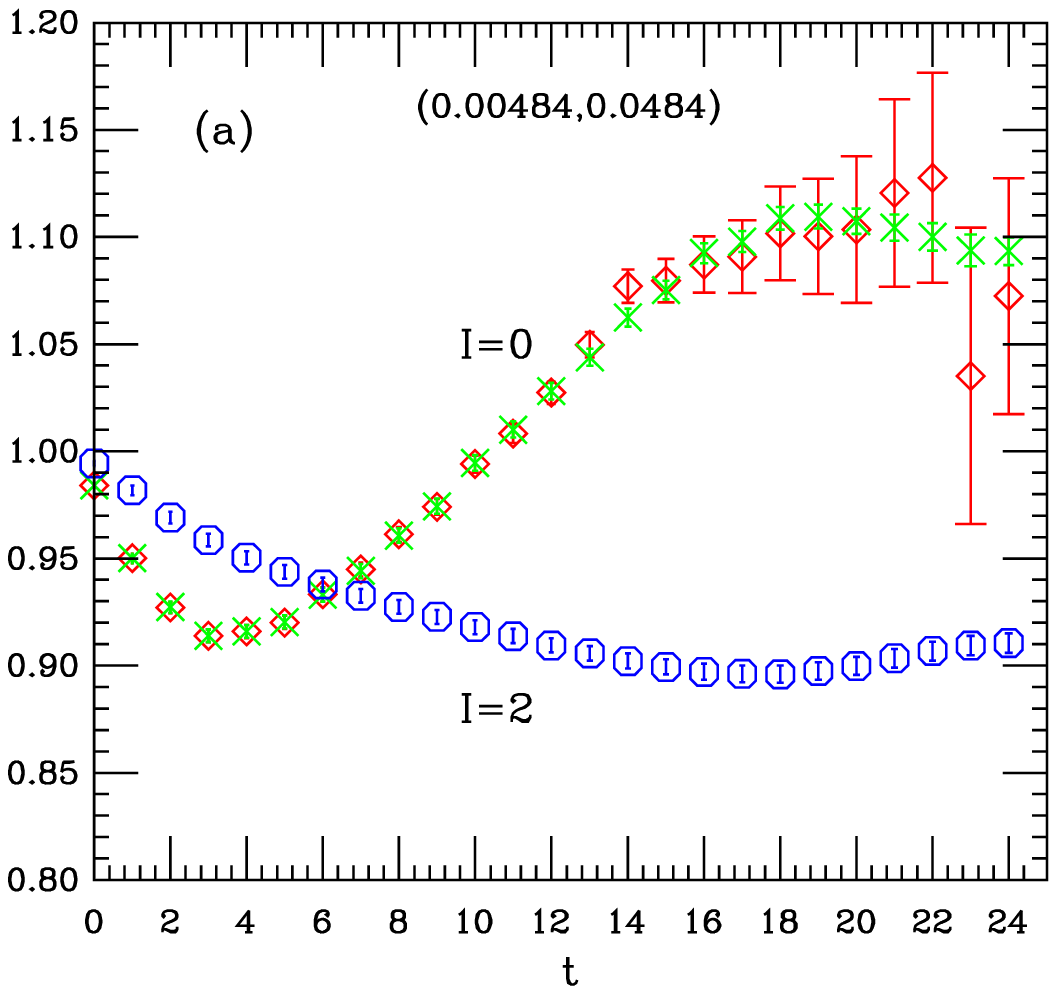}
\includegraphics[width=8.0cm,clip]{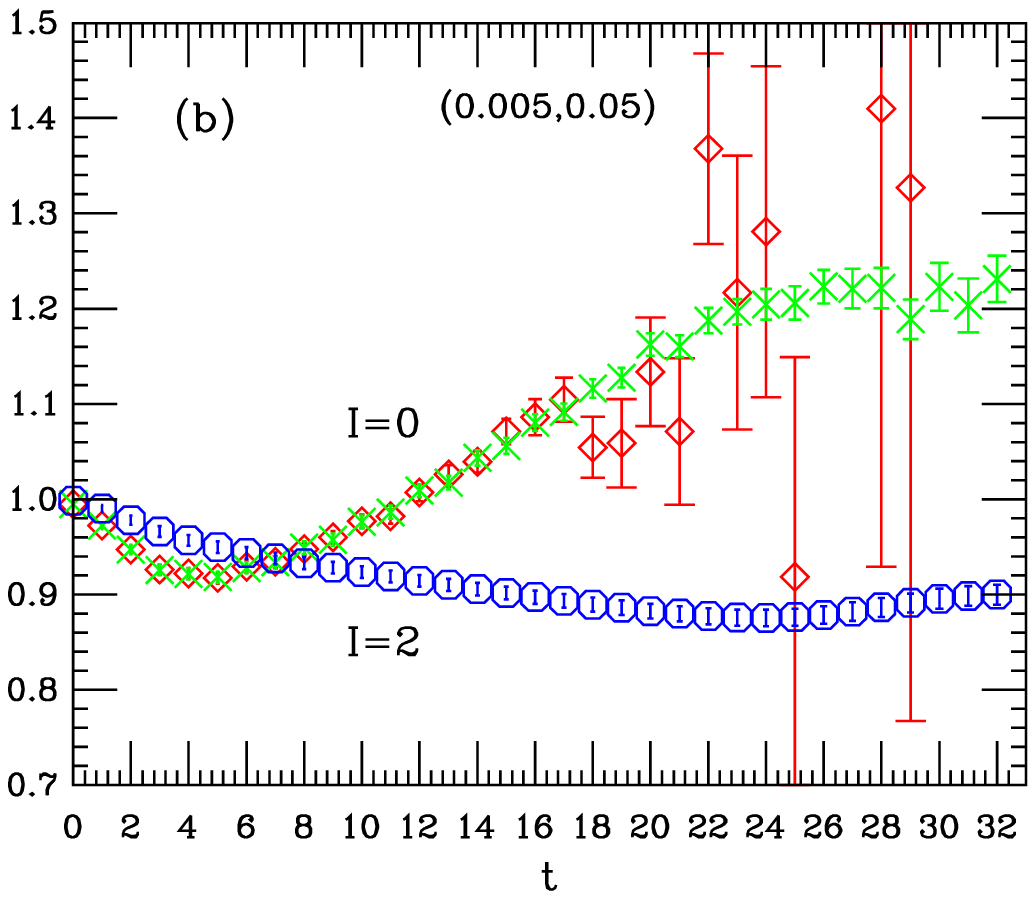}
\caption{(color online).
$R_I(t)$ ($I=0$ and $2$) for the $\pi\pi$ four-point function
at zero momenta calculated by the moving wall source without gauge fixing
as the functions of $t$  for the MILC lattice
(a) ($0.00484, 0.0484$) ensemble; and
(b) ($0.005, 0.05$) ensemble.
The cross yellow points indicate the ratio $R_I(t)$
in the $I = 0$  channel where the vacuum diagram is turned off.
\label{fig:R_I_02}
}
\end{figure}

The ratios $R_I(t)$ projected onto the isospin $I = 0$ and $2$ eigenchannels
for the MILC ($0.00484, 0.0484$) and ($0.005, 0.05$) ensembles,
are demonstrated in Fig.~\ref{fig:R_I_02}.
A decrease of the ratio $R_{I=2}(t)$ indicates
a repulsive interaction among two pions for the $I = 2$ channel,
on the other hand, an increase of the ratio $R_{I=0}(t)$
suggests an attractive interaction for the $I = 0$ channel.
In the $I = 0$ channel, a dip at $t=3$ for the ($0.00484, 0.0484$) and  $t=5$
for the ($0.005, 0.05$)
can be clearly observed, and its physical origin
is not clear to us as well~\cite{Kuramashi:1993ka,Fukugita:1994ve}.

Due to the rather small quark mass of two lattice ensembles,
the systematically oscillating behavior for the $I=0$ channel
in large $t$ is not clearly observed,
which is a typical characteristic of
the Kogut-Susskind formulation of lattice fermions~\cite{Barkai:1985gy}.
Addtionally, this oscillating feature is
hardly noticed  for the $I=2$ channel.

In order to present the contribution from the vacuum term more intuitively,
we employ the yellow cross points to indicate the ratio
$R_I(t)$ for the $\pi\pi$ four-point function in the $I = 0$  channel
without the presence of the disconnected diagram.
From Fig.~\ref{fig:R_I_02},
we can clearly notice that the contribution from the disconnected diagram
is only obvious when $t \ge 20$ for the ($0.00484, 0.0484$) ensemble,
and $t \ge 16$ for the  ($0.005, 0.05$) ensemble.

\subsection{Lattice artifact}
\label{sec:Lat_artifact}
From Fig.~\ref{fig:ratio}, we observe that
there exists  a pollution from the ``wraparound''
effects~\cite{Gupta:1993rn,Umeda:2007hy,Feng:2009ij}
approximately starting as early as at $t = 12 \sim 18$ for the
MILC ``medium-coarse'' ensembles and
$t=22\sim25$ for the ``coarse'' ensembles.
As discussed in Refs.~\cite{Gupta:1993rn,Yagi:2011jn,Umeda:2007hy,Feng:2009ij},
one of two pions can propagate $T - t$ time steps backwards
due to the periodic boundary condition in the temporal direction.
This operates as a constant contribution and
deforms the $\pi\pi$ four-point functions in large $t$
(especially around $T/2$),
according to the discussions in
Refs.~\cite{Gupta:1993rn,Yagi:2011jn,Umeda:2007hy,Feng:2009ij},
it is roughly suppressed by
$$
\exp \left( -  m_\pi  T \right) / \exp \left( - 2m_\pi t \right)
$$
as compared with forward propagation of the $\pi\pi$ state.
We can opt the fitting ranges
satisfying $t_{\rm max} \ll T/2$ to reduce this effect~\cite{Gupta:1993rn}.
However, according to the arguments in Refs.~\cite{Gupta:1993rn,Yagi:2011jn,Umeda:2007hy,Feng:2009ij},
if pion mass is small enough (e.g. the ($0.00484, 0.0484$) ensemble),
the wrap-around pollution can not be suppressed even for $t \ll T/2$,
we should include this term for the successful fit,~\footnote{
It turned out that a five-parameter cosh-fit of $C$, $A_{\pi\pi}$,
$E_{\pi\pi}^I$, $A_{\pi\pi}^{\prime}$, and $E_{\pi\pi}^{I \prime}$
yields a satisfactory result with an pretty acceptable $\chi^2$.
Moreover, the excited states will be taken into account as one of the important
sources of systematical error in this work.
}
\begin{eqnarray}
\label{eq:E_pi_pi}
\hspace{-0.7cm} C_{\pi\pi}^I(t)  &=& C+
A_{\pi\pi}\cosh\left[E_{\pi\pi}^I\left(t - \tfrac{1}{2}T\right)\right] \cr
&& +
(-1)^t A_{\pi\pi}^{\prime}\cosh \left[E_{\pi\pi}^{I \prime} \left(t-\tfrac{1}{2}T\right)\right] + \cdots.
\end{eqnarray}
where $C$ is a constant corresponding to the wrap-around term.
This can be easily understood by evaluating the contribution of two fake diagrams in
Fig.~2 of Ref.~\cite{Yagi:2011jn}, and the $C$ can be expressed as
\begin{equation}
\label{eq:C_fake_diagram}
C = 2A_\pi^2 e^{-m_\pi T}.
\end{equation}
For easy notation the superscript WP in $A_{\pi}$ is omitted in the rest of analyses.

In this work, we accurately extract the overlapping amplitudes
$A_\pi$ and pion masses $m_\pi$ corresponding to
pion correlators~\cite{Nagata:2008wk}
which are listed in Table~\ref{tab:m_pi_pi},
and these values are sufficiently precise
to estimate the wraparound terms with Eq.~(\ref{eq:C_fake_diagram}),
which are listed in Table~\ref{tab:C_C}.
\begin{table}[h]	
\caption{\label{tab:C_C}
Summaries of the calculated wraparound contributions
from overlapping amplitude $A_\pi$ and pion mass $am_\pi$.
The second block  shows the wraparound  contributions calculated from
Eq.~(\ref{eq:C_fake_diagram}),
where its errors are roughly estimated from
the statistical errors of $A_\pi$ and $m_\pi$.
}
\begin{ruledtabular}
\begin{tabular}{ll}
${\rm Ensemble} $     &$C$  \\
\hline
$(0.00484,0.0484)$  &$451.36(2.63)$  \\
$(0.005,0.05)$      &$43.23(51)$     \\
$(0.007,0.05)$      &$1.815(29)$     \\
$(0.0097,0.0484) $  &$2.362(25)$     \\
$(0.01,0.05) $      &$0.1533(31)$    \\
$(0.0194,0.0484) $  &$0.01424(15)$   \\
\end{tabular}
\end{ruledtabular}
\end{table}

We note that Ref.~\cite{Dudek:2012gj} has recently taken
the similar definition as
\begin{equation}
\label{eq:Dudek_C}
C = 2A_{\pi\pi} e^{-m_\pi T} ,
\end{equation}
where $A_{\pi\pi}$ is defined in Eq.~(\ref{eq:E_pi_pi}).
Additionally,  there exists the similar general form which contains
the two-particle as well as one-particle eigenvalues,
and gives a not rigorous proof in Ref.~\cite{Prelovsek:2010kg}.
Furthermore, when studying $K\to \pi\pi$ decay amplitudes,
there is an analogous fitting functional form  in Ref.~\cite{Blum:2011pu}.

\begin{figure}[ht]
\includegraphics[width=8.0cm,clip]{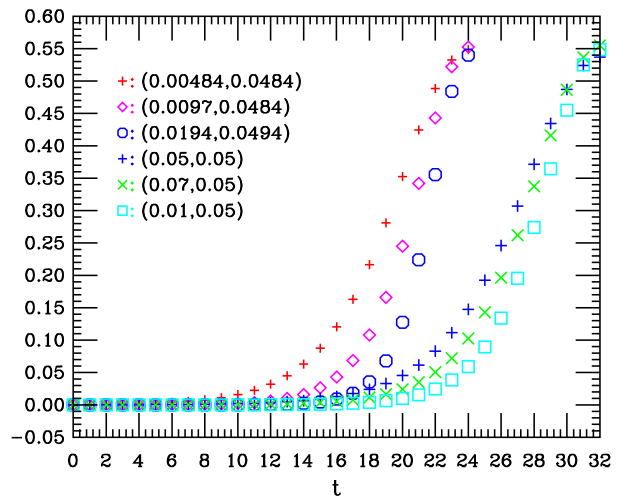}
\caption{(color online).
The ratios of the wraparound terms
to the corresponding $\pi\pi$ four-point functions
for six MILC lattice ensembles calculated by Eq.~(\ref{eq:ratio_WC}).
All of these ratios make a significant contribution
and are approximately close to $1/2$  as  $t$ approaches to
$T/2$ as expected from the arguments in
Refs.~\cite{Gupta:1993rn,Yagi:2011jn,Umeda:2007hy,Feng:2009ij}.
\label{fig:ratio_WC}
}
\end{figure}

In order to comprehend this wraparound effects at a quantitative level,
we denote a quantity
\begin{equation}
\label{eq:ratio_WC}
R_{WC}(t) = \frac{C}{C_{\pi\pi}^{I=2}(t)},
\end{equation}
which is the ratio of the wraparound pollution
to the $\pi\pi$ four-point function in the $I=2$ channel.
In fact, we exploited the data of the wraparound contribution $C$
listed in Table~\ref{tab:C_C} and
$C_{\pi\pi}^{I=2}(t)$ calculated from Eq.~(\ref{EQ:phy_I0_2})
to approximately evaluate these ratios.
The ratios
for six MILC lattice ensembles are illustrated in Fig.~\ref{fig:ratio_WC}.
All of these ratios make a significant contribution
and are approximately close to $1/2$  as  $t$ approaches to
$T/2$ as expected from the arguments in Refs.~\cite{Gupta:1993rn,Yagi:2011jn,Umeda:2007hy,Feng:2009ij}.
We can note that
as the pion mass of the lattice ensemble becomes smaller, the wraparound contribution $C$
is clearly observed even at small $t$.
For example,  we can keenly notice the wrap-around term even
as early as at $t=10\sim12$ satisfying $t \ll T/2$
for the ($0.00484, 0.0484$) ensemble.
It is, therefore, absolutely necessary for us to explicitly
consider the wrap-around term, especially for lattice ensembles with small pion masses
when we extract the energy $E$ of the $\pi\pi$ system~\cite{Yagi:2011jn}.

To get rid of this pollution, Xu et al.~\cite{Feng:2009ij}
employed a derivative method
and denoted a modified ratio~\cite{Umeda:2007hy}.
By ignoring terms suppressed relative to the leading contribution,
the energy shift $\delta E$ can be obtained
from the asymptotic form of the modified ratio.
To identify the time-separations where the ground state dominates,
Yagi et al. used their self-defined ratios~\cite{Yagi:2011jn}.
Moreover, Dudek et al.~\cite{Dudek:2012gj} recently eliminated
this unwanted pollution term by the shifted correlator.

In principle, we can use one of the three above-mentioned methods to
process our $\pi\pi$ scattering data for isospin-$2$.
However, for those of the $I=0$ channel,
there is a further complication introduced by staggered fermions,
the oscillating term is appreciable, we must consider the oscillating term
and modify the corresponding functional forms,
as a consequence, it is not convenient to use these methods.
Additionally, for the pion-kaon ($\pi K$) scattering~\cite{Nagata:2008wk,Fu:2011wc},
and the $\pi\pi$ correlators ``in flight''~\cite{Dudek:2012gj},
the wrap-around term is not a constant,
and these methods are not suitable.

Nagata et al. solved this problem by subtracting the wrap-around
term numerically from the obtained quantities~\cite{Nagata:2008wk},
since the lattice-measured data are sufficiently precise
to allow such subtraction.
Dudek et al. eliminated this term
by means of the shifted correlator for the $\pi\pi$ scattering at rest
and weighted-shifted correlator for that  ``in flight''~\cite{Dudek:2012gj}.
We already exhibited that the overlapping amplitude $A_\pi$
and pion mass $m_\pi$ can be calculated with high accuracy,
so, it is natural to borrow these methods
to our case~\cite{Dudek:2012gj,Nagata:2008wk}.
As a consistency check, we numerically compared these results
calculated from the above-mentioned methods for lattice data
in the $I=2$ channel, and found that they are well consistent with each other
within  errors.
Therefore, in this work we only present the results
from the last method, namely,
using equation~(\ref{eq:E_pi_pi}) to
extract the energy  $E$ of  the $\pi\pi$ system
in a conceptually clean way.

\section{Fitting analyses}
\label{sec:fittingResults}

As already explained in previous sections,
we will use Eq.~(\ref{eq:E_pi_pi}) to
get the energies  $aE$ of the $\pi\pi$ system,
which are inserted into the L\"uscher formula (\ref{eq:luscher})
to obtain the corresponding $s$-wave $\pi\pi$ scattering lengths.
Hence, appropriately extracting the energies is
a central step to our ultimate results.
A persuasive way to process our lattice data is
the  resort to the ``effective energy'' plot,
which is a variant of the effective mass plot,
and  very similar to the  ``effective scattering length''
plot~\cite{Beane:2005rj,Beane:2007xs}.

\subsection{$I=2$ channel}
\begin{figure}[h!]
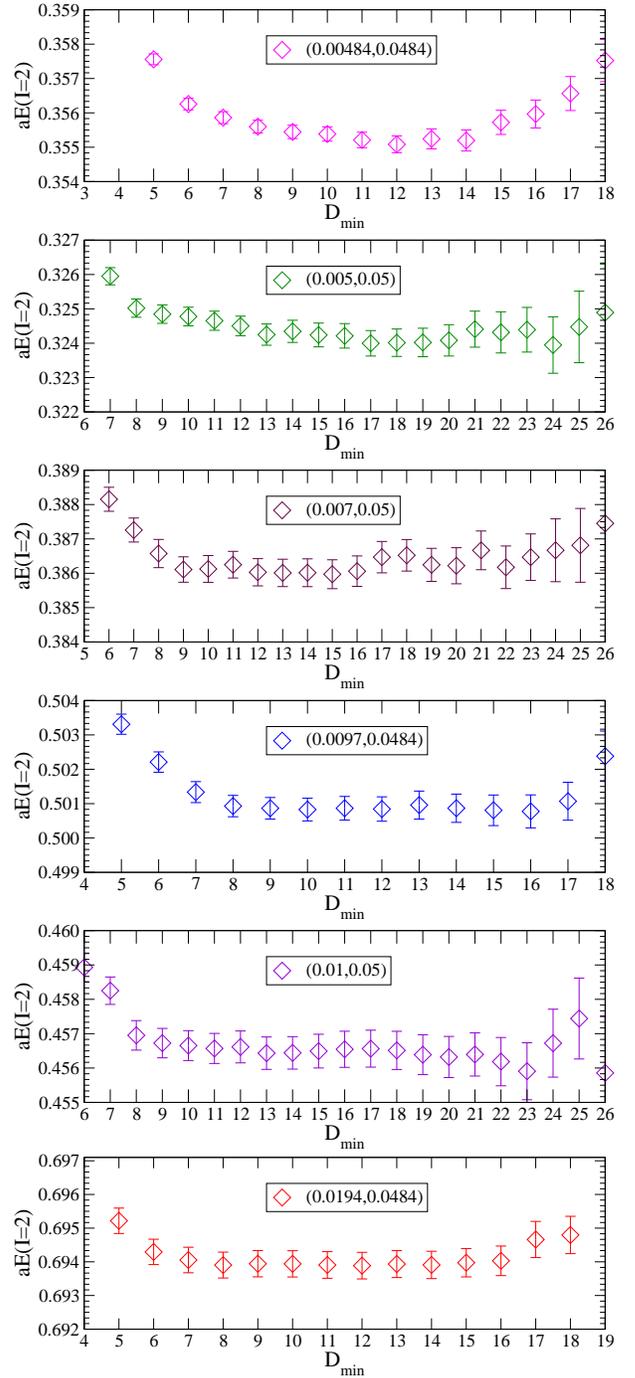

\includegraphics[width=8.0cm,clip]{I2_00484.eps}
\includegraphics[width=8.0cm,clip]{I2_005.eps}
\includegraphics[width=8.0cm,clip]{I2_007.eps}
\includegraphics[width=8.0cm,clip]{I2_0097.eps}
\includegraphics[width=8.0cm,clip]{I2_01.eps}
\includegraphics[width=8.0cm,clip]{I2_0194.eps}
\caption{\label{fig:eff_eng_I2}
(color online). The ``effective energy'' plots
as the functions of $\rm D_{min}$
for the $\pi\pi$ scattering in the $I=2$ channel in lattice units.
The ``effective  energy''  plots
have small errors within a broad minimum fitting distance region.
The estimates of the systematic uncertainty due to fitting
are not displayed in this figure.
}
\end{figure}

In practice,  $\pi\pi$ four-point functions were fit by
varying the minimum fitting distances $\rm D_{min}$,
and with the maximum distance $\rm D_{max}$ either at $T/2$
or where the fractional statistical errors surpassed about $20\%$
for two sequential time slices~\cite{Bernard:2001av,Aubin:2004wf,DeGrand:2006zz}.
Additionally, the fitting parameter $C$
was constrained by priors to conform the lattice-calculated
wraparound contribution $C$ listed in Tables~\ref{tab:C_C}~\cite{Lepage:2001ym}.
For each ensemble,  the ``effective energy'' plots as a function
of $\rm D_{min}$ are illustrated in Fig.~\ref{fig:eff_eng_I2}.
The central value and statistical error at each time slice were
evaluated by the Levenberg-Marquardt algorithm~\cite{Press:1992zz}.
To make these fits more robust,
we double-check them with SNOBFIT,
which is  a soft constrained noisy optimization~\cite{Waltraud:2008}.
From Fig.~\ref{fig:eff_eng_I2}, we also observed that the effective energies
have larger statistical errors near $t \sim T/2$
because of the wrap-around effect
as it is discussed in detail in Ref.~\cite{Yagi:2011jn}.

We should remark that the physical model in Eq.~(\ref{eq:E_pi_pi})
just include the ground state~\cite{Beane:2005rj,Beane:2007xs,Feng:2009ij}.
In fact, we can fit with the inclusion of the first excited state,
and the difference between these procedures,  as well as
the difference arising from the arbitrary choice of ${\rm D_{max}}$,
is incorporated in the systematic error for $a E$
at each time slice.

\begin{figure}[h!]
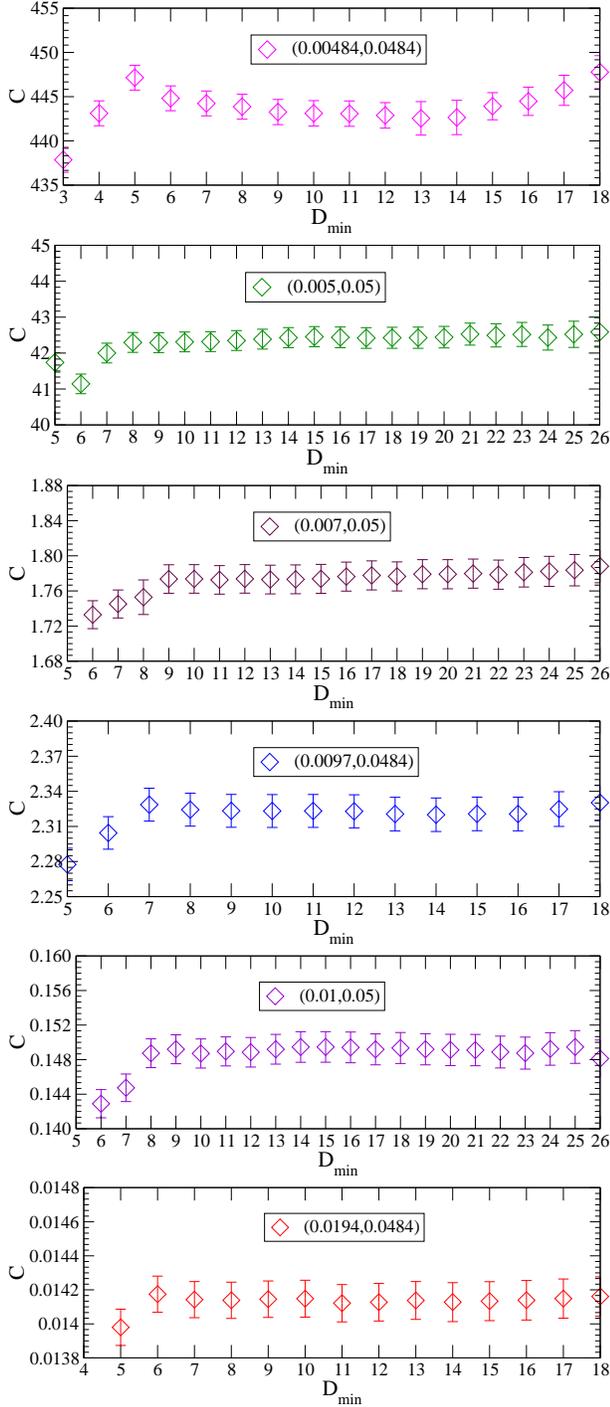

\includegraphics[width=8.0cm,clip]{C_00484.eps}
\includegraphics[width=8.0cm,clip]{C_005.eps}
\includegraphics[width=8.0cm,clip]{C_007.eps}
\includegraphics[width=8.0cm,clip]{C_0097.eps}
\includegraphics[width=8.0cm,clip]{C_010.eps}
\includegraphics[width=8.0cm,clip]{C_0194.eps}
\caption{\label{fig:eff_eng_C}
(color online). The ``effective wraparound constant'' plots
as the functions of $\rm D_{min}$
for the $\pi\pi$ scattering in the $I=2$ channel in lattice units.
}
\end{figure}

In this work, the energies $aE$ of the $\pi\pi$ system
in the $I = 2$ channel were secured
from the ``effective energy'' plots
for each of the MILC lattice ensembles,
and they were strenuously opted by looking for a combination of the
``plateau'' in the energy as the function of  $\rm D_{min}$,
good fit quality~\cite{Beane:2005rj,Beane:2007xs,Feng:2009ij},
and  $\rm D_{min}$ large enough to suppress the excited states.
We found that the effective energy of the $\pi\pi$ system
for the $I=2$ channel have relatively
small errors within a broad minimum time distance region.

For the same fitting range, analogously we secured the wraparound term $C$
from the corresponding ``effective wraparound constant'' plots
in Fig.~\ref{fig:eff_eng_C}.
It is worth mentioning that the fitted wraparound pollution $C$
are in fair agreement with the calculated wraparound pollution $C$ within errors.
Nonetheless, it is interesting to note that there exist about $1\%$ differences,
and the physical indication is not clear to us, which are highly needed to be
further investigated in the future work.

The fitted values of the energies $aE$ of the $\pi\pi$ system with isospin-$2$,
fitting range and fit quality ($\chi^2/{\rm dof}$)
are tabulated  in Table~\ref{tab:pp_I2},
together with the fitted values of the wraparound contribution $C$.
We note that the fitted values of $C$ is a statistically significant
constant term de facto for the lattice ensembles with small pion masses.
Additionally, we clearly found that these fitted values of $C$
are  close to our estimated values listed in Table~\ref{tab:C_C} as
already noticed in Ref.~\cite{Dudek:2012gj}.
\begin{table*}[ht]	
\caption{\label{tab:pp_I2}
Summaries of lattice results of the $s$-wave scattering lengths for the $I=2$ channel.
The second block presents the energies in lattice units,
where the first uncertainties are statistical,
the second ones are the estimates of the systematic
uncertainties due to fitting,
Column three shows the fitted values of the wraparound term $C$,
Column four indicates the time range for the chosen fit,
and Column five gives the fit quality $\chi^2\hspace{-0.05cm}/{\rm dof}$.
Column seven gives the center-of-mass scattering momentum $k^2$ in GeV,
and Column eight presents the product of the pion mass
and scattering length: $m_\pi a^{I=2}_{\pi\pi}$,
where the first uncertainty is statistical and, the second one is systematic.
}
\begin{ruledtabular}
\begin{tabular}{lllllll}
${\rm Ensemble}$ & $aE$ & $C$  & {\rm Range} & $\chi^2\hspace{-0.05cm}/{\rm dof}$
& $k^2$[${\rm GeV}^2$]  & $m_\pi a^{I=2}_{\pi\pi}$ \\
\hline
$(0.00484,0.0484)$   & $0.35520(25)(20)$  & $438.05(1.94)$
& $14\hspace{-0.05cm}-\hspace{-0.1cm}24$
& $4.21/6$  & $0.00167(10)(7)$      & $-0.0915(52)(35)$    \\
$(0.005,0.05)$       & $0.32424(35)(33)$  & $42.44(29)$
& $16\hspace{-0.05cm}-\hspace{-0.1cm}32$
& $10.6/12$ & $0.00220(21)(15)$     & $-0.125(11)(8)$   \\
$(0.007,0.05)$       & $0.38606(44)(37)$  & $1.776(17)$
& $16\hspace{-0.05cm}-\hspace{-0.1cm}32$
& $11.5/12$ & $0.00444(31)(20)$      & $-0.167(10)(7)$   \\
$(0.0097,0.0484)$    & $0.50087(41)(38)$  & $2.320(14)$
& $14\hspace{-0.05cm}-\hspace{-0.1cm}24$
& $6.1/6$  & $0.00437(25)(17)$      & $-0.167(9)(6)$   \\
$(0.01,0.05)$        & $0.45648(56)(41)$  & $0.1493(18)$
& $18\hspace{-0.05cm}-\hspace{-0.1cm}32$
& $8.7/10$ & $0.00464(48)(26)$     & $-0.209(19)(10)$ \\
$(0.0194,0.0484)$    & $0.69392(40)(34)$  & $0.01414(11)$
& $13\hspace{-0.05cm}-\hspace{-0.1cm}24$
& $6.6/7$ & $0.00527(35)(22)$     & $-0.277(16)(10)$ \\
\end{tabular}
\end{ruledtabular}
\end{table*}

It is well-known that the interaction among
two pions in the $I=2$ channel is pretty weak such that
the energy difference between the interacting and non-interacting $\pi\pi$ states
is a quite small fraction of total energy of the $\pi\pi$ system,\footnote{
In this work, the ratio of the energy shift to total energy is about $2\%$.
For other lattice studies~\cite{Sharpe:1992pp,Kuramashi:1993ka,
Fukugita:1994ve,Gupta:1993rn,Li:2007ey,Aoki:2002in,Du:2004ib,
Yamazaki:2004qb,Beane:2005rj,Beane:2007xs,Feng:2009ij,Dudek:2012gj,Yagi:2011jn,
Beane:2011sc,Dudek:2010ew},
it is actually close to this number.
On the other hand, this ratio for the $I=0$ channel is
around $5\%$~\cite{Kuramashi:1993ka,Fukugita:1994ve,Liu:2009uw},
and the ratio of this work is approximately to $5\%$ as well.
}
which can be estimated from the data in Tables~\ref{tab:m_pi_pi} and~\ref{tab:pp_I2}.
This forces us to make the rigorous measurements of
both the energies spectrum of the $\pi\pi$ system and pion masses,
and even seriously account for various small systematic effects
to resolve the rather small differences.
We have indeed extracted the energies of the $\pi\pi$ system
and pion masses with significantly high precision
which are shown in Tables~\ref{tab:m_pi_pi} and~\ref{tab:pp_I2}.

Now it is straightforward to substitute these energies $aE$  into L\"uscher formula~(\ref{eq:luscher})
and secure the relevant $s$-wave scattering lengths $a_{\pi\pi}^{I=2}$, where we plugged the pion masses in Column three in Table~\ref{tab:m_pi_pi}.
The center-of-mass scattering momentum $k^2$
is computed by Eq.~(\ref{eq:MF_k_e})
with pion masses given in Table~\ref{tab:m_pi_pi}.
However, to get rid of the scale-setting uncertainties,
it turns out to be more customary to adopt the dimensionless quantity:
$m_\pi a_{\pi\pi}^{I=2}$~\cite{Beane:2005rj,Beane:2007xs}.
All of these values for each lattice ensemble
are summarized in Table~\ref{tab:pp_I2},
where the statistical errors of $k^2$ are calculated from
the statistical errors of $aE$
and $am_\pi$, and
its systematic errors are only estimated from
the systematic errors of $aE$.
Likewise, the statistical errors for $m_\pi a_{\pi\pi}^{I=2}$ are computed
from the systematic errors of $k^2$ and $am_\pi$,
while its systematic errors are estimated from
the systematic errors of $k^2$ and the subsequently-mentioned
two finite volume effects.

Since the periodic boundary condition is imposed in the spacial directions
of the lattice, there has an exponentially small finite volume (FV) correction  to
the $s$-wave $\pi\pi$ scattering length in the $I=2$ channel,
which has been determined in the vicinity of the threshold in Ref.~\cite{Bedaque:2006yi}.
The consequent finite volume correction $\Delta_{FV}$ is provided here as~\cite{Bedaque:2006yi},
\begin{equation}
(m_\pi a_{\pi\pi}^{I=2})_{L}=(m_\pi a_{\pi\pi}^{I=2})_{\infty}+\Delta_{FV}
\end{equation}
where
\begin{eqnarray}
\hspace{-0.6cm}\label{eq:finite_volume} \Delta_{FV}
\hspace{-0.15cm}&=&\hspace{-0.15cm}
\frac{1}{2^{13/2}\pi^{5/2}}\left(\frac{m_\pi}{f_\pi}\right)^4
\hspace{-0.1cm}\sum_{{\bf n} \in \mathbb{Z}^3}\!\!\rule{0mm}{1em}^{\prime}
\frac{e^{-|{\bf{n}}|m_\pi L}}{\sqrt{|{\bf{n}}|m_\pi L}} \cr
&&  \hspace{-0.1cm}\times \hspace{-0.1cm}\left\{
\hspace{-0.06cm}1\hspace{-0.06cm}-\hspace{-0.06cm}\frac{17}{8}\frac{1}{|{\bf{n}}|m_\pi L}
\hspace{-0.06cm}+\hspace{-0.06cm}
\frac{169}{128}\frac{1}{|{\bf n}|^2 m_{\pi}^2 L^2}
\hspace{-0.06cm}+\hspace{-0.06cm}
{\cal O}(L^{-3})\hspace{-0.06cm}
\right\} \hspace{-0.06cm} ,
\end{eqnarray}
here $\sum_{ {\bf n} \in \mathbb{Z}^3 }^{\prime}$ indicates
a summation without ${\bf n} = {\bf 0}$.
Using this formula, we compute the finite volume corrections
to $m_\pi a^{I=2}_{\pi\pi}$, which are listed in Table~\ref{tab:FVC},
where we insert the values of $m_\pi L$ and $m_\pi/f_\pi$
listed in Table~\ref{tab:m_pi_pi}.
\begin{table}[ht]	
\caption{\label{tab:FVC}
Summaries of the finite volume corrections ${\rm \Delta_{FV}}$.
Column two shows the finite volume corrections to the $I=2$
$\pi\pi$ scattering length, and
Column three gives ratios of the finite volume corrections
to the corresponding statistical error.
Here we use the pion masses, pion decay constants
and $m_\pi L$ values listed in Table~\ref{tab:m_pi_pi}.
}
\begin{ruledtabular}
\begin{tabular}{lll}
${\rm Ensemble} $  & ${\rm \Delta_{FV}}$  & ${\rm Ratios}$  \\
\hline
$(0.00484,0.0484)$ & $0.000258$   &$0.049$       \\
$(0.005,0.05)$     & $0.000341$   &$0.031$       \\
$(0.007,0.05)$     & $0.000625$   &$0.060$       \\
$(0.0097,0.0484) $ & $0.000529$   &$0.062$       \\
$(0.01,0.05) $     & $0.000453$   &$0.023$       \\
$(0.0194,0.0484) $ & $0.000257$   &$0.016$
\end{tabular}
\end{ruledtabular}
\end{table}

From Table~\ref{tab:FVC}, we note that these corrections
are more and more important for the lattice ensembles with pion masses
smaller and smaller~\cite{Bedaque:2006yi}.
Since we use the lattice ensembles with small pion masses,
we should consider these effects,
although they are slight, and never more than $7\%$ of
the corresponding statistical errors~\cite{Feng:2009ij}.

Another important finite volume effect stems from effective range approximation,
$k \cot\delta(k)=1/a_{\pi\pi}^{I=2}+\frac{1}{2}r k^2$~\cite{Beane:2007xs}.
While the dependence on the effective range $r$ is small,
and the range truncation actually leads to the correction at $O(L^{-6})$
in L\"uscher formula (\ref{eq:luscher})~\cite{Beane:2007xs}.
In practice, we compute this correction for each lattice ensemble
as suggested in Ref.~\cite{Beane:2007xs}.

These two finite volume corrections have been also added in quadrature to
the systematical errors listed in Table~\ref{tab:pp_I2}.
Other sources of systematic uncertainty like: isospin violation,
finite volume effect due to the fixed global topology,
pion mass correction~\cite{Beane:2007xs,Yagi:2011jn}, etc.
are believed to be very small or we currently do not
have enough computational resources to fulfil it.
These effects should be incorporated into the more sophisticated
lattice computation at some points in the future.

\subsection{$I=0$ channel}

\begin{figure}[h!]
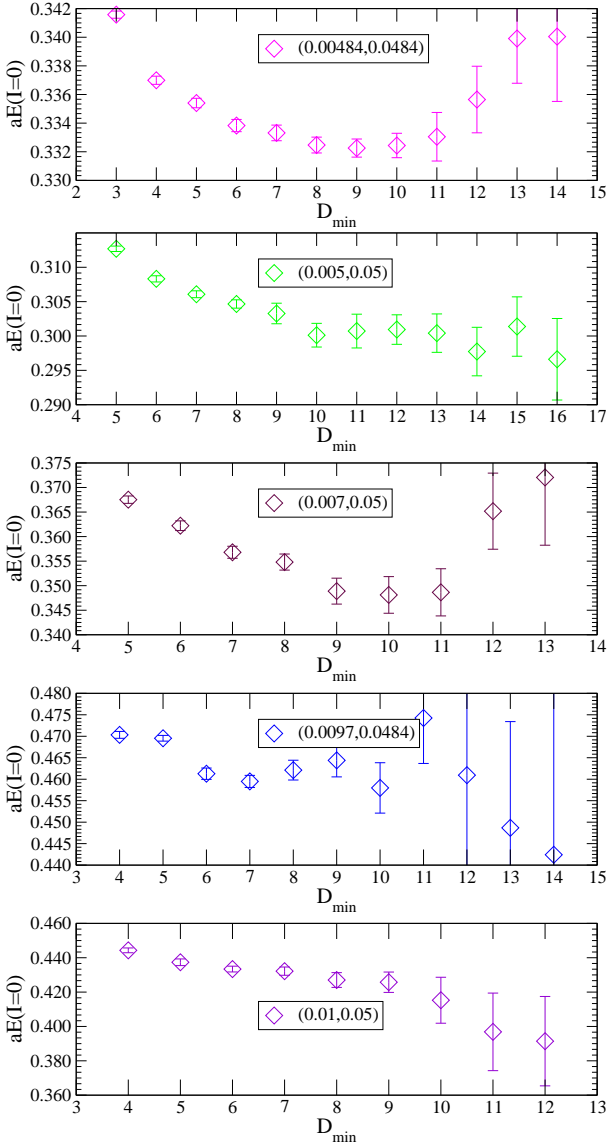

\includegraphics[width=8.0cm,clip]{I0_00484.eps}
\includegraphics[width=8.0cm,clip]{I0_005.eps}
\includegraphics[width=8.0cm,clip]{I0_007.eps}
\includegraphics[width=8.0cm,clip]{I0_0097.eps}
\includegraphics[width=8.0cm,clip]{I0_01.eps}
\caption{\label{fig:eff_eng_I0} (color online).
The ``effective energy'' plots
as the functions of $\rm D_{min}$
for the $\pi\pi$ scattering in the $I=0$ channel in lattice units.
}
\end{figure}

As already performed for the $I=2$ channel,
we analyze our lattice data
with the ``effective energy'' plot.
We should stress that
when using physical fitting model~(\ref{eq:E_pi_pi})
to extract the desired energies $aE$ of the $\pi\pi$ system,
we fix the fitting parameters of wraparound contribution $C$
with the estimated values listed in Tables~\ref{tab:C_C}.~\footnote{
For rectangular (R), and vacuum (V ) diagrams, there is no the wraparound pollution.
So the wraparound contribution for the $I=0$ channel is the same with
that in the $I=2$ channel.
It is reasonable to fix the  wraparound contribution $C$.
}
In practice, the $\pi\pi$ four-point functions were fit by
altering the minimum fitting distances $\rm D_{min}$,
and putting the maximum distance $\rm D_{max}$ either at $T/2$
or where the fractional statistical errors exceeded about $20\%$
for two sequential time slices~\cite{Bernard:2001av,DeGrand:2006zz}.
The ``effective energy'' plots as the functions of $\rm D_{min}$
are illustrated in Fig.~\ref{fig:eff_eng_I0}.
The central value and statistical error at each time slice were
evaluated by Levenberg-Marquardt method~\cite{Press:1992zz}.
To make these fits robust,
we double-check them with SNOBFIT~\cite{Waltraud:2008}.
We do not show the result of $(0.0194, 0.0484)$ ensemble
in Fig.~\ref{fig:eff_eng_I0} since it is too noisy.

For each  lattice ensemble, the energies $a E$ of the $\pi\pi$ system
for the $I = 0$ channel are secured from the ``effective energy'' plots,
and chosen by looking for a combination of a ``plateau'' in the energy
as the function of  $\rm D_{min}$, a good confidence level
and  $\rm D_{min}$ large enough to suppress the excited states~\cite{Beane:2005rj,Beane:2007xs,Feng:2009ij}.
In addition, as performed for isospin-$2$,
we approximately estimate the systematic errors
owing to fitting~\cite{Beane:2005rj,Beane:2007xs,Feng:2009ij},
which are not displayed in Fig.~\ref{fig:eff_eng_I0}.

\begin{table*}[t]	
\caption{\label{tab:pp_I0}
Summaries of the lattice results for the fitted energies of
the $\pi\pi$ system for the $I=0$ channel.
The second block shows the energies in lattice units,
where the first uncertainties are statistical, the second ones are
the estimates of the systematic uncertainties.
Column three shows the fitting range,
and Column four shows the number of degrees of freedom (dof) for the fit.
The six block shows the center-of-mass scattering momentum $k^2$ in GeV,
and Column seven gives the product of pion mass
and scattering length: $m_\pi a^{I=0}_{\pi\pi}$,
where the first uncertainty is statistical and, the second one is systematic.
}
\begin{ruledtabular}
\begin{tabular}{llllll}
${\rm Ensemble}$   & $a E$  &  {\rm Range} & $\chi^2/{\rm dof}$
& $k^2$[${\rm GeV}^2$] & $m_\pi a^{I=0}_{\pi\pi}$\\
\hline
$(0.00484,0.0484)$ & $0.33226(63)(78)$     & $9-24$               & $13.3/12$
                   & $-0.00572(21)(24)$    & $0.476(25)(29)$ \\
$(0.005,0.05)$     & $0.3013(16)(18)$      & $11-24$              & $14.6/10$
                   & $-0.00791(71)(76)$    & $0.811(123)(133)$\\
$(0.007,0.05)$     & $0.3499(23)(26)$      & $9-32$               & $28.3/20$
                   & $-0.0140(12)(13)$     & $1.181(202)(223)$
\end{tabular}
\end{ruledtabular}
\end{table*}

The fitted values of the energies $a E$ of the $\pi\pi$ system,
fit range and fit quality ($\chi^2/{\rm dof}$)
are summarized in Table~\ref{tab:pp_I0}.
The fit quality $\chi^2/{\rm dof}$ is reasonable for the $I=0$ channel.
It is well-known that the disconnected term give rise to the considerable fluctuations
to the $\pi\pi$ four-point function, and
it is pretty hard to reliably calculate this term.
In reality, only the lattice ensembles with small pion mass have a good signal.
From Fig.~\ref{fig:eff_eng_I0}, we found that, for the
$(0.097, 0.0484)$ and $(0.01, 0.05)$ ensembles, the plateaus are
not too obvious, so in this work, we don't include these results.

Nevertheless, one thing greatly comforting us is that
the interaction among two pions in this channel
is not too weak such that the discrete energies in a torus
are shifted relatively bigger than that of the  $I=2$ channel
from the values relevant for noninteracting pions,
and as we can see from Tables~\ref{tab:m_pi_pi} and~\ref{tab:pp_I0},
the energy shift between the interacting and non-interacting $\pi\pi$ states
is not a  too small fraction of total energy.
This indicates that the rigorous calculation of disconnected diagrams
is at present the most important thing.

The center-of-mass scattering momentum $k^2$
is calculated by Eq.~(\ref{eq:MF_k_e})
with pion masses listed in Table~\ref{tab:m_pi_pi},
and then the corresponding $s$-wave scattering lengths
$m_\pi a^{I=0}_{\pi\pi}$ can be obtained through Eq.~(\ref{eq:luscher}).
All of these values are summarized in Table~\ref{tab:pp_I0},
where the statistical errors of $k^2$ are calculated from
the statistical errors of the energies $aE$
and pion mass $am_\pi$, and
its systematic errors are only estimated from
the systematic errors of $aE$.
Likewise, the statistical for $m_\pi a_{\pi\pi}^{I=0}$ are computed
from the statistical errors of $k^2$ and $am_\pi$,
and its systematic errors are estimated from
the systematic errors of $k^2$
and one finite volume effect~\cite{Feng:2009ij}.

As already explained in the previous section,
the dependence on the effective range $r$ is small,
and the range truncation actually leads to
the finite volume correction at $O(L^{-6})$
in L\"uscher formula (\ref{eq:luscher})~\cite{Beane:2007xs}.
In practice, we compute this correction for each lattice ensemble
as suggested in Ref.~\cite{Beane:2007xs}.
This finite volume corrections have been combined in quadrature to
the systematical errors listed in Table~\ref{tab:pp_I0}.
Other sources of systematic uncertainty like: nonuniversal exponentially
suppressed corrections\cite{Bedaque:2006yi},
pion mass correction~\cite{Beane:2007xs,Bedaque:2006yi,Yagi:2011jn}, etc.
are believed to be very small as compared with the rather large systematic
error of the energies $aE$ or we currently do not
have enough computational resources to fulfil it.
With more reliable calculation of the energies $aE$ of
the $\pi\pi$ system in the $I=0$ channel in the future,
these effects should be eventually incorporated into
the more sophisticated lattice computation.

We should point out that, in this work,
we do not quote our results for
the $(0.01, 0.05)$, $(0.097, 0.0484)$ and $(0.0194, 0.0484)$ ensembles
due to two considerations:
First, the vacuum contributions of these ensembles are
noisy (see Fig.~\ref{fig:ratio}), and it is pretty hard to
see the clear plateau (see Fig.~\ref{fig:eff_eng_I0})
in the ``effective energy'' plots.
Second, the presence of the $\sigma$ resonance is clearly presented in low energy~\cite{Sasaki:2010zz,Fu:2011wc},
and thus it should be necessary for us to
map out ``avoided level crossings''
between $\sigma$ resonances and $\pi\pi$ states with isospin-$0$
to secure the reliable scattering
length as investigated in the $\pi K$ scattering
in Refs.~\cite{Sasaki:2010zz,Fu:2011wc}.
Luckily, as studied in Refs.~\cite{Sasaki:2010zz,Fu:2011wc,Albaladejo:2012te},
the contaminations from $\sigma$ meson for three lattice ensembles
with small pion masses are negligible.
Therefore, we only consider these results
in the rest of the analysis.

\section{Chiral extrapolations}
\label{sec:chiralExtrapolation}
In this work, we employed the rather small pion masses
ranging from $240$~MeV to $463$~MeV,
which are still larger than the physical one.
Therefore, $\chi$PT is needed to carry out a chiral extrapolation of
the scattering lengths to the physical pion mass.
The resulting NLO $\chi$PT formulas, which can be directly built
from the results in Ref.~\cite{Bijnens:1997vq}
(see Appendix~\ref{app:ChPT NNLO} for details),
are described as~\cite{Fu:2011bz,Feng:2009ij}
\begin{eqnarray}
\label{eq:fit}
\label{eq:m_pipi_I0}
\hspace{-0.6cm} m_\pi a^{I=0}_{\pi\pi}
 &=& \;\, \frac{7m_\pi^2}{16\pi f_\pi^2}
 \left\{1-\frac{m_\pi^2}{16\pi^2 f^2_\pi} \right. \cr
&&\times \left.\left[9\ln \frac{m_\pi^2}{f_\pi^2}-
5 - l_{\pi\pi}^{I=0}(\mu=f_{\pi,\mathrm{phy}})
\right]\right\} , \\
\label{eq:m_pipi_I2}
\hspace{-0.6cm} m_\pi a^{I=2}_{\pi\pi}
&=& -\frac{m_\pi^2}{8\pi f_\pi^2}
\left\{1+\frac{m_\pi^2}{16\pi^2 f^2_\pi} \right.\cr
&&\times\left.\left[3\ln \frac{m_\pi^2}{f_\pi^2}-1-
l_{\pi\pi}^{I=2}(\mu=f_{\pi,\mathrm{phy}})\right] \right\} ,
\end{eqnarray}
where the values of $m_\pi$ and $f_\pi$
listed in Table~\ref{tab:m_pi_pi} are inserted into,
and the $\chi$PT renormalization scale is fixed at the physical pion decay constant
$\mu=f_{\pi,\mathrm{phy}}$. Where and whereafter
a quantity with a ``phys'' subscript
are referred to as the value of that quantity in the physical case.
The $l_{\pi\pi}^{I=0}(\mu)$ and $l_{\pi\pi}^{I=2}(\mu)$
are the combinations of the LEC's in $\chi$PT at
a quark-mass independent scale $\mu$~\cite{Beane:2005rj,Beane:2007xs,Feng:2009ij}.
From the discussions in Appendix~\ref{app:ChPT NNLO}, the $l_{\pi\pi}^{I=0}(\mu)$
and $l_{\pi\pi}^{I=2}(\mu)$
are connected to the LEC's $\bar{l}_n$
as~\cite{Gasser:1984gg,Bijnens:1997vq}
\begin{eqnarray}
\label{eq:a0_pipi_I0}
l_{\pi\pi}^{I=0} &=&
\frac{40}{21}\bar{l}_1+\frac{80}{21}\bar{l}_2-\frac{5}{7}\bar{l}_3+4\bar{l}_4
+ 9\ln\frac{m_\pi^2}{f_{\pi,{\rm phy}}^2},\\
\label{eq:a0_pipi_I2}
l_{\pi\pi}^{I=2} &=&
\frac{8}{3}\bar{l}_1+\frac{16}{3}\bar{l}_2-\bar{l}_3-4\bar{l}_4
+ 3\ln\frac{m_\pi^2}{f_{\pi,{\rm phy}}^2}.
\end{eqnarray}
It should be noted that the Eqs.~(\ref{eq:m_pipi_I0}) and (\ref{eq:m_pipi_I2})
are expressed in terms of the full $f_\pi$ computed on the lattice,
and not the physical value $f_{\pi,\mathrm{phy}}$.
In reality, in the chiral expansion, the difference between
utilizing $f_\pi$ and $f_{\pi,{\rm phy}}$
in the argument of the logarithm only alters scattering lengths
at higher orders~\cite{Beane:2005rj,Beane:2007xs}.

As recommended in Refs.~\cite{Beane:2005rj,Beane:2007xs,Feng:2009ij},
we will carry out the extrapolation of the products $m_\pi a^{I=2}_{\pi\pi}$
and $m_\pi a^{I=0}_{\pi\pi}$  by means of the ratio $m_\pi/f_\pi$
in place of $m_\pi$.
From Appendix~\ref{app:ChPT NNLO},
we  note that extrapolating in $m_\pi/f_\pi$ in lieu of $m_\pi$
does transform the representations for
$m_\pi a^{I=2}_{\pi\pi}$ and $m_\pi a^{I=0}_{\pi\pi}$
but only at NNLO or higher.
Additionally, since $m_\pi/f_\pi$ is a dimensionless quantity,
there is no systematic error arising from the scale
setting~\cite{Beane:2005rj,Beane:2007xs,Feng:2009ij}.

We should remark that the lattice calculations
reported here used two lattice spacing of $0.15$~fm and $0.12$~fm.
Thus, it is meaningless to directly compare the energies $aE$ of the $\pi\pi$ system.
However, on the assumption that the L\"uscher technique
properly explains for the finite volume dependence of the energies $aE$
for these lattice ensembles, we can compare
$m_\pi a^{I=2}_{\pi\pi}$ and $m_\pi a^{I=0}_{\pi\pi}$
for two lattice spacings~\cite{Feng:2009ij}, and we observe such  agreement
with statistical error in Table~\ref{tab:pp_I2}.

\subsection{$I=2$ channel}
\begin{figure}[h]
\includegraphics[width=8.5cm,clip]{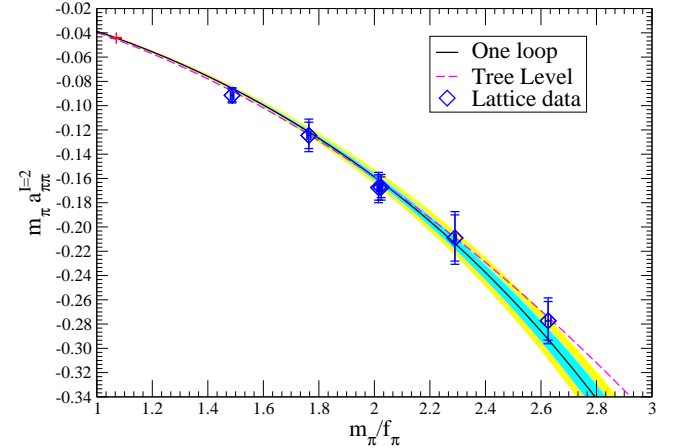}
\caption{\label{fig:ChPT-fit_I2} (color online).
The lattice-measured values of $m_\pi a^{I=2}_{\pi\pi}$
as a function of $m_\pi/f_\pi$.
The red plus point indicates the scattering length
at the physical limit: $(m_\pi a^{I=2}_{\pi\pi})_{\rm phys}$.
The shaded bands correspond to statistical (inner-cyan) errors
and statistical and systematic errors
combined in quadrature (outer-yellow).
The solid (black) curve is the central value of the NLO $\chi$PT fit.
The dashed (magenta) line is the tree-level $\chi$PT prediction.
}
\end{figure}

We are now in a position to fit lattice results of
$m_\pi a^{I=2}_{\pi\pi}$ in Table~\ref{tab:pp_I2}
to the NLO $\chi$PT functional form~(\ref{eq:m_pipi_I2})
to obtain low energy constant
$l_{\pi\pi}^{I=2}(\mu=f_{\pi,{\rm phy}})$,
then the extrapolated value at the physical point
$(m_\pi a^{I=2}_{\pi\pi})_{\rm phys}$ can be obtained.
The lattice-calculated values of $m_\pi a^{I=2}_{\pi\pi}$
as the function of $m_\pi/f_\pi$ are shown in Fig.~\ref{fig:ChPT-fit_I2},
and the outer error on the extrapolated result represents the
systematic error and statistical error combined in quadrature.
The one-loop $\chi$PT fit curve is displayed by the black solid line,
and the red plus point indicates
its physical $s$-wave scattering length: $(m_\pi a^{I=2}_{\pi\pi})_{\rm phys}$,
which is the chiral extrapolation of the $m_\pi a^{I=2}_{\pi\pi}$
at the physical limit.
In the same figure, we present the tree-level prediction as well.
It is important to note that lattice data manifests
pretty small displacement from the tree-level forecast.
Additionally, we  notice that
our lattice results for $m_\pi a^{I=2}_{\pi\pi}$
are in general agreement with the one-loop formula.
In fact, the deviation of $(m_\pi a^{I=2}_{\pi\pi})_{\rm phys}$
from tree-level prediction is a natural
aftermath of NLO $\chi$PT fitting~\cite{Feng:2009ij}.

\begin{table*}[ht!]
\caption{
A compilation of the various theoretical (or phenomenological), experimental
and lattice QCD determinations of $m_\pi a_{\pi\pi}^{I=2}$
extracted from the literature.
Together with every reference, for an easier comparison, the first author name
or the collaborations are given.
The first uncertainty is statistical and
second one is systematic if provided.
}
\label{tab:vardet_I_2}
\begin{ruledtabular}
\begin{tabular}{lll}
{\rm Ref.} &  $m_\pi a_{\pi\pi}^{I=2}$ & {\rm Remarks} \\
\hline
This work                        & $-0.04430 \pm 0.00025 \pm 0.00040$  &
{\rm  The calcualtion made in this paper. }      \\
Yagi(2011)~\cite{Yagi:2011jn}    & $-0.04410 \pm 0.00069 \pm 0.00018$  &
{\rm Extrapolation  with NNLO $\chi$PT. }  \\
Xu(2010)~\cite{Feng:2009ij}      & $-0.04385 \pm 0.00028 \pm 0.00038$ &
{\rm Using two flavors of maximally twisted mass fermions.} \\
NPLQCD(2007)~\cite{Beane:2007xs} & $-0.04330 \pm 0.00042$  &
{\rm Error combines statistical \& systematic errors in quadrature.} \\
CLQCD(2007)~\cite{Li:2007ey}     & $-0.0399  \pm 0.0070$ &
{\rm The result from Scheme I of anisotropic lattices. }  \\
NPLQCD(2005)~\cite{Beane:2005rj} & $-0.0426  \pm 0.0006 \pm 0.0003$ &
{\rm With fully-dynamical domain-wall valence-quark propagators. } \\
Du(2004)~\cite{Du:2004ib}        & $-0.0467  \pm 0.0045$  &
{\rm Using anisotropic lattices in an asymmetric box.}  \\
CP-PACS(2004)~\cite{Yamazaki:2004qb}   & $-0.0413  \pm 0.0029$  &
{\rm Compensating the mass dependence of the scattering length.}  \\
JLQCD(2002)~\cite{Aoki:2002in}   & $-0.0410  \pm 0.0069$  &
{\rm Selecting the result from EXP which employs a single exponential. }  \\
\hline
Albaladejo(2012)~\cite{Albaladejo:2012te} & $-0.0424  \pm 0.0012$ &
{\rm Employing unitary chiral perturbation theory. }\\
Guo(2009)~\cite{Guo:2009hi}      & $-0.0444  \pm 0.0011$ &
{\rm Providing full results for all the contributing ${\cal O}(p^6)$ couplings.} \\
Sasaki(2008)~\cite{Sasaki:2008sv}   & $-0.0431  \pm 0.0015$ &
{\rm Obtaining directly from the $\pi\pi$   wave function.} \\
MILC(2004)~\cite{Aubin:2004fs}   & $-0.0433  \pm 0.0009$ &
{\rm Using MILC's determinations of LECs along with Roy equations.}      \\
Zhou(2004)~\cite{Zhou:2004ms}    & $-0.0440  \pm 0.0011$
&{\rm Chiral unitarization with crossing symmetry \& phase shift data. } \\
CGL(2001)~\cite{Colangelo:2001df}& $-0.0444  \pm 0.0010$ &
{\rm Two loop accuracy. } \\
Weinberg(1966)~\cite{Weinberg:1966kf} & $-0.04557 \pm 0.00014$
& {\rm Tree level prediction.} \\
\hline
NA48/2(2011)~\cite{Bizzeti:2011zza,Wanke:2011zz}&$-0.0429\pm0.0044\pm0.0016$ &
{\rm With independent experimental errors \& different theoretical inputs.} \\
E865(2003)~\cite{Pislak:2003sv}                 &$-0.0454\pm0.0031\pm0.0010$ &
{\rm With the $\chi$PT constraints in the analysis.}  \\
\end{tabular}
\end{ruledtabular}
\end{table*}

In principle,  we can fit our lattice-calculated data to the NNLO $\chi$PT
form for $m_\pi a^{I=2}_{\pi\pi}$~\cite{Colangelo:2001df,Bijnens:1997vq}
(see the concrete form in Eq.~(\ref{neq:m_pipi_I2}))
as it is done in Ref.~\cite{Yagi:2011jn},
since we have six lattice data at our disposal.
In the meantime, we can make an estimate of NNLO LECs
with the careful analysis for the chiral extrapolation of $m_\pi a^{I=2}_{\pi\pi}$,
since we have lattice data points at the lighter quark masses.
However, the NNLO fit has larger errors in both $l_{\pi\pi}^{I=2}$
and $m_\pi a^{I=2}_{\pi\pi}$ than those with NLO fit as shown in Refs.~\cite{Beane:2005rj,Beane:2007xs,Feng:2009ij,Yagi:2011jn},
and the errors of the LECs $l_{\pi\pi,I=2}^{(2)}$ and $l_{\pi\pi,I=2}^{(3)}$ are
rather large like the corresponding obtained values in Ref.~\cite{Yagi:2011jn}.
Therefore, the calculations of NNLO LECs with physical meaning
can not be obtained in this work.
A rigorous NNLO $\chi$PT fit should wait for more lattice
data at pion masses further closer to the physical point
than we presently have.
Admittedly, the resulting NNLO extrapolated value of $m_\pi a^{I=2}_{\pi\pi}$
is indeed in harmony with NLO fit as we expect~\cite{Feng:2009ij}.
Actually,  we  use the NNLO $\chi$PT functional form
to estimate systematic errors due to
truncating the $\chi$PT series to NLO form~\cite{Beane:2005rj}.

As  practiced in Ref.~\cite{Feng:2009ij},
we only consider three major sources of systematic uncertainty
on the extrapolated value of $m_\pi a^{I=2}_{\pi\pi}$ and
$l_{\pi\pi}^{I=2}$.
First, the lattice-calculated systematic errors of
$m_\pi a^{I=2}_{\pi\pi}$ per ensemble are spread
by the chiral extrapolation~\cite{Feng:2009ij}.
Second, the systematic error inherently stems
from NLO $\chi$PT fit~\cite{Beane:2005rj,Beane:2007xs},
which can be roughly calculated by taking the  discrepancy
between the NLO $\chi$PT extrapolated value from all six data sets
and that from ``pruning'' the heaviest data set~\cite{Feng:2009ij}.
Third, the experimental errors on $m_\pi$ and $f_\pi$~\cite{Beringer:1900zz}
give rise to another important source of systematic error~\cite{Feng:2009ij}.
All three components are combined in quadrature to
make up the entire systematic error.
Taking the latest PDG data~\cite{Beringer:1900zz}
for the most accurate charged pion mass $m_{\pi}=m_{\pi^+}=139.57018(35)$~MeV
and pion decay constant $f_{\pi} = f_{\pi^+}=130.41(20)$~MeV,
where a couple of experimental errors are added in quadrature,
hence $m_{\pi}/f_{\pi}=1.07024(166)$,
we finally secure the upshots
\begin{eqnarray}
m_\pi a^{I=2}_{\pi\pi}                    &=& -0.04430(25)(40); \cr
l_{\pi\pi}^{I=2}(\mu=f_{\pi, {\rm phys}}) &=& 3.27(.77)(1.12),
\end{eqnarray}
where the first uncertainty is statistical
and the second one is an estimate of the systematic error.

These outcomes are comparable with the aforementioned results of theoretical
(or phenomenological) studies~\cite{Weinberg:1966kf,Zhou:2004ms,
Guo:2009hi,Albaladejo:2012te,Colangelo:2001df},
experimental determinations~\cite{Bizzeti:2011zza,Wanke:2011zz,Pislak:2003sv}
and lattice calculations~\cite{Yagi:2011jn,Feng:2009ij,Beane:2007xs,Beane:2005rj}
within statistical errors.
The relevant results for $m_\pi a_{\pi\pi}^{I=2}$
are courteously compiled in Table~\ref{tab:vardet_I_2}.
The first group is lattice QCD results.
The second one is theoretical (or phenomenological) studies.
Also included are two experimental values in the third group.

\begin{figure}[h]
\includegraphics[width=8.5cm,clip]{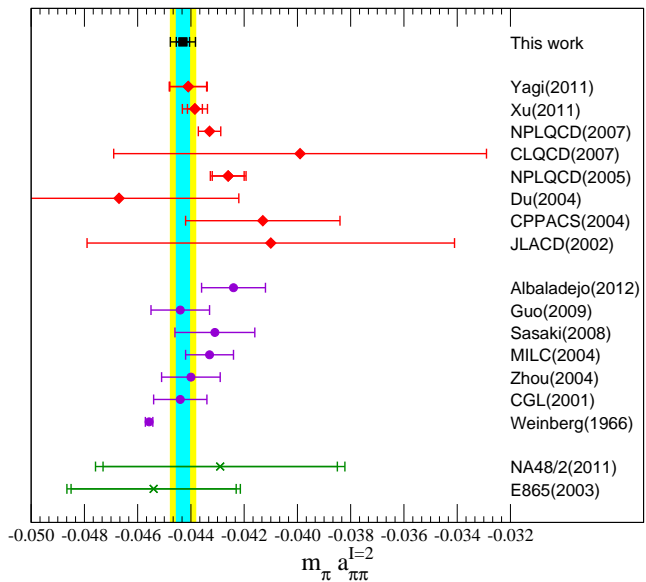}
\caption{\label{fig:compare_I2} (color online).
A collection of  various lattice QCD, theoretical (or phenomenological) and
experimental results of $m_\pi a_{\pi\pi}^{I=2}$
listed in Table~\ref{tab:vardet_I_2}.
The lattice studies are shown by red diamonds,
purple circles are theoretical (or phenomenological) predictions,
and the experimental ones are given by green crosses.
Our result is indicated by a black square. For an easier comparison,
the (cyan) inner strip corresponds to the statistical error
whereas the (yellow) outer strip represents
the statistical and systematic errors added in quadrature.
}
\end{figure}

To make our demonstrations of these results more intuitive,
they are offered  graphically in Fig.~\ref{fig:compare_I2},
where we clearly note that
the various results for every $m_\pi a_{\pi\pi}^{I=2}$
are fairly compatible with each other within errors.

Our calculation of the LEC $l_{\pi\pi}^{I=2}$ is satisfactory as well,
although it is just about $25\%$ precision,
while it can be comparable with relevant results obtained
by phenomenological predictions~\cite{Colangelo:2001df},
experimental determinations~\cite{Bizzeti:2011zza,Wanke:2011zz,Pislak:2003sv},
and lattice calculations~\cite{Yagi:2011jn,Feng:2009ij,Beane:2007xs,Beane:2005rj}.
The relevant values of $l_{\pi\pi}^{I=2}$ are
collected in Table~\ref{tab:vardet_L2}.
The first group is lattice results.
The second one is phenomenological and experimental determinations,
which are transformed directly from the experimental and
phenomenological results of
$m_\pi a_{\pi\pi}^{I=2}$ into $l_{\pi\pi}^{I=2}$ at NLO $\chi$PT
as preformed in Ref.~\cite{Feng:2009ij}.

\begin{table}[h]
\caption{
Some values of $l_{\pi\pi}^{I=2}$ extracted from the literature.
The first uncertainty is statistical,
and the second one is systematic if present.
The first group is lattice QCD results.
The second one is phenomenological and experimental determinations,
which are transformed directly from the corresponding results
of $m_\pi a_{\pi\pi}^{I=2}$ into $l_{\pi\pi}^{I=2}$
at NLO $\chi$PT~\cite{Feng:2009ij}.
}
\label{tab:vardet_L2}
\begin{ruledtabular}
\begin{tabular}{ll}
{\rm Ref.} &  $l_{\pi\pi}^{I=2}$  \\
\hline
This work                          & $3.27 \pm 0.77 \pm 1.12$  \\
Yagi(2011)~\cite{Yagi:2011jn}      & $5.8  \pm 1.2$   \\
Xu(2010)~\cite{Feng:2009ij}        & $4.65 \pm 0.85 \pm 1.07$  \\
NPLQCD(2007)~\cite{Beane:2007xs}   & $6.2  \pm 1.2$  \\
NPLQCD(2005)~\cite{Beane:2005rj}   & $3.3  \pm 0.6 \pm 0.3$  \\
\hline
CGL(2001)~\cite{Colangelo:2001df}  & $3.0  \pm 3.1$~\footnote{
It is interesting to note that if we make use of Eq.~(\ref{eq:a0_pipi_I2})
with the values of $\bar{l}_i$ reported in Ref.~\cite{Colangelo:2001df},
and necessary PDG values, we obtain $l_{\pi\pi}^{I=2}=2.0 \pm 3.1$.
} \\
NA48/2(2011)~\cite{Bizzeti:2011zza,Wanke:2011zz}
                                   &$7.5\pm13.3\pm4.8$~\footnote{
Using the data from Ref.~\cite{Bizzeti:2011zza}.
}  \\
E865(2003)~\cite{Pislak:2003sv}    & $0.0  \pm 9.4 \pm 3.0$ \\
\end{tabular}
\end{ruledtabular}
\end{table}

The reason why we made a significant improvement in precision
over our previous work~\cite{Fu:2011bz}
is the recent comprehension of various lattice-spacing artifacts
(in particular wraparound effect).
In fact, approximate $0.5\%$ accuracy of
our ultimate result for $m_\pi a_{\pi\pi}^{I=2}$
is typically joint efforts
from lattice QCD and $\chi$PT.
This can be understood from two aspects:
First, we have lattice data closer to the physical point,
which have relatively smaller uncertainties for $m_\pi a_{\pi\pi}^{I=2}$.
Second, the chiral extrapolation of $m_\pi a_{\pi\pi}^{I=2}$ is considerably
restricted by $\chi$PT
and $m_\pi a_{\pi\pi}^{I=2}$ is solely predicted
in terms of $m_\pi/f_\pi$ at LO
and depends uniquely upon a LEC, $l_{\pi\pi}^{I=2}$, at NLO.
This means that the statistical error of NLO $\chi$PT extrapolation of
$m_\pi a_{\pi\pi}^{I=2}$ solely rests on the statistical error of $l_{\pi\pi}^{I=2}$.
Consequently, although our lattice-calculated results of $m_\pi a_{\pi\pi}^{I=2}$
are only with about $6\% \sim 10\%$ accuracy,
we still obtain less than $1\%$ precise determination of
$(m_\pi a^{I=2}_{\pi\pi})_{\rm phys}$.

\subsection{$I=0$ channel}
\begin{figure}[h]
\includegraphics[width=8.0cm,clip]{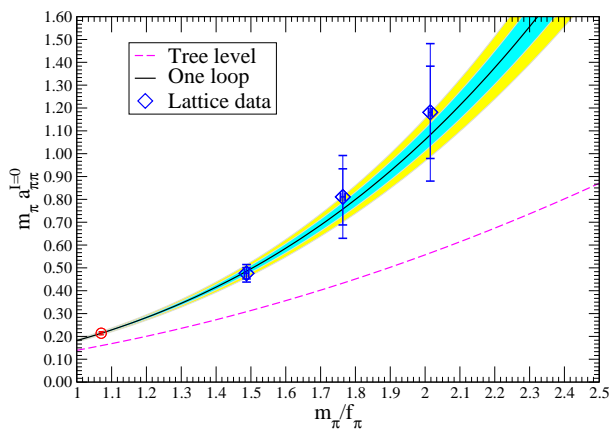}
\caption{\label{fig:ChPT-fit_I0} (color online).
The lattice-measured values of $m_\pi a^{I=0}_{\pi\pi}$
as a function of $m_\pi/f_\pi$.
The red circle point indicates the scattering length
at the physical limit: $(m_\pi a^{I=0}_{\pi\pi})_{\rm phys}$.
The shaded bands correspond to statistical (inner-cyan) errors
and statistical and systematic errors combined in quadrature (outer-yellow).
The solid (black) curve is the central value of the NLO $\chi$PT fit.
The dashed (magenta) line is the tree-level $\chi$PT prediction.
}
\end{figure}

We are now in a position for fitting lattice results of
$m_\pi a^{I=0}_{\pi\pi}$ in Table~\ref{tab:pp_I0}
to NLO $\chi$PT functional form~(\ref{eq:m_pipi_I0})
to secure the low energy constant
$l_{\pi\pi}^{I=0}(\mu=f_{\pi,{\rm phy}})$,
then obtain the extrapolated value at the physical point
$(m_\pi a^{I=0}_{\pi\pi})_{\rm phys}$.
The lattice-measured values of $m_\pi a^{I=0}_{\pi\pi}$
as the function of $m_\pi/f_\pi$ are demonstrated in Fig.~\ref{fig:ChPT-fit_I0}.
The one-loop $\chi$PT fit curve is displayed by the black solid line,
and the red circle point indicates
its physical $s$-wave scattering length: $(m_\pi a^{I=0}_{\pi\pi})_{\rm phys}$,
which is the chiral extrapolation of the $m_\pi a^{I=0}_{\pi\pi}$
at the physical point.
In the same figure, we present the tree-level prediction as well.
It is important to note that lattice data manifests
a rather large displacement from the tree-level $\chi$PT forecast,
which is well consistent with one conclusion of Ref.~\cite{Colangelo:2001df}
that the NLO corrections make about
$25\%$ modification to tree-level prediction.
Additionally, we can notice that
our lattice results for $m_\pi a^{I=0}_{\pi\pi}$
generally agree with the one-loop formula.
The large deviation of $(m_\pi a^{I=0}_{\pi\pi})_{\rm phys}$
from tree-level prediction is totally a
natural aftermath of the NLO $\chi$PT fitting.

\begin{table*}[!ht]
\caption{
A compilation of the various theoretical (or phenomenological), experimental
and lattice QCD determinations of $m_\pi a_{\pi\pi}^{I=0}$
extracted from the literature.
Together with every reference, for an easier comparison, the first author
or the collaborations are given.
The first uncertainty is statistical and
the second one is systematic if given.
}
\label{tab:vardet_I_0}
\begin{ruledtabular}
\begin{tabular}{lll}
{\rm Ref.} &  $m_\pi a_{\pi\pi}^{I=0}$  & {\rm Remark} \\
\hline
This work                                  & $0.214  \pm 0.004 \pm 0.007$&
{\rm Full QCD.} \\
Fu(2012)~\cite{Fu:2011bz}                  & $0.186  \pm 0.002$  &
{\rm Partially quenched QCD.} \\
\hline
Albaladejo(2012)~\cite{Albaladejo:2012te}  & $0.219  \pm 0.005$ &
{\rm Employing unitary chiral perturbation theory. }\\
Guo(2009)~\cite{Guo:2009hi}                & $0.220  \pm 0.005$  &
{\rm Full results for all the contributing ${\cal O}(p^6)$ couplings.} \\
Caprini(2008)~\cite{Caprini:2008fc}        & $0.218  \pm 0.014$ &
{\rm Using a large class of analytic parameterizations.} \\
Yndurain(2007)~\cite{Yndurain:2007qm}      & $0.233  \pm 0.013$ &
{\rm Extrapolating to the pole of the  sigma resonance.} \\
Zhou(2004)~\cite{Zhou:2004ms}              & $0.211  \pm 0.011$ &
{\rm Using chiral unitarization with crossing symmetry \& phase shift data. } \\
CGL(2001)~\cite{Colangelo:2001df}          & $0.220  \pm 0.005$ &
{\rm Two loop accuracy. } \\
Weinberg(1966)~\cite{Weinberg:1966kf}      & $0.1595 \pm 0.0005$ &
{\rm Tree level prediction. }  \\
\hline
NA48/2(2011)~\cite{Bizzeti:2011zza,Wanke:2011zz}& $0.2210 \pm 0.0047 \pm 0.0015$  &
{\rm With independent experimental errors \& different theoretical inputs.} \\
E865(2003)~\cite{Pislak:2003sv}               & $0.216  \pm 0.013 \pm 0.002$ &
{\rm With the $\chi$PT constraints in the analysis.}
\end{tabular}
\end{ruledtabular}
\end{table*}

In this work we can not fit our
lattice-calculated data  to the NNLO $\chi$PT functional
form~(\ref{eq:m_pipi_I0})~\cite{Colangelo:2001df,Bijnens:1997vq}
(see concrete form~(\ref{neq:m_pipi_I2})),
since we just have three lattice data at our disposal.
A NNLO $\chi$PT determination should wait for more lattice
data at pion masses further closer to the physical point
than we now have.
Admittedly, as explained in detail in Ref.~\cite{Colangelo:2001df},
the NLO correction increases the LO prediction by about $25\%$,
and NNLO correction further raises the LO prediction more than $10\%$,
this mean that the NNLO $\chi$PT determination of $m_\pi a^{I=0}_{\pi\pi}$
should be significantly more rigorous than the  NLO $\chi$PT determination
of $(m_\pi a^{I=0}_{\pi\pi})_{\rm phys}$, and this can in part explain
the relative large error for our final NLO $\chi$PT extrapolated
result of $(m_\pi a^{I=0}_{\pi\pi})_{\rm phys}$.
As a consequence, the systematic error from truncating the $\chi$PT series
to the NLO form should be considered.

In this work, we only consider two major sources of systematic errors
in the extrapolated value of $m_\pi a^{I=0}_{\pi\pi}$
since the systematic error from the experimental errors on $m_\pi$ and $f_\pi$ is
found to be pretty small as compared with its statistical error.
First, the lattice-computed systematic errors of $m_\pi a^{I=0}_{\pi\pi}$
per ensemble are propagated via the chiral extrapolation.
Second, the systematic error inherently stems from NLO  fit,
which is computed by taking the difference between
the extrapolated values from NLO fit to all three data set
and that from `cropping'' the heaviest data set~\cite{Feng:2009ij}.
All two parts are added in quadrature to
give the whole computed systematic error.
We get the final upshot
\begin{eqnarray}
m_\pi a^{I=0}_{\pi\pi} &=& 0.214(4)(7); \cr
l_{\pi\pi}^{I=0}(\mu=f_{\pi, {\rm phys}}) &=& 43.2\pm  3.5 \pm 5.6,
\end{eqnarray}
where the numbers in the first and second parentheses
are the statistical and systematic uncertainty, respectively.

These results can be fairly comparable with the above-mentioned results
by theoretical (or phenomenological) studies~\cite{Weinberg:1966kf,Zhou:2004ms,Guo:2009hi,Albaladejo:2012te,
Colangelo:2001df,Caprini:2008fc,Yndurain:2007qm} (except the tree-level prediction),
and experimental determinations~\cite{Bizzeti:2011zza,Wanke:2011zz,Pislak:2003sv}.
The relevant results for $m_\pi a_{\pi\pi}^{I=0}$
are compiled in Table~\ref{tab:vardet_I_0}.
The first group is lattice results.
The second one is theoretical (or phenomenological) studies.
Also contained are two experimental values in third group.

\begin{figure}[ht]
\includegraphics[width=8.0cm,clip]{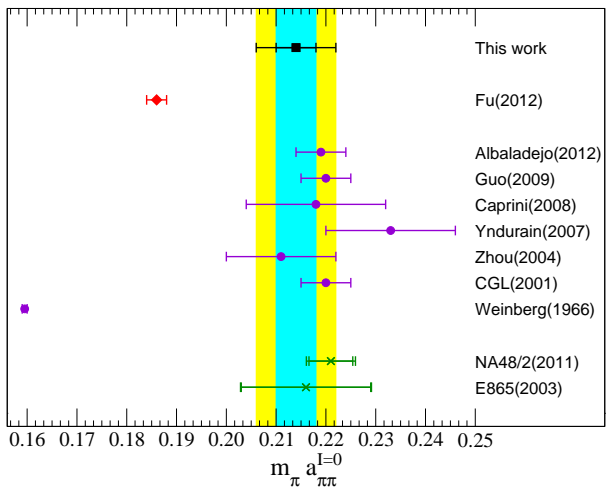}
\caption{\label{fig:compare_I0} (color online).
A collection of  various lattice QCD, theoretical (or phenomenological) and
experimental results of $m_\pi a_{\pi\pi}^{I=0}$
listed in Table~\ref{tab:vardet_I_0}.
The red diamonds are lattice determinations,
purple circles are theoretical (or phenomenological) studies,
and experimental ones are represented by green crosses.
Our result is shown by a black square.
For an easier comparison,
the (cyan) inner strip corresponds to the statistical error
whereas the (yellow) outer strip represents
the statistical and systematic errors added in quadrature.
}
\end{figure}

To make our report of these results more intuitive,
these results are given graphically in Fig.~\ref{fig:compare_I0} as well,
where the various results of $m_\pi a_{\pi\pi}^{I=0}$
are compatible with each other within errors
except the tree-level prediction~\cite{Weinberg:1966kf}
and our crude lattice study~\cite{Fu:2011bz}.

Our calculation of the LEC $l_{\pi\pi}^{I=0}$ is satisfactory as well,
although it is about $5\%$ accuracy,
nevertheless, as we show soon,
this result can be comparable with the results
of phenomenological~\cite{Colangelo:2001df} and
experimental determinations~\cite{Bizzeti:2011zza,Wanke:2011zz,Pislak:2003sv},
and lattice studies~\cite{Yagi:2011jn,Feng:2009ij,Beane:2007xs,Beane:2005rj}.
The relevant values of $l_{\pi\pi}^{I=0}$ are
collected in Table~\ref{tab:vardet_L0}.
The first group is lattice results.
The second one is phenomenological and experimental determinations,
which are converted directly from the experimental  phenomenological
results of $m_\pi a_{\pi\pi}^{I=0}$ into $l_{\pi\pi}^{I=0}$ at NLO $\chi$PT
as conducted by Xu et al. in Ref.~\cite{Feng:2009ij} for the $I=2$ channel.

\begin{table}[ht]
\caption{
Some values of $l_{\pi\pi}^{I=0}$ extracted from the literature.
The first uncertainty is statistical,
and the second one is systematic if present.
The first group is lattice simulation results.
The second one is phenomenological and experimental
determinations, which are directly transformed
from the corresponding results of $m_\pi a_{\pi\pi}^{I=0}$
into $l_{\pi\pi}^{I=0}$ at NLO $\chi$PT~\cite{Feng:2009ij}.
}
\label{tab:vardet_L0}
\begin{ruledtabular}
\begin{tabular}{ll}
{} &  $l_{\pi\pi}^{I=0}$ \\
\hline
This work                          & $43.2 \pm 3.5 \pm 5.6$ \\
Fu(2012)~\cite{Fu:2011bz}          & $18.7 \pm 1.2$ \\
\hline
CGL(2001)~\cite{Colangelo:2001df}  & $48.5 \pm 4.3$~\footnote{
If we make use of Eq.~(\ref{eq:a0_pipi_I0}) with
the values of $\bar{l}_i$ reported in Ref.~\cite{Colangelo:2001df},
and required PDG values, we get $l_{\pi\pi}^{I=0}=32.4 \pm 2.3$.
}\\
NA48/2(2011)~\cite{Bizzeti:2011zza}& $49.3 \pm 4.1  \pm 1.3$ \\
E865(2003)~\cite{Pislak:2003sv}    & $45.0 \pm 11.2 \pm 3.5$
\end{tabular}
\end{ruledtabular}
\end{table}

The remarkable improvement in accuracy over
our previous results~\cite{Fu:2011bz} is
an joint effort from lattice QCD and $\chi$PT
First, we have measured the lattice data closer to the physical point,
which have smaller uncertainties.
Second, chiral perturbation theory considerably constrains
the chiral extrapolation of the product $m_\pi a_{\pi\pi}^{I=0}$,
which is uniquely predicted
in terms of $m_\pi/f_\pi$ at LO
and relies solely on a LEC, $l_{\pi\pi}^{I=0}$, at NLO.
This suggests that the statistical error of NLO $\chi$PT extrapolation of
$m_\pi a_{\pi\pi}^{I=0}$ solely rests on the statistical error of $l_{\pi\pi}^{I=0}$.
Consequently, although our lattice-calculated results of $m_\pi a_{\pi\pi}^{I=0}$
are only with $5\% \sim 17\%$ accuracy,
we still obtain about $2\%$ precise determination of
$(m_\pi a^{I=0}_{\pi\pi})_{\rm phys}$.

\section{Summary and outlook}
\label{sec:conclude}
We have reported results of a lattice QCD calculation
of the $s$-wave $\pi\pi$ scattering lengths
for both $I = 0$ and $2$ channels on
the MILC ``medium coarse'' ($a\approx0.15$ fm) and
``coarse'' ($a\approx0.12$ fm) lattice
ensembles with the $2+1$ flavors of
the Asqtad-improved staggered sea quarks.
We exploited the moving wall sources
without the gauge fixing~\cite{Kuramashi:1993ka,Fukugita:1994ve}
to compute all the four diagrams assorted
in Refs.~\cite{Sharpe:1992pp,Kuramashi:1993ka},
and viewed a clear attractive signal for the $I=0$ channel
and a good repulsive one for the $I=2$ channel, respectively.
Moveover, extrapolating our lattice data of
the $s$-wave scattering lengths for both isospin eigenstates
to the physical pion mass  gives the scattering lengths:
$m_\pi  a_{\pi\pi}^{I=2} = -0.04430(25)(40)$ and
$m_\pi  a_{\pi\pi}^{I=0} = 0.214(4)(7)$
for the $I=2$ and $0$ channels, respectively,
which are in fair agreement with the current theoretical (or phenomenological)
predictions to one-loop levels and the present experimental reports,
and can be comparable with other lattice studies.

After our extremely crude estimation of the $\pi\pi$ scattering length
in the $I=0$ channel in Ref.~\cite{Fu:2011bz},
and this relatively more sophisticated computation,
we can fairly claim that even with the limited computing
resources, the lattice calculation of
the $\pi\pi$ scattering length in the $I=0$ channel is
feasible although this work is absolutely needed to
be further improved,
and the various sources of systematic error
yet need to be clarified thoroughly.
Most of all,  from this work, we found that the rule of thumb estimation of lattice ensemble with the Goldstone pion mass eligible to study the $s$-wave
$\pi\pi$ scattering in the $I=0$ channel
should be less than about $300$~MeV (the smaller, the better),
which is very helpful for the people to pursue this fascinating enterprise.
We view it as one of the important results of this work.

As we revealed, a reasonable signal can be gained for
the $(0.00484, 0.0484)$ and $(0.005, 0.05)$ ensembles
in the vacuum diagram of the $\pi\pi$ scattering.
In principle, the signal-to-noise ratio can be further enhanced
by launching the same calculation on the lattice ensembles with a smaller
pion mass (of course we can also improve the statistics by using
more lattice gauge configurations or performing the calculation on a larger volume).
In addition, the behavior close the physical point is intensely influenced
by the chiral logarithm term, so an extraction of the
$\pi\pi$ scattering lengths without a long extrapolation
is still highly needed to guarantee the convergence of the chiral expansion.
Fortunately, the MILC collaboration has generated
enough lattice ensembles whose Goldstone pion masses
are smaller than or close to $240$~MeV~\cite{Colangelo:2010et,Bazavov:2009fk}
(e.g., the fine $(0.00155, 0.0310)$ ensemble,
whose Goldstone pion mass is about $177$~MeV).
Furthermore, as we explained early, the NNLO $\chi$PT
will be the proper physical functional form to fit the lattice data
(at least four data point) of
the $I=0$ $\pi\pi$ scattering length
and it needs more lattice data near the physical point.
I have an impetus to do these works.
However,  it is beyond the scope of this work
since this will need an astronomical amount of computing allocations.
We will enthusiastically
appeal for all the possible computational resources
to accomplish this peculiar and challenging task.

It is well-known that $\pi\pi$ scattering in the $I=0$ channel is
challenging and stimulating phenomenologically
due to the existence of the $\sigma$ resonance.
In the work, we have exhibited that the $\pi\pi$ scattering
for the $I=0$ channel can be reliably calculated
by the moving wall sources without gauge
fixing~\cite{Kuramashi:1993ka,Fukugita:1994ve}.
It gives us an anticipation that this technique can be successfully
exploited to study the $\sigma$ resonance.
In our previous work~\cite{Fu:2011zzh},
we have evaluated the $\sigma$ mass from lattice QCD,
and found that the decay $\sigma \to \pi\pi$ is
allowed kinematically only for enough small $u$ quark mass.
This work and our lattice calculation
for the $\pi\pi$ scattering lengths delivered in this paper
will encourage the researchers to study the $\sigma$ resonance.
We have been investigating the $\sigma$ resonance parameters
with the isospin representation of $(I, I_z) = (0, 0)$,
and the preliminary lattice results are already reported
in Ref.~\cite{Fu:2012gf}.

Additionally, for $\pi\pi$ scattering in the $I = 0$ channel,
we realize an important issue that the presence of $\sigma$ resonance
is possible in the low-energy, and thus it should be necessary
for us to employ the variational
method~\cite{Luscher:1990ck} to
secure the rigorous scattering length
as investigated in the $\pi K$ scattering in the $I=1/2$ channel
in Refs.~\cite{Sasaki:2010zz,Fu:2011wc}.
Since we only make use of the relative small quark masses to
study the $\pi\pi$ scattering in the $I=0$ channel,
we can temporarily  and reasonably overlook this contamination
in the present study as already explained in Ref.~\cite{Fu:2011wc}.
However, we should bear in mind that
this issue should be settled in the  more sophisticated lattice examination.
It will be very interesting to systematically study
this pollution for the $\pi\pi$ scattering in the $I=0$ channel.

Admittedly, due to the intense theoretical and experimental efforts put into the
scalar-isoscalar and scalar-isovector sector of the meson-meson scattering recently,
studying the $K \overline{K}$ scattering length on the lattice
is gradually becoming a very interesting enterprise.
As pointed out in Ref.~\cite{Savage:2011xk},
the robust calculation of the $\pi\pi$ scattering  lengths
(in particular for the $I=0$ channel) will naturally encourage us to
study other challenging systems like $K \overline{K}$, etc.
Physically,  as explained in Ref.~\cite{Fu:2012jb,Doring:2011vk,Oller:1998hw},
studying $K \overline{K}$ is very important, and
the calculation of the $s$-wave $K \overline{K}$ scattering length
in the $I=0$ channel is
a genuine two coupled-channel problem~\cite{Doring:2011vk,multi_channel},
where the system can be approximately described
by only $\pi\pi$ and $K\overline{K}$ two channels
(we refer to $\pi\pi$ as channel $1$ and $K \overline{K}$ channel $2$),
then the $S$-matrix is a $2\times 2$ unitary matrix
which contains $3$ real parameters~\cite{Doring:2011vk,multi_channel}
(the $s$-wave $K \overline{K}$ scattering in the $I = 1$ channel can be treated analogously~\cite{Fu:2012ng},
and the lattice study of $\pi\eta$ scattering is then highly desirable.).
Therefore, it is absolutely  necessary to
incorporate the $s$-wave $I=0, \pi\pi$-channel for a physical
calculation of the $s$-wave $K \overline{K}$ scattering length in the $I=0$ channel.
The generalized L\"uscher's formula in this case gives a relation among
these three parameters, all of which are functions of the
energy~\cite{multi_channel}.
since some of these parameters
are still poorly measured in the present experiments,
the lattice calculation is valuable and highly desirable.
With our lattice efforts on channel $2$ in Ref.~\cite{Fu:2012jb}, at present,
if we can compute the $s$-wave $\pi\pi \to K\overline{K}$ scattering in the
$I=0$ channel, in principle, we can solve this problem.
We are launching a series of lattice studies for these aims.

\section*{Acknowledgments}
We feel indebted a lot  to the MILC collaboration
for using the Asqtad lattice ensembles and MILC codes.
We should thank NERSC for providing us the platform to
download MILC gauge configurations,
and Dr. Massimo Di Pierro for his wonderful Python toolkit.
The author  profoundly thanks Carleton DeTar
for indoctrinating me the necessary theoretical knowledge and
computational skills for this work during my Ph.D study in University of Utah,
and supplying us with the required software.
The author deeply appreciates Paul Kienzle for teaching me
computer skills during my work in NIST, Gaithersburg, USA.
The author heartily thanks Grace Development Team for the use of xmgrace.
We especially thank Eulogio Oset for his enlightening and constructive
comments and corrections.
The author must express my respect to Sasa Prelovsek and Liu chuan
for reading this manuscript and giving some valuable comments.
The author would like to express his gratitude to the Institute of
Nuclear Science and Technology, Sichuan University where the computer resources
and technical support are provided
(in particular Hou Qing,~\footnote{
The numerical calculations of this work are unceasingly
carried out for more than two years.
We should especially thank Prof. Hou qing's continuous encouragements
and comprehensive support,
without his kind and selfless help,
it is not possible for us to launch this work, and
have an opportunity to do it.
}
Zhu An, Ning Liu and Jun Wang).
Numerical calculations for this paper were carried out at AMAX,
CENTOS and HP workstations.
Parts of numerical calculations were conducted
at the Utah Center for High Performance Computing,
University of Utah, during my Ph.D study.

\appendix
\section{Scattering length of $\pi\pi$  in $\chi$PT at NNLO}
\label{app:ChPT NNLO}
In Ref.~\cite{Fu:2011bz}, we provided the compact continuum
three-flavor $\chi$PT form  for the $s$-wave $\pi\pi$ scattering
length for isospin-$0$ at the NLO
in the continuum limit of QCD by constructing from Appendix C in Ref.~\cite{Bijnens:1997vq}.
Here we follow the original derivations and
notations in Refs.~\cite{Yagi:2011jn,
Bijnens:1997vq,Colangelo:2001df,Colangelo:2000jc}
to derive its compact form at the NNLO.

The $\pi\pi$ scattering lengths are provided at the NNLO of $\chi$PT
in Refs.~\cite{Bijnens:1997vq,Colangelo:2001df,Colangelo:2000jc},
and the $s$-wave $\pi\pi$ scattering length in the $I=0$ channel
is described as~\cite{Bijnens:1997vq,Colangelo:2001df,Colangelo:2000jc}
\begin{eqnarray}
m_{\pi}a_{\pi\pi}^{I=0} \hspace{-0.1cm}&=&\hspace{-0.1cm}
\frac{7 m_\pi^2}{32\pi f_\pi^2}  \bigg\{
1 + \frac{x}{7}
[49 + 5\bar{b}_{1}+12\bar{b}_{2}+48\bar{b}_{3}+32\bar{b}_{4}]  \cr
&& \hspace{-0.1cm} + x^2
\left[	
\frac{7045}{63} - \frac{215\pi^{2}}{126} +
10\bar{b}_{1}+24\bar{b}_{2}+96\bar{b}_{3} \right.\cr
&& \left. \hspace{-0.1cm} + 64\bar{b}_{4}
+\frac{192}{7}\bar{b}_{5}
\right]
\bigg\}+{\cal O}(x^4), \label{aChPT}
\end{eqnarray}
where $\displaystyle x = m_\pi^2/(16\pi^2 f_\pi^2) $ is
the chiral expansion parameter
and $\bar{b}_i$'s are dimensionless combinations of the coupling constants
introduced in Refs.~\cite{Bijnens:1997vq,Colangelo:2001df}
to parameterize the pion scattering amplitude.
After some strenuous algebraic manipulations,
we can check that
\begin{eqnarray}
5\bar{b}_{1} + 12\bar{b}_{2} + 48\bar{b}_{3} + 32\bar{b}_{4}
&=& \frac{63}{2}\tilde{L} - \frac{503}{6} x\tilde{L}^2 \cr
&& \hspace{-3.0cm}
-\frac{20}{3}\tilde{l}_1  + \frac{40}{3}\tilde{l}_2
-\frac{5}{2}\tilde{l}_3   + 14\tilde{l}_4 - \frac{63}{2} \cr
&&\hspace{-3.0cm} + x \tilde{L}
\left( -\frac{388}{3}\tilde{l}_1 - \frac{472}{3}\tilde{l}_2
-35\tilde{l}_3  + 154\tilde{l}_4
+ \frac{1405}{12}
\right) \cr
&&\hspace{-3.0cm} + x \bigg(
\frac{80}{3}\tilde{l}_1\tilde{l}_4+
\frac{160}{3}\tilde{l}_2\tilde{l}_4
-15\tilde{l}_3\tilde{l}_4
+35\tilde{l}_4^{\ 2}
-\frac{5}{2}\tilde{l}_3^{\ 2} \cr
&&\hspace{-3.0cm}+ \frac{364}{3}\tilde{l}_1
+\frac{1336}{9}\tilde{l}_2
+\frac{141}{4}\tilde{l}_3 - 126\tilde{l}_4 +
\frac{162719}{432} \cr
&&\hspace{-3.0cm} - \frac{373}{18}\pi^2  + 5\tilde{r}_1+12\tilde{r}_2+
48\tilde{r}_3 + 32\tilde{r}_4 \bigg),  \label{LECb}
\end{eqnarray}
where the low-energy constants $\tilde{l}_i$, $\tilde{r}_i$ are
the quark mass independent couplings
from the subleading orders ${\cal L}_4$, ${\cal L}_6$ of the effective Lagrangian,
respectively~\cite{Colangelo:2001df},
and renormalized at the running scale $\mu$, and
$\displaystyle
\tilde{L}=\ln(\mu^2/m_\pi^2).
$

Inserting Eq.~(\ref{LECb}) into Eq.~(\ref{aChPT}) and
considering expression B.~3 of Ref.~\cite{Colangelo:2001df}
\begin{eqnarray}
\bar{b}_{5}  &=& \frac{85}{72} \tilde{L}^2 +
\tilde{L} \left\{
\frac{7}{8} \tilde{l}_1  +  \frac{107}{72} \tilde{l}_2
-\frac{625}{288} \right\} + \tilde{r}_5 \cr
&&
+\frac{7}{54}\pi^2 - \frac{66029}{20736}
+{\cal O}(x^4), \nonumber
\end{eqnarray}
we recast the result in the order of $x$ as
\begin{eqnarray}
\hspace{-1.2cm} m_{\pi} a_{\pi\pi}^{I=0} &=&
\frac{7m_\pi^2}{32 \pi f_\pi^2}
\bigg\{ 1 + \frac{x}{2} \left( 9\tilde{L}  + l_{a}^0 \right) \cr
&&+ x^2\bigg( \frac{857}{42}\tilde{L}^2
  + l_{b}^0 \tilde{L}+ l_{c}^0 \bigg)
\bigg\}+{\cal O}(x^4), \label{aII}
\end{eqnarray}
with
\begin{eqnarray}
\label{eq:a0_I0_coeff}
l_{a}^0 &=&
\frac{40}{21}\tilde{l}_1  + \frac{80}{21}\tilde{l}_2
-\frac{5}{7}\tilde{l}_3   + 4\tilde{l}_4 + 5 , \cr
l_{b}^0 &=&
\frac{116}{21}\tilde{l}_{1} +
\frac{128}{7}\tilde{l}_{2} -
5\tilde{l}_{3}  - 22 \tilde{l}_{4}
-\frac{3595}{84}, \cr
l_{c}^0 &=& \frac{5}{7}\tilde{r}_1 + \frac{12}{7}\tilde{r}_2+
\frac{48}{7}\tilde{r}_3 + \frac{32}{7}\tilde{r}_4  + \frac{192}{7}\tilde{r}_5 \cr
&&+\frac{80}{21}\tilde{l}_1\tilde{l}_4+
\frac{160}{21}\tilde{l}_2\tilde{l}_4
-\frac{15}{7}\tilde{l}_3\tilde{l}_4
+5\tilde{l}_4^{\ 2} \cr
&&-\frac{5}{14}\tilde{l}_3^{\ 2}
+ \frac{148}{21}\tilde{l}_{1}
+ \frac{232}{21}\tilde{l}_{2}
+ \frac{1}{28}\tilde{l}_{3} \cr
&&+ 10\tilde{l}_{4}
-\frac{17561}{504} + \frac{394}{63}\pi^2 .
\end{eqnarray}
The right-hand side (rhs) of Eq.~(\ref{aChPT}) is
scale independent~\cite{Colangelo:2001df}.
On the whole, the rhs of Eq.~(\ref{aII}) is scale invariant as well.
Therefore, in principle, we can select the running scale $\mu$ stochastically.
However, when fitting our lattice-obtained scattering lengths as a function of $x$,
it is highly desired for us to settle
the fitting parameters quark-mass independent.
As it is done in Ref.~\cite{Yagi:2011jn},
we select $\mu = 4\pi f_{\pi,{\rm phy}}$.
Using this scale, we can see
\begin{eqnarray}
\tilde{L}(\mu=4\pi f_{\pi,{\rm phy}}) & =&
-\ln x -2 x \tilde{l}_4(\mu=4\pi f_{\pi,{\rm phy}})  \cr
&&+ 2x \ln x +{\cal O}(x^2),
\label{ChLn}
\end{eqnarray}
where we exploit the chiral expansion of the pion decay constant
$f_\pi=f_{\pi,{\rm phy}} \{ 1+ x \bar{l}_4 +{\cal O}(x^2) \}$~\cite{Colangelo:2001df}.

Plugging Eq.~(\ref{ChLn}) into Eq.~(\ref{aII}),
and rearranging the result in the order of $x$,
we achieve the $\pi\pi$ scattering length in the $I=0$ channel as
\begin{eqnarray}
\label{eq:fzw_I0}
m_{\pi} a_{\pi\pi}^{I=0}
\hspace{-0.2cm}&=&\hspace{-0.2cm} \frac{7 m_\pi^2}{32\pi f_\pi^2}
\Bigg\{1 +  \frac{ m_\pi^2}{32\pi^2 f_\pi^2} \left[
  -9\ln \left(\frac{ m_\pi^2}{16\pi^2 f_\pi^2}\right)  + l_{a}^0 \right] \cr
\hspace{-0.2cm}&&\hspace{-0.2cm}+ x^2\left[\frac{857}{42}(\ln x)^2
  -(l_{b}^0+9) \ln x+ (l_{c}^0+9\tilde{l}_4) \right]
\Bigg\}\cr
\hspace{-0.2cm}&&\hspace{-0.2cm} +{\cal O}(x^4). \label{aIII}
\end{eqnarray}

The continuum $\chi$PT forms for the $s$-wave $\pi\pi$ scattering
length in the $I=2$ channel $a_{\pi\pi}^{I=2}$ at the NNLO was presented by Yagi et. al.
in Ref.~\cite{Yagi:2011jn}, namely,
\begin{eqnarray}
\label{eq:yagi_I2}
m_{\pi} a_{\pi\pi}^{I=2}
\hspace{-0.2cm}&=&\hspace{-0.2cm} -\frac{m_\pi^2}{16\pi f_\pi^2}
\Bigg\{1 + \frac{ m_\pi^2}{32\pi^2 f_\pi^2} \left[
  3\ln \left(\frac{ m_\pi^2}{16\pi^2 f_\pi^2}\right) + l_{a}^2 \right] \cr
\hspace{-0.2cm}&&\hspace{-0.2cm}+
x^2\left[ -\frac{31}{6}(\ln x)^2 -
(l_{b}^2+3) \ln x + (l_{c}^2+3\tilde{l}_4) \right]
\Bigg\}\cr
\hspace{-0.2cm}&&\hspace{-0.2cm} +{\cal O}(x^4). \label{aIIII}
\end{eqnarray}
with
\begin{eqnarray}
\label{eq:a0_I2_coeff}
l_{a}^2 &=&
 -\frac{8}{3}\tilde{l}_{1} -\frac{16}{3}\tilde{l}_{2} + \tilde{l}_{3} +
 4\tilde{l}_{4} - 1, \cr
l_{b}^2 &=&
{-}\frac{4}{3}\tilde{l}_{1} {-}
	8\tilde{l}_{2} {+}
	\tilde{l}_{3}  {-}
	2\tilde{l}_{4} + \frac{119}{12}, \cr
l_{c}^2 &=& {\frac{1}{2}\tilde{l}_3^{\ 2}-}
		\left(
			\frac{16}{3}\tilde{l}_{1} + \frac{32}{3}\tilde{l}_{2}
			-3\tilde{l}_{3} -5\tilde{l}_{4}
		\right)	\tilde{l}_{4} +
	\frac{4}{3}\tilde{l}_{1}  +
	\frac{16}{3}\tilde{l}_{2} \cr
    &&+
	\frac{7}{4}\tilde{l}_{3} -
	2\tilde{l}_{4} +{\frac{163}{16}}
	- \frac{22}{9}\pi^2 {-} \tilde{r}_1-16 \tilde{r}_4.
\end{eqnarray}

In the above equations, $f_\pi$ is the pion decay constant, which is originally
written as $F_\pi$ (around $92.4$~MeV)~\cite{Bijnens:1997vq,Colangelo:2001df,Colangelo:2000jc}.
In the present work,  $\sqrt{2} F_\pi$ is denoted by $f_\pi$ (about $130$~MeV)
for the convenience of the fitting our lattice data.
Then the above equations are recast as
\begin{eqnarray}
\label{neq:m_pipi_I0}
m_{\pi} a_{\pi\pi}^{I=0}
&=& \frac{7 m_\pi^2}{16\pi f_\pi^2}
\Bigg\{1 - \frac{ m_\pi^2}{16\pi^2 f_\pi^2} \left[
9\ln \frac{ m_\pi^2}{f_\pi^2} - 5 -
l_{\pi\pi}^{I=0} \right] \cr
&&+\frac{m_\pi^4}{64\pi^2 f_\pi^4} \Bigg[ \frac{857}{42}
\left(\ln\frac{ m_\pi^2}{f_\pi^2}\right)^2 \cr
&&+l_{\pi\pi,I=0}^{(2)}
\ln \frac{ m_\pi^2}{f_\pi^2}
+l_{\pi\pi,I=0}^{(3)}  \Bigg]
\Bigg\} , \\
\label{neq:m_pipi_I2}
m_{\pi} a_{\pi\pi}^{I=2}
&=& -\frac{m_\pi^2}{8\pi f_\pi^2}
\Bigg\{1 + \frac{m_\pi^2}{16\pi^2 f_\pi^2} \left[
3 \ln \frac{ m_\pi^2}{f_\pi^2} - 1 -
l_{\pi\pi}^{I=2} \right] \cr
&&+ \frac{m_\pi^4}{64\pi^4 f_\pi^4} \bigg[
-\frac{31}{6} \left( \ln\frac{ m_\pi^2}{f_\pi^2}\right)^2
\cr
&&+l_{\pi\pi,I=2}^{(2)} \ln \frac{ m_\pi^2}{f_\pi^2} +
l_{\pi\pi,I=2}^{(3)} \bigg]
\Bigg\}  ,
\end{eqnarray}
where $l_{\pi\pi}^{(i)}$s
are the combinations of LEC's in $\chi$PT at
a quark-mass independent running scale
since all the LEC's are  independent of quark mass,
therefore, we can regard them as the fitting parameters
in the chiral extrapolation of the $s$-wave $\pi\pi$ scattering
lengths~\cite{Yagi:2011jn}.

From Eqs.~(\ref{eq:a0_I0_coeff}),
(\ref{eq:fzw_I0}), (\ref{eq:yagi_I2}) and (\ref{eq:a0_I2_coeff}),
we can easily get its specific forms of $l_{\pi\pi}^{I=0} $
and $l_{\pi\pi}^{I=2}$, which are related to the Gasser-Leutwyler
coefficients $\tilde{l}_i$ as~\cite{Gasser:1984gg,Bijnens:1997vq}
\begin{eqnarray}
\label{neq:l_pipi_I0}
\hspace{-1.4cm} l_{\pi\pi}^{I=0} &=&
\frac{40}{21}\bar{l}_1+\frac{80}{21}\bar{l}_2-\frac{5}{7}\bar{l}_3+4\bar{l}_4
+ 9\ln\frac{m_\pi^2}{f_{\pi,{\rm phy}}^2},\\
\label{neq:l_pipi_I2}
\hspace{-1.4cm} l_{\pi\pi}^{I=2} &=&
\frac{8}{3}\bar{l}_1+\frac{16}{3}\bar{l}_2-\bar{l}_3-4\bar{l}_4
+ 3\ln\frac{m_\pi^2}{f_{\pi,{\rm phy}}^2}.
\end{eqnarray}
where we consider the equality
$\displaystyle \bar{l}_n = \tilde{l}_n +
\ln(m_\pi^2/\mu^2)$~\cite{Colangelo:2001df}.

These are the forms what we used in our previous work~\cite{Fu:2011bz}.
For the other $l_{\pi\pi}^{(i)}$s, its explicit forms are given
or can be inferred from Eqs.~(\ref{eq:a0_I0_coeff}),
(\ref{eq:fzw_I0}), (\ref{eq:yagi_I2}) and (\ref{eq:a0_I2_coeff}).
It is interesting and important to note that
if we opt the running scale $\mu = f_{\pi,{\rm phy}}$,
we obtain the same expressions.



\begin{thebibliography}{90}
\bibitem{Weinberg:1966kf} S.~Weinberg,
Phys.\ Rev.\ Lett.\  {\bf 17}, 616 (1966).

\bibitem{Bijnens:1997vq} J.~Bijnens, G.~Colangelo, G.~Ecker, J.~Gasser and M.~E.~Sainio,
Nucl.\ Phys.\  B {\bf 508}, 263 (1997) [arXiv:hep-ph/9707291].

\bibitem{Colangelo:2001df} G.~Colangelo, J.~Gasser and H.~Leutwyler,
Nucl.\ Phys.\  B {\bf 603}, 125 (2001) [arXiv:hep-ph/0103088].

\bibitem{Colangelo:2000jc} G.~Colangelo, J.~Gasser and H.~Leutwyler,
Phys.\ Lett.\ B {\bf 488}, 261 (2000)  [hep-ph/0007112].

\bibitem{Roy:1971tc} S.~M.~Roy,
Phys.\ Lett.\  B {\bf 36}, 353 (1971).

\bibitem{Ananthanarayan:2000ht} B.~Ananthanarayan, G.~Colangelo, J.~Gasser and H.~Leutwyler,
Phys.\ Rept.\  {\bf 353}, 207 (2001) [arXiv:hep-ph/0005297].

\bibitem{Zhou:2004ms} Z.~Y.~Zhou, G.~Y.~Qin, P.~Zhang, Z.~Xiao, H.~Q.~Zheng and N.~Wu,
JHEP {\bf 0502}, 043 (2005)  [hep-ph/0406271].

\bibitem{Sasaki:2008sv} K.~Sasaki and N.~Ishizuka,
Phys.\ Rev.\ D {\bf 78}, 014511 (2008).

\bibitem{Guo:2009hi} Z.~H.~Guo and J.~J.~Sanz-Cillero,
Phys.\ Rev.\ D {\bf 79}, 096006 (2009)  [arXiv:0903.0782 [hep-ph]].

\bibitem{Albaladejo:2012te} M.~Albaladejo and J.~A.~Oller,
Phys.\ Rev.\ D {\bf 86}, 034003 (2012)  [arXiv:1205.6606 [hep-ph]].

\bibitem{Pislak:2003sv} S.~Pislak, R.~Appel, G.~S.~Atoyan, B.~Bassalleck, D.~R.~Bergman,
N.~Cheung, S.~Dhawan and H.~Do {\it et al.},
Phys.\ Rev.\ D {\bf 67}, 072004 (2003)  [hep-ex/0301040].

\bibitem{GarciaMartin:2011cn} R.~Garcia-Martin, R.~Kaminski, J.~R.~Pelaez,
 J.~Ruiz de Elvira, F.~J.~Yndurain,
Phys.\ Rev.\  {\bf D 83}, 074004 (2011) [arXiv:1102.2183 [hep-ph]].

\bibitem{Batley:2007zz} J.~R.~Batley {\it et al.}  [NA48/2 Collaboration],
Eur.\ Phys.\ J.\  C {\bf 54}, 411 (2008).

\bibitem{Batley:2010zza} J.~R.~Batley {\it et al.}  [NA48-2 Collaboration],
Eur.\ Phys.\ J.\ C {\bf 70}, 635 (2010).

\bibitem{Bizzeti:2011zza} A.~Bizzeti,
AIP Conf.\ Proc.\  {\bf 1374}, 639 (2011).

\bibitem{Wanke:2011zz} R.~Wanke,
Nucl.\ Phys.\ Proc.\ Suppl.\  {\bf 210-211}, 193 (2011).

\bibitem{Sharpe:1992pp} S.~R.~Sharpe, R.~Gupta and G.~W.~Kilcup,
Nucl.\ Phys.\  B {\bf 383}, 309 (1992).

\bibitem{Kuramashi:1993ka} Y.~Kuramashi, M.~Fukugita, H.~Mino,
M.~Okawa and A.~Ukawa, Phys. Rev. Lett.  {\bf 71} 2387 (1993).

\bibitem{Fukugita:1994ve} M.~Fukugita, Y.~Kuramashi, M.~Okawa, H.~Mino and A.~Ukawa,
Phys.\ Rev.\  D {\bf 52}, 3003 (1995) [arXiv:hep-lat/9501024].

\bibitem{Li:2007ey} X.~Li {\it et al.}  [CLQCD Collaboration],
JHEP {\bf 0706}, 053 (2007)  [hep-lat/0703015].

\bibitem{Gupta:1993rn} R.~Gupta, A.~Patel and S.~R.~Sharpe,
Phys.\ Rev.\  D {\bf 48}, 388 (1993) [arXiv:hep-lat/9301016].

\bibitem{Aoki:2002in} S.~Aoki {\it et al.}  [JLQCD Collaboration],
Phys.\ Rev.\ D {\bf 66}, 077501 (2002) [arXiv:hep-lat/0206011].

\bibitem{Du:2004ib} X.~Du, G.~W.~Meng, C.~Miao and C.~Liu,
Int.\ J.\ Mod.\ Phys.\ A {\bf 19}, 5609 (2004)  [hep-lat/0404017].

\bibitem{Yamazaki:2004qb} T.~Yamazaki {\it et al.}  [CP-PACS Collaboration],
Phys.\ Rev.\  D {\bf 70}, 074513 (2004) [arXiv:hep-lat/0402025].

\bibitem{Beane:2005rj} S.~R.~Beane, P.~R.~Bedaque, K.~Orginos and M.~J.~Savage,
Phys.\ Rev.\  D {\bf 73}, 054503 (2006) [arXiv:hep-lat/0506013].

\bibitem{Beane:2007xs} S.~R.~Beane, T.~C.~Luu, K.~Orginos, A.~Parreno,
M.~J.~Savage, A.~Torok and A.~Walker-Loud,
Phys.\ Rev.\  D {\bf 77}, 014505 (2008) [arXiv:0706.3026 [hep-lat]].

\bibitem{Feng:2009ij} X.~Feng, K.~Jansen and D.~B.~Renner,
Phys.\ Lett.\  B {\bf 684}, 268 (2010) [arXiv:0909.3255 [hep-lat]].

\bibitem{Beane:2011sc} S.~R.~Beane {\it et al.}  [NPLQCD Collaboration],
Phys.\ Rev.\  D {\bf 85}, 034505 (2012) [arXiv:1107.5023 [hep-lat]].



\bibitem{Dudek:2010ew} J.~J.~Dudek, R.~G.~Edwards, M.~J.~Peardon,
D.~G.~Richards and C.~E.~Thomas,
Phys.\ Rev.\ D {\bf 83}, 071504 (2011)  [arXiv:1011.6352 [hep-ph]].

\bibitem{Dudek:2012gj} J.~J.~Dudek, R.~G.~Edwards and C.~E.~Thomas,
Phys.\ Rev.\ D {\bf 86}, 034031 (2012)  [arXiv:1203.6041 [hep-ph]].

\bibitem{Yagi:2011jn} T.~Yagi, S.~Hashimoto, O.~Morimatsu and M.~Ohtani,
arXiv:1108.2970 [hep-lat].

\bibitem{Liu:2009uw} Q.~Liu,
PoS {\bf LAT2009}, 101 (2009) [arXiv:0910.2658 [hep-lat]].

\bibitem{Fu:2011bz} Z.~Fu,
Commun.\ Theor.\ Phys.\  {\bf 57}, 78 (2012) [arXiv:1110.3918 [hep-lat]].

\bibitem{Lepage:1989hd} G. P. Lepage,
in Proceedings of TASI'89 Summer School, edited by T. DeGrand and D. Toussaint (World
Scientific, Singapore, 1990), p. 97.

\bibitem{Maiani:1990ca} L.~Maiani and M.~Testa,
Phys.\ Lett.\ B {\bf 245}, 585 (1990).

\bibitem{Luscher:1986pf} M.~L\"uscher,
Commun.\ Math.\ Phys.\  {\bf 105} (1986) 153.

\bibitem{Luscher:1990ux} M.~L\"uscher,
Nucl.\ Phys.\  B {\bf 354}, 531 (1991).

\bibitem{Luscher:1990ck} M.~Luscher and U.~Wolff,
Nucl.\ Phys.\  B {\bf 339}, 222 (1990).

\bibitem{Rummukainen:1995vs} K.~Rummukainen and S.~A.~Gottlieb,
Nucl.\ Phys.\  B {\bf 450}, 397 (1995) [arXiv:hep-lat/9503028].

\bibitem{Beane:2003da} S.~R.~Beane, P.~F.~Bedaque, A.~Parre\~no and M.~J.~Savage,
Phys.\ Lett.\  B {\bf 585}, 106 (2004) [arXiv:hep-lat/0312004].

\bibitem{Kim:2005gf} C.~h.~Kim, C.~T.~Sachrajda and S.~R.~Sharpe,
Nucl.\ Phys.\  B {\bf 727}, 218 (2005) [arXiv:hep-lat/0507006].

\bibitem{Christ:2005gi} N.~H.~Christ, C.~Kim and T.~Yamazaki,
Phys.\ Rev.\  D {\bf 72}, 114506 (2005) [arXiv:hep-lat/0507009].

\bibitem{Feng:2004ua} X.~Feng, X.~Li and C.~Liu,
Phys.\ Rev.\ D {\bf 70}, 014505 (2004)  [hep-lat/0404001].

\bibitem{Lang:2011mn} C.~B.~Lang, D.~Mohler, S.~Prelovsek, M.~Vidmar,
Phys. Rev. {\bf D  84}, 054503 (2011)  [arXiv:1105.5636 [hep-lat]].

\bibitem{Fu:2011xz} Z.~Fu,
Phys.\ Rev.\  {\bf D 85}, 014506 (2012)  [arXiv:1110.0319 [hep-lat]].

\bibitem{Leskovec:2012gb} L.~Leskovec and S.~Prelovsek,
Phys.\ Rev.\  {\bf D 85}, 114507 (2012)  [arXiv:1202.2145 [hep-lat]].

\bibitem{Doring:2012eu} M.~Doring, U.~G.~Meissner, E.~Oset and A.~Rusetsky,
Eur.\ Phys.\ J.\ A {\bf 48}, 114 (2012)  [arXiv:1205.4838 [hep-lat]].

\bibitem{Briceno:2012rv} R.~A.~Briceno and Z.~Davoudi,
arXiv:1212.3398 [hep-lat].

\bibitem{Guo:2012hv} P.~Guo, J.~Dudek, R.~Edwards and A.~P.~Szczepaniak,
arXiv:1211.0929 [hep-lat].

\bibitem{Bernard:2001av} C.~W.~Bernard {\it et al.},
Phys.\ Rev.\  D {\bf 64}, 054506 (2001) [arXiv:hep-lat/0104002].

\bibitem{Aubin:2004wf} C.~Aubin {\it et al.},
Phys.\ Rev.\  D {\bf 70}, 094505 (2004) [arXiv:hep-lat/0402030].

\bibitem{stag_fermion}
K.~Orginos and D.~Toussaint,
Phys.\ Rev.\  D {\bf 59}, 014501 (1998) [arXiv:hep-lat/9805009];
K.~Orginos, D.~Toussaint and R.~L.~Sugar,
Phys.\ Rev.\  D {\bf 60}, 054503 (1999) [arXiv:hep-lat/9903032];
T.~Blum {\it et al.},
Phys.\ Rev.\ D {\bf 55}, R1133 (1997) [arXiv:hep-lat/9609036];
J.~F.~Laga\"e and D.~K.~Sinclair,
Phys.\ Rev.\ D {\bf 59}, 014511 (1998) [arXiv:hep-lat/9806014];
G.~P.~Lepage,
Phys.\ Rev.\ D {\bf 59}, 074502 (1999) [arXiv:hep-lat/9809157];
C.~W.~Bernard {\it et al.}  [MILC Collaboration],
Phys.\ Rev.\ D {\bf 61}, 111502(R) (2000)[arXiv:hep-lat/9912018].

\bibitem{Fu:2011wc} Z.~Fu,
Phys.\ Rev.\  D {\bf 85}, 074501 (2012) [arXiv:1110.1422 [hep-lat]].

\bibitem{Fu:2012gf} Z.~Fu,
JHEP {\bf 1207}, 142 (2012)  [arXiv:1202.5834 [hep-lat]].

\bibitem{Lang:2012sv}
C.~B.~Lang, L.~Leskovec, D.~Mohler and S.~Prelovsek,
Phys.\ Rev.\ D {\bf 86}, 054508 (2012)  [arXiv:1207.3204 [hep-lat]].

\bibitem{Blum:2011pu} T.~Blum {\it et al.},
Phys.\ Rev.\  D {\bf 84}, 114503 (2011) [arXiv:1106.2714 [hep-lat]].

\bibitem{Fu:2011xw} Z.~Fu,
JHEP {\bf 1201}, 017 (2012)  [arXiv:1110.5975 [hep-lat]].

\bibitem{Fu:2012tj}Z.~Fu and K.~Fu,
Phys.\ Rev.\  D {\bf 86}, 094507 (2012) [ arXiv:1209.0350 [hep-lat]].

\bibitem{Fu:2012ng} Z.~Fu,
Eur.\ Phys.\ J.\ C \ {\bf 72}, 2159 (2012)  [arXiv:1201.3708 [hep-lat]].

\bibitem{Fu:2012jb} Z.~Fu,
arXiv:1210.5185 [hep-lat].

\bibitem{DeGrand:2006zz} T.~DeGrand and C.~E.~Detar,
{\it Lattice methods for quantum chromodynamics,}
New Jersey, USA: World Scientific (2006) 345 p.

\bibitem{Durr:2004as} S.~D\"urr, C.~Hoelbling and U.~Wenger,
Phys.\ Rev.\ D {\bf 70}, 094502 (2004);
C.~Bernard, Phys.\ Rev.\ D {\bf 73}, 114503 (2006);
C.~Bernard, M.~Golterman, Y.~Shamir and S.~R.~Sharpe,
Phys.\ Lett.\  B {\bf 649}, 235 (2007);
C.~Bernard, M.~Golterman and Y.~Shamir,
Phys.\ Rev.\  D {\bf 73}, 114511 (2006);
M.~Creutz, Phys.\ Lett.\  B {\bf 649}, 241 (2007);
S.~D\"urr and C.~Hoelbling, Phys.\ Rev.\  D {\bf 74}, 014513 (2006);
A.~Hasenfratz and R.~Hoffmann, Phys.\ Rev.\  D {\bf 74}, 014511 (2006).

\bibitem{Albaladejo:2008qa} M.~Albaladejo and J.~A.~Oller,
Phys.\ Rev.\ Lett.\  {\bf 101}, 252002 (2008)  [arXiv:0801.4929 [hep-ph]].

\bibitem{Hanhart:2012wi} C.~Hanhart,
Phys.\ Lett.\ B {\bf 715}, 170 (2012)  [arXiv:1203.6839 [hep-ph]].

\bibitem{Doring:2011vk} M.~Doring, U.~-G.~Meissner, E.~Oset and A.~Rusetsky,
Eur.\ Phys.\ J.\ A {\bf 47}, 139 (2011)  [arXiv:1107.3988 [hep-lat]].

\bibitem{Barkai:1985gy} D.~Barkai, K.~J.~M.~Moriarty and C.~Rebbi,
Phys.\ Lett.\  B {\bf 156}, 385 (1985);
A.~Mihaly, H.~R.~Fiebig, H.~Markum and K.~Rabitsch,
Phys.\ Rev.\  D {\bf 55}, 3077 (1997);
A.~Mih\'aly,
Ph.D thesis, Lajos Kossuth University, Debrecen, 1998.

\bibitem{Umeda:2007hy} T.~Umeda,
Phys.\ Rev.\  D {\bf 75}, 094502 (2007) [arXiv:hep-lat/0701005].

\bibitem{Aubin:2004fs} C.~Aubin {\it et al.}  [MILC Collaboration],
Phys.\ Rev.\  D {\bf 70}, 114501 (2004) [arXiv:hep-lat/0407028].

\bibitem{Alford:1995hw} M.~G.~Alford, W.~Dimm, G.~P.~Lepage,
G.~Hockney and P.~B.~Mackenzie,
Phys.\ Lett.\  B {\bf 361}, 87 (1995) [arXiv:hep-lat/9507010].


\bibitem{Bazavov:2009bb} A.~Bazavov {\it et al.},
Rev.\ Mod.\ Phys.\  {\bf 82}, 1349 (2010) [arXiv:0903.3598 [hep-lat]].

\bibitem{Kaplan:1992bt}
D.~B.~Kaplan, Phys. Lett. B {\bf 288}, 342 (1992);
Y.~Shamir, Nucl. Phys. {\bf B406}, 90 (1993);
Y.~Shamir, Phys. Rev. D {\bf 59}, 054506 (1999).

\bibitem{Hasenfratz:2001hp}
A.~Hasenfratz and F.~Knechtli, Phys. Rev.  D {\bf 64}, 034504 (2001);
T.~A.~DeGrand, A.~Hasenfratz and T.~G.~Kovacs, Phys. Rev.  D {\bf 67}, 054501 (2003);
T.~DeGrand, Phys. Rev. D {\bf 69}, 014504 (2004);
S.~Durr, C.~Hoelbling and U.~Wenger, Phys. Rev.  D {\bf 70}, 094502 (2004).

\bibitem{Renner:2004ck} D.~B.~Renner {\it et al.}  [LHP Collaboration],
Nucl. Phys. Proc. Suppl.  {\bf 140}, 255 (2005);
R.~G.~Edwards {\it et al.} [LHPC Collaboration],
PoS {\bf LAT2005}, 056 (2006).

\bibitem{Lepage:1992xa} G.~P.~Lepage and P.~B.~Mackenzie,
Phys.\ Rev.\  D {\bf 48}, 2250 (1993) [arXiv:hep-lat/9209022].

\bibitem{Bernard:2000gd} C.~W.~Bernard, T.~Burch, K.~Orginos, D.~Toussaint,
T.~A.~DeGrand, C.~E.~DeTar, S.~A.~Gottlieb and U.~M.~Heller {\it et al.},
Phys.\ Rev.\ D {\bf 62}, 034503 (2000)  [hep-lat/0002028].

\bibitem{Bernard:2007ps} C.~Bernard {\it et al.},
PoS {\bf LAT2007}, 090 (2007) [arXiv:0710.1118 [hep-lat]].

\bibitem{Fu:2011zzh} Z.~W.~Fu,
Chin.\ Phys.\ Lett.\  {\bf 28} (2011) 081202;
Z.~W.~Fu and C.~DeTar,
Chin. Phys. C {\bf 35} 896 (2011);
C.~Bernard, C.~E.~DeTar, Z.~Fu and S.~Prelovsek,
Phys.\ Rev.\  D {\bf 76}, 094504 (2007) [arXiv:0707.2402 [hep-lat]];
Z.~Fu,
2006, UMI-32-34073  [arXiv:1103.1541 [hep-lat]].


\bibitem{Nagata:2008wk} J.~Nagata, S.~Muroya, A.~Nakamura,
Phys.\ Rev.\ {\bf C 80}, 045203 (2009) [arXiv:0812.1753 [hep-lat]].

\bibitem{Prelovsek:2010kg} S.~Prelovsek, T.~Draper, C.~B.~Lang, M.~Limmer,
K.~-F.~Liu, N.~Mathur and D.~Mohler,
Phys.\ Rev.\ D {\bf 82}, 094507 (2010)  [arXiv:1005.0948 [hep-lat]].


\bibitem{Lepage:2001ym} G.~P.~Lepage, B.~Clark, C.~T.~H.~Davies,
K.~Hornbostel, P.~B.~Mackenzie, C.~Morningstar and H.~Trottier,
Nucl.\ Phys.\ Proc.\ Suppl.\  {\bf 106}, 12 (2002) [arXiv:hep-lat/0110175];
C.~Morningstar,
Nucl.\ Phys.\ Proc.\ Suppl.\  {\bf 109A}, 185 (2002) [arXiv:hep-lat/0112023];
M.~Wingate, J.~Shigemitsu, C.~T.~H.~Davies,
G.~P.~Lepage and H.~D.~Trottier,

\bibitem{Press:1992zz} W.~H.~Press, S.~A.~Teukolsky,
W.~T.~Vetterling and B.~P.~Flannery,
{\it Numerical recipes in C (2nd ed.): the art of scientific computing,}
New York, NY, USA: Cambridge University Press, 1992.

\bibitem{Waltraud:2008} W. Huyer and A. Neumaier,
ACM Transactions on Mathematical Software (TOMS) Volume 35 Issue 2,
(July 2008) Article No. 9;
matlab implementation: http://www.mat.univie.ac.at/\~{}neum/software/snobfit/;
python: http://reflectometry.org/danse/docs/snobfit/.


\bibitem{Bedaque:2006yi} P.~F.~Bedaque, I.~Sato and A.~Walker-Loud,
Phys.\ Rev.\  D {\bf 73}, 074501 (2006)[arXiv:hep-lat/0601033].

\bibitem{Sasaki:2010zz} K.~Sasaki, N.~Ishizuka, T.~Yamazaki and M.~Oka
[PACS-CS Collaboration],
Prog.\ Theor.\ Phys.\ Suppl.\  {\bf 186} (2010) 187.

\bibitem{Gasser:1984gg} J.~Gasser and H.~Leutwyler,
Nucl.\ Phys.\  B {\bf 250}, 465 (1985).

\bibitem{Beringer:1900zz} J.~Beringer {\it et al.}
[Particle Data Group Collaboration],
Phys.\ Rev.\ D {\bf 86}, 010001 (2012).

\bibitem{Yndurain:2007qm}  R.~Garcia-Martin, J.~R.~Pelaez and F.~J.~Yndurain,
Phys.\ Rev.\ D {\bf 76}, 074034 (2007)  [hep-ph/0701025].

\bibitem{Caprini:2008fc} I.~Caprini,
Phys.\ Rev.\ D {\bf 77}, 114019 (2008)

\bibitem{Colangelo:2010et} G.~Colangelo, S.~Durr, A.~Juttner, L.~Lellouch,
H.~Leutwyler, V.~Lubicz, S.~Necco and C.~T.~Sachrajda {\it et al.},
Eur.\ Phys.\ J.\ C {\bf 71}, 1695 (2011)  [arXiv:1011.4408 [hep-lat]].

\bibitem{Bazavov:2009fk} A.~Bazavov {\it et al.}  [MILC Collaboration],
PoS CD {\bf 09}, 007 (2009)  [arXiv:0910.2966 [hep-ph]].

\bibitem{Savage:2011xk} M.~J.~Savage,
Prog.\ Part.\ Nucl.\ Phys.\  {\bf 67}, 140 (2012)  [arXiv:1110.5943 [nucl-th]].

\bibitem{Oller:1998hw}
J.~A.~Oller, E.~Oset and J.~R.~Pelaez,
Phys.\ Rev.\ D {\bf 59}, 074001 (1999) [hep-ph/9804209];
F.~Guerrero and J.~A.~Oller,
Nucl.\ Phys.\ B {\bf 537}, 459 (1999)  [hep-ph/9805334].

\bibitem{multi_channel}
S.~He, X.~Feng and C.~Liu,
JHEP {\bf 0507}, 011 (2005)  [hep-lat/0504019];
M.~Doring and U.~G.~Meissner,
JHEP {\bf 1201}, 009 (2012)  [arXiv:1111.0616 [hep-lat]];
M.~T.~Hansen and S.~R.~Sharpe,
Phys.\ Rev.\ D {\bf 86}, 016007 (2012)  [arXiv:1204.0826 [hep-lat]];
C.~Liu, X.~Feng and S.~He,
Int.\ J.\ Mod.\ Phys.\ A {\bf 21}, 847 (2006)  [hep-lat/0508022];
N.~Li and C.~Liu,
Phys.\ Rev.\ D {\bf 87}, 014502 (2013)  [arXiv:1209.2201 [hep-lat]].


\end{thebibliography}
\end{document}